\documentclass{aa}  

\usepackage{graphicx}
\usepackage{txfonts}
\usepackage[]{hyperref}
\usepackage{breakurl}
\usepackage{lscape}
\usepackage{subcaption}
\usepackage{placeins}

\renewcommand{\linenumbers}{} 
\renewcommand{\nolinenumbers}{} 

\nolinenumbers

\begin{document}

   \title{Discovery of carbon monoxide emission from five debris disks around young A-type stars}

   \subtitle{}

   \author{A. Mo\'or
           \inst{1}
	   \and
	   P. \'Abrah\'am
           \inst{1,2,3}
	   \and
	   \'A. K\'osp\'al
	   \inst{1,2,4}
	   \and
	   G. Cataldi
	   \inst{5}
	   \and
	   A. M. Hughes
	   \inst{6}
	   \and
	   S. Marino
	   \inst{7}
	  \and
	   Q. Kral
	   \inst{8} 
          \and
	   J. Milli
	   \inst{9}
	   \and
	   N. Pawellek
	   \inst{3,1}
          }

   \institute{HUN-REN Research Centre for Astronomy and Earth Sciences, Konkoly Observatory, MTA Centre of Excellence, \\ 
              Konkoly-Thege Mikl\'os \'ut 15-17, H-1121 Budapest, Hungary \\
              \email{moor.attila@csfk.org}
	 \and 
	     ELTE E\"otv\"os Lor\'and University, Institute of Physics and Astronomy, P\'azm\'any P\'eter s\'et\'any 1/A, H-1117 Budapest, Hungary\\     
         \and
             Department of Astrophysics, University of Vienna, T\"urkenschanzstra{\ss}e 17, 1180, Vienna, Austria\\
	 \and
	     Max-Planck-Institut f\"ur Astronomie, K\"onigstuhl 17, D-69117 Heidelberg, Germany\\
	 \and 
	     National Astronomical Observatory of Japan, Osawa 2-21-1, Mitaka, Tokyo 181-8588, Japan \\
	 \and     
	     Astronomy Department and Van Vleck Observatory, Wesleyan University, 96 Foss Hill Drive, Middletown, CT 06459, USA \\
	 \and 
	     Department of Physics and Astronomy, University of Exeter, Stocker Road, Exeter, EX4 4QL, UK \\
	 \and LIRA, Observatoire de Paris, Universit\'e PSL, Sorbonne Universit\'e, Universit\'e Paris Cit\'e, CY Cergy Paris Universit\'e, \\ 
	     CNRS, 92190 Meudon, France \\  
	 \and 
	     Univ. Grenoble Alpes, CNRS, IPAG, F-38000 Grenoble, France \\         
             }

   \date{Received September 15, 1996; accepted March 16, 1997}

 
  \abstract
   {Over the past fifteen years, surveys mainly at millimeter wavelengths have led to the discovery of $\sim$20 gas-bearing 
debris disks, most of them surrounding young intermediate-mass stars. Exploring the properties and origin of this gas could 
be fundamental to better understanding the transition between the protoplanetary and debris disk phases, the evolution of icy 
planetesimal belts, and the formation of planetary atmospheres.}
   {To expand the list of known CO-bearing debris disks and to improve our knowledge of the environmental conditions 
under which they can form, we targeted twelve dust-rich debris disks around young ($<$50\,Myr) intermediate-mass stars.}
   {Using the ALMA 12-m Array we performed millimeter continuum and CO line observations to search for dust and gas and 
to measure their quantity and spatial distribution.}
   {We discovered CO gas in five disks. Two of them have a low CO content of a few times 10$^{-5}$\,M$_\oplus$, similar 
to that of $\beta$\,Pic. The other three disks, however, are CO-rich with $M_\mathrm{CO}>10^{-3}$\,M$_\oplus$. By combining 
our results with those of other studies we concluded, in agreement with previous findings, that the detection rate of CO gas 
is significantly higher for disks around stars with $6.5~L_\odot<L_*<21.9~L_\odot$ ($\sim$A8--A0 spectral type) than for disks 
around less luminous stars ($0.18~L_\odot<L_*<6.4~L_\odot$, K7--A9).
A comparison of the measured CO masses and the estimated mass loss rates of solids in disks with low CO content ($<$10$^{-4}$\,M$_\oplus$) 
suggests that collisions may play a role in CO gas production in such systems. Interestingly, however, the estimated mass loss 
rates of CO-rich debris disks are not higher than those of systems with low CO content. In light of this finding, we speculate 
on what could lead to the formation of CO-rich debris disks.
}
   {}


   \maketitle
%

\section{Introduction}
Circumstellar debris dust rings, identified mainly based on their excess emission at infrared (IR) wavelengths, are common 
components of exoplanetary systems around main-sequence stars \citep{hughes2018}. Their presence indicates that the system 
contains a belt or belts of planetesimals, whose erosion through mutual collisions, continuously replenishes the dust particles 
that are removed on short timescales by interactions with radiative forces \citep{wyatt2008,krivov2010}.
Planetesimals are the remnants of the epoch of planet formation that took place in gas-rich, massive protoplanetary disks in 
an earlier stage of evolution. If formed in a cold enough region far from the star, these small bodies can contain significant 
amounts of ice alongside rocky material. This means that the erosion processes in the debris disk phase can release not only 
dust particles, but also gases \citep{kral2017}. 

While the study of the dust component has been at the forefront of research since the discovery of the first debris disks 
\citep{aumann1984}, more extensive studies of the potential gaseous material were delayed until the 2010s due to the lack of 
sufficiently sensitive instruments, and were given a major boost mainly by the inauguration of the ALMA interferometer. While 
before 2010 only one debris disk with millimeter CO emission was identified \citep[49\,Cet,][]{zuckerman1995}, the number of 
such systems has increased to $\sim$20 in the last 15 years 
\citep{dent2014,moor2011b,moor2015b,moor2017,greaves2016,lieman-sifry2016,marino2016,matra2017a,matra2019,kral2020,lovell2021,rebollido2022}.
Additional atomic gas components in emission have also been detected in some of these disks 
\citep{riviere-marichalar2012,riviere-marichalar2014,roberge2013,cataldi2018,cataldi2020,cataldi2023,kral2019,worthen2024,samland2025}. 
As high angular resolution ALMA images show, in most of these systems the bulk of the detected CO gas lies at radial distances 
of tens of au, more or less co-located with the outer planetesimal belts. The estimated CO masses show a large variety and 
their distribution implies two well distinguishable groups within the sample. Some of them contain an unexpectedly large amount 
of CO gas of $>$10$^{-3}$\,M$_\oplus$, the most gas-rich ones have CO content comparable to that of protoplanetary 
disks \citep{moor2017}. These so-called CO-rich disks are dust-rich as well 
($f_\mathrm{d} = L_\mathrm{disk}/L_\mathrm{*} > 5\times10^{-4}$), and all are found around A-type stars younger than 50\,Myr. For 
the rest of the sample, the estimated CO gas content is $<$10$^{-4}$\,M$_\oplus$. In addition to A-type stars, this subsample 
also includes later spectral type host stars, and although these systems are also typically young, two of them, Fomalhaut and 
$\eta$\,Crv, are substantially older than 50\,Myr \citep{marino2018,matra2017a}.  

The detection of gas in exosolar systems is not limited to the outer planetary regions. Hot gas, derived from the sublimation of 
minor bodies approaching their host stars, has been detected as various absorption features in the spectra of dozens of stars 
including some that also host colder CO gas  
\citep[e.g.][]{ferlet1987,welsh2013,kiefer2014,iglesias2018,rebollido2020}. This phenomenon is also most commonly associated 
with young early-type (A and B) stars \citep{rebollido2020,bendahan-west2025}.

It is not surprising that debris disks contain not only secondary dust but also gas, as planetesimals heated by mutual
collisions, stellar irradiation, and possibly decay of radioactive nuclides can outgas their volatiles 
\citep{zuckerman2012,kral2017,davidsson2021,bonsor2023}. Since the tenuous, optically thin dust material of these disks 
does not provide significant shielding, the stellar and interstellar UV photons reach the released molecules without any 
hindrance, causing their photodissociation. This limits the lifetime of even the most resistant molecule, CO, to $\sim$120\,yr 
\citep[even considering only the interstellar UV field,][]{visser2009,heays2017}. Despite this very short lifetime, the 
amount of gas observed in low CO content disks ($<$10$^{-4}$\,M$_\oplus$) can persist for a very long time, because 
the secondary gas production mechanisms mentioned above are able to ensure a sufficient rate of continuous replenishment 
\citep{kral2017,marino2020}. 

However, this is not the case for CO-rich systems, whose long-term survival - over 
millions or even tens of millions of years - would require unfeasibly high gas production rates. Their existence can be 
explained if there is some material in these systems that effectively shields the CO molecules from UV photons, thereby 
significantly prolonging their photodissociation lifetime. There are two proposals for the shielding, which can be 
attributed to very different gas origins. According to the shielded secondary gas scenario, assuming sufficiently high, but still 
realistic CO production rates, neutral carbon atoms produced by the photodissociation of mainly CO and CO$_2$ can become 
optically thick for the UV photons that threaten CO molecules. The increased lifetime allows the pile up of molecules 
to such an extent that, after a while, CO self-shielding can also begin to manifest, further reducing the photodissociation
rate and thereby increasing the efficiency of the accumulation process \citep{kral2019,moor2019,marino2020}. 

However, because of the young age of CO-rich systems, it cannot be excluded that their gas disk is predominantly composed of 
long-lived primordial gas, in which case the unseen H$_2$ molecules can provide efficient shielding for CO 
\citep{kospal2013,pericaud2017,difolco2020,nakatani2021}. Since dust is considered to be of secondary origin, such disks would 
be hybrid in nature. Indeed, recent theoretical results suggest that primordial disks, especially those around A stars, which 
are depleted in small dust particles (making far-UV photoevaporation less efficient), may persist for tens of millions of years 
\citep{nakatani2021,nakatani2023,ooyama2024}. For the sake of completeness, we note that although the gas disk in this model 
is dominantly primordial, the observed CO could be partly or even largely of second generation origin.

The primordial and secondary gas scenarios predict very different gas compositions for CO-rich disks, with H$_2$ molecules 
being the dominant component in the former case, and CO molecules together with photodissociation end products 
(H, C, C$^{+}$, and O) of the most common molecules released from icy bodies (mainly H$_2$O, CO, and CO$_2$, in analogy to 
Solar System comets) being the main constituents in the latter case. This may have interesting implications for the evolution of 
the planets in the system as they can accrete this gas material leading to significant changes in the composition of their 
atmospheres \citep{kral2020b}.

At present, we do not know for sure which scenario might prevail for CO-rich disks, or what the ratio of primordial-to-secondary 
systems is within this sample. Constraining the amount of H$_2$ could clarify the situation, as little of this component is 
expected in secondary gas disks, but due to the difficulty of detecting this molecule in the typically low temperature 
environment of debris disks, this is not a viable option. Recent observations of \ion{C}{I} in CO-rich debris disks have 
revealed less carbon gas than current secondary models would require for sufficient shielding \citep{cataldi2023,brennan2024}. 
Furthermore, it has been suggested that if the vertical mixing of the gas is sufficiently strong, the shielding efficiency 
of carbon gas may be substantially weaker than suggested by the model \citep{cataldi2020,marino2022}. These do not rule out 
a secondary origin, but indicate that the current model needs refinement. 
The development of the hybrid disk model is ongoing, but further improvements, for example the inclusion of thermochemistry, is 
needed to allow a direct comparison with observations \citep{ooyama2024}.

Despite the observational and theoretical results achieved so far, we still do not clearly understand all aspects of 
the CO-rich, or alternatively the shielded, debris disk phenomenon. For further in-depth studies and to refine the proposed 
formation scenarios, it would be important to clarify the type of systems in which these disks can form, and under what 
environmental conditions. This requires not only detailed studies of known systems, but also the search for
additional gas-rich debris disks in systems that may represent different environments than those we have explored so far.
For example, the current sample may be biased, as previous surveys have typically focused on systems where earlier far-IR 
measurements have revealed the presence of cold ($\lesssim$130--140\,K) debris disks \citep[e.g.][]{moor2017}. 
As a result, although population synthesis of secondary gas around A-type stars predicts that warmer planetesimal rings 
at 10--30\,au can also form CO-rich shielded disks \citep{marino2020}, currently this 
cannot be tested due to the sparse observations of such targets. 

With the motivation to discover new CO-rich debris disks and to better explore the possible environmental conditions 
favorable for their formation, we initiated an ALMA mini-survey to observe a sample of 12 carefully selected, dust-rich 
($f_\mathrm{d} > 5\times10^{-4}$), 
young debris disks (Sect.~\ref{sec:sampleselection}). Here we present the outcomes of the observations, in which we discovered 
five new CO-bearing debris disks including three CO-rich systems. After presenting the observations 
(Sect.~\ref{sec:obsdanddatareduction}) and analysing the continuum and line data (Sect.~\ref{sec:continuum}--\ref{sec:co}), 
we merge our results with existing literature data to investigate the detection rate of CO gas in dust-rich young debris 
disks and to compare the newly identified CO-bearing disks with the previously known ones.
Using the new, unified sample, we investigate the possible origin of the CO gas by confronting the results with the 
predictions of both secondary and primary models (Sect.~\ref{sec:discussion}). For those systems from our sample 
whose continuum emission was spatially resolved, the position of the planetesimal belt can be estimated.
Using this information we discuss the possible mechanisms for the dynamic excitation of these disks and 
whether the result of this analysis could indicate the presence of larger planet(s) in these
systems (Sect.~\ref{sec:stirring}). The paper is closed with a summary of our results and conclusions. 

\section{Sample selection} \label{sec:sampleselection}
In our survey, we focus on dust-rich debris disks around intermediate-mass stars with spectral types between F1 and B9.5. 
The choice of the lower limit of the spectral range was motivated by the fact that while previous observations 
well established that stars later than F1-type are unlikely to possess CO-rich debris disks \citep{lieman-sifry2016,moor2017}, 
the F1--A7 spectral range has been poorly sampled, making the lower threshold for the existence of such disks unclear. Candidates 
were selected using data from the Gaia~DR2 catalog \citep{brown2018,lindegren2018}. The spectral type was estimated based on 
the $B_\mathrm{p}-R_\mathrm{p}$ color index, using the updated online 
database\footnote{\url{https://www.pas.rochester.edu/~emamajek/EEM_dwarf_UBVIJHK_colors_Teff.txt}} 
of stellar parameters of dwarf stars \citep{pecaut2013}. The $B_\mathrm{p}-R_\mathrm{p}$ colors were dereddened using data 
from the \texttt{Stilism} 3D database\footnote{\url{https://stilism.obspm.fr}} \citep{capitanio2017,lallement2018}. 

We selected stars located within 200\,pc, thus extending the outer range with respect to previous surveys, which typically 
focused on objects within 150\,pc \citep{moor2017}. To ensure good observability from the ALMA site, we constrained declination 
between $-$78{\degr} and $+$32{\degr} and we removed those targets for which ALMA observations of sufficient quality are already 
available, and also those that are classified in the literature as Herbig\,Ae systems with protoplanetary disks.
To identify systems showing excess at mid- and/or far-infrared (far-IR) wavelengths we used data from the point source 
catalogs of the {\sl IRAS}, {\sl AKARI}, {\sl WISE}, {\sl Spitzer}, and {\sl Herschel} space telescopes 
\citep{moshir1990,ishihara2010,yamamura2010,wright2010,herschel}. 
We fitted the observed excess by one or two modified blackbody components to derive the characteristic dust temperatures 
and fractional luminosities of the disks. In contrast to previous strategies, no criteria for dust temperature were 
applied in the target selection, but we restricted the sample to disks with $f_\mathrm{d}$ $>$5$\times10^{-4}$, as this proved 
to be a good indicator of the presence of CO gas \citep[e.g.,][]{moor2017}. 
After these steps we were left with a list of 28 dust-rich debris disks.

In the framework of our ALMA project (2021.1.01487.S, PI: A. Mo\'or) observations were performed for 12 of the 28 
targets\footnote{Since our ALMA program received a C-grade priority, only a fraction of the requested observations 
were ultimately carried out.}. 
Table~\ref{tab:targets} lists their main parameters including the fundamental stellar properties as well as the ages of 
these 12 systems. Appendix~\ref{sec:props} describes in more detail how these parameters were obtained. The sample includes 
11 A-type and 1 F-type stars, whose distances range between 93 and 188\,pc. The ages of the selected systems range from 
2.5 to 44\,Myr. For 11 of the targets, there were already previous indications in the literature regarding their IR 
excess (see Table~\ref{tab:targets}), but the disk of HD\,152989 is a new discovery.

\begin{table*}                                                                  
\setlength{\tabcolsep}{1.6mm}                                             
\begin{center}                                                                                                                                       
\caption{Overview of the sample.  
 \label{tab:targets} }
\begin{tabular}{lccccccccccc}                                                     
\hline\hline
Target name & SpT  & Distance               &  $v_\mathrm{rad}$   &  Group   & $T_{\rm eff}$ & $L_*$       & $M_*$       &  Age     &  Mult.  &  Ref. & Label \\
            &      &  (pc)                  &     (km~s$^{-1}$)   &          &   (K)         & (L$_\odot$) & (M$_\odot$) & (Myr)    &         &       &       \\             
\hline
HD\,9985    &  A2  & 158.2$^{+1.0}_{-1.0}$ & $+$10.2$\pm$3.7 &  \ldots       &  8900$\pm$200 &  14.18$\pm$0.61  & 1.88$\pm$0.04          &  44$_{-16}^{+31}$    &  \ldots &  3 &   1 \\
HD\,31305   &  A3  & 142.6$^{+2.6}_{-2.7}$ & $-2.1\pm3.3$    & TAU           &  8600$\pm$170 &  16.95$\pm$0.64	& 1.97$_{-0.07}^{+0.06}$ &  2.5$_{-1.5}^{+3.8}$	&   Y	  &  1 & \ldots \\
HD\,112532  &  A2  & 119.5$^{+0.4}_{-0.4}$ & $+12.4\pm4.4$   & LCC           &  8850$\pm$200 &  13.87$\pm$0.62	& 1.87$\pm$0.04          & 15$\pm$3             &  \ldots &  2 &  2  \\
HD\,131960  &  F1  & 146.6$^{+0.5}_{-0.5}$ & $+$2.58$\pm$1.47& UCL           &  7000$\pm$150 &   4.22$\pm$0.18  & 1.41$\pm$0.04          & 16$\pm$2             &  \ldots &  2 &  \ldots  \\
HD\,141960  &  A9  & 142.5$^{+0.5}_{-0.4}$ & $+$0.90$\pm$0.72&  US           &  7400$\pm$130 &   6.46$\pm$0.23  & 1.57$\pm$0.04          & 10$\pm$3             &  \ldots &  2 & 3   \\
HD\,144277  &  A3  & 143.4$^{+0.6}_{-0.7}$ & $+$1.72$\pm$0.49& UCL           &  8550$\pm$170 &  11.83$\pm$0.47  & 1.80$\pm$0.05          & 16$\pm$2             &   Y	  &  2 &  \ldots  \\
HD\,145101  &  A6  & 137.7$^{+0.8}_{-0.9}$ & $-$5.07$\pm$0.35& US            &  7950$\pm$200 &  18.58$\pm$0.68  & 1.89$_{-0.05}^{+0.14}$ & 6.7$\pm$0.5          &  \ldots &  2 & 4   \\
HD\,152989  &  A6  & 114.9$^{+0.4}_{-0.5}$ & $-$7.49$\pm$0.80& UCL           &  7950$\pm$150 &   7.21$\pm$0.22  & 1.59$_{-0.03}^{+0.02}$ & 16$\pm$2             &   Y	  &  4 & 5   \\
HD\,155853  &  A4  & 136.1$^{+0.7}_{-0.6}$ & $+$0.39$\pm$1.51& UCL           &  8250$\pm$150 &  11.24$\pm$0.41  & 1.79$_{-0.04}^{+0.03}$ & 16$\pm$2             &  \ldots &  2 & 6   \\
HD\,159595  &  A3  &  93.0$^{+0.3}_{-0.2}$ & $-$9.08$\pm$0.65& BPMG?         &  8450$\pm$140 &  10.58$\pm$0.35  & 1.75$\pm$0.03          & 21$_{-5}^{+4}$       &  \ldots &  2 & 7    \\
HD\,170116  &  A4  & 187.8$^{+0.8}_{-0.8}$ & $-$17.83$\pm$1.33& \ldots       &  8250$\pm$250 &  11.63$\pm$0.60  & 1.79$\pm$0.04          & 31$_{-7}^{+9}$       &  \ldots &  3 & 8    \\
HD\,176497  &  A4  & 153.3$^{+0.6}_{-0.6}$ & $-$5.14$\pm$1.64& UCRA          &  8200$\pm$170 &   8.90$\pm$0.37  & 1.68$\pm$0.03          & 10                   &  \ldots &  2 & 9   \\
\hline
\end{tabular}
\tablefoot{Distance estimates are taken from \citet{bailerjones2021} using their geometric approach that is based only on 
the Gaia~EDR3 parallaxes. Where available, the listed heliocentric radial velocity ($v_\mathrm{rad}$) data were taken from the 
Gaia DR3 catalogue. For HD\,9985 the quoted $v_\mathrm{rad}$ is from \citet{gontcharov2006}, while for HD\,31305 and HD\,112532 we list the 
optimal radial velocities provided by the BANYAN $\Sigma$ tool \citep{gagne2018}, assuming that the star is indeed a member of the stellar association 
determined from astrometric data alone (Appendix~\ref{sec:membership}).
For a more detailed description of how the listed stellar properties ($T_{\rm eff}$, $L_*$, $M_*$, group membership, age, and 
multiplicity) are determined, see Appendix~\ref{sec:props}. Abbreviations in the group column are as follows -- COL: Columba moving group; LCC: Lower Centaurus Crux association; TAU: Taurus; UCRA: Upper CrA; UCL: Upper Centaurus Lupus association; US: Upper Scorpius association. 
Spectral type are derived from the obtained $T_\mathrm{eff}$ estimates using the updated online table\footnote{\url{https://www.pas.rochester.edu/~emamajek/EEM\_dwarf\_UBVIJHK\_colors\_Teff.txt}} of stellar parameters compiled by \citet[][]{pecaut2013}. 
The listed references show those studies in which the infrared excess of the stars was first reported: 
(1): \citet{clark1991}; (2): \citet{cotten2016}; (3): \citet{mcdonald2012}; (4): this work. 
The last column shows which label we use to mark the given object in Figs.~\ref{fig:rdiskls}--\ref{fig:primordial}.}
\end{center}
\end{table*} 

\section{Observations and data reduction} \label{sec:obsdanddatareduction}
All observations of our targets were carried out with the ALMA~12-m Array in Cycle~8. An important lesson from previous 
observations of CO-bearing debris disks is that not only $^{12}$CO but even the $^{13}$CO line can be optically thick. 
Thus, reliable CO mass estimate requires the measurement of $^{12}$CO, $^{13}$CO, and C$^{18}$O lines. Considering the 
typical CO excitation temperatures of 8--30\,K, measured in known gas-bearing debris disks 
\citep{kospal2013,matra2017b,higuchi2020,difolco2020}, the 
J=2--1 transition of CO is the best choice for gas detection, that is observable for all three isotopologs in Band~6. 
Therefore, we performed both our line and continuum measurements in this band. The correlator setup included six spectral 
windows.  Two time division mode (TDM) spectral windows, each providing a total bandwidth of 2.0~GHz, split into 128 channels 
and centered at frequencies of 216.5 and 234\,GHz, were used to probe the dust continuum. Three frequency division mode (FDM) 
spectral windows centered at 219.45, 220.399, and 230.538\,GHz, each with 960 channels, were tuned to cover the C$^{18}$O, 
$^{13}$CO, and $^{12}$CO lines, respectively. The $^{12}$CO line was observed with spectral resolution of 244.14~kHz 
(0.32\,km~s$^{-1}$), while for the other two lines, we applied the coarsest spectral resolution of 976.56\,kHz (1.33\,km~s$^{-1}$). 
An additional FDM window -- in the same baseband as the $^{12}$CO measurement -- centered at 232\,GHz and having a bandwidth 
of 0.234\,GHz and 960 channels was utilized to observe the continuum. Table~\ref{tab:obslog} summarizes the main parameters 
of our observations including their dates, the number of antennas (N$_\mathrm{ant}$), the baseline ranges of the array 
configurations (L$_\mathrm{B}$), as well as 
the name of the sources used as flux, bandpass, and phase calibrators. 
Assuming a source with a Gaussian brightness distribution, with the shortest baseline length of 15\,m, our observations have a maximum 
recoverable scale (MRS) of $\sim$8{\arcsec} at a 50\% flux level, i.e. for a Gaussian source with such a 
full width at half maximum (FWHM), 50\% 
of the true peak flux is expected to be lost \citep[][Appendix~A, eq. A8]{wilner1994}. 
If we assume a uniform disk instead, we obtain an MRS of 8\farcs9 for the same flux loss level  
\citep[][Appendix~A, eq. A11]{wilner1994}.

 
\begin{table}
\footnotesize                                                                  
\setlength{\tabcolsep}{1.0mm}                                                   
\begin{center}                                                                                                                                       
\caption{Observational parameters \label{tab:obslog} }
\begin{tabular}{lcccccc}                                                     
\hline\hline
Target      & Obs. date & N$_\mathrm{ant}$ & L$_\mathrm{B}$ (m) &   \multicolumn{2}{c}{Calibrators} \\
            &           &                  &                    &   {\scriptsize Flux/Bandpass}   &  {\scriptsize Phase}   \\    
\hline
HD\,9985   & 2022-09-01 &  46 & 15--784    &  J0238+1636  &  J0152+2207  \\
HD\,9985   & 2022-09-07 &  43 & 15--784    &  J0006-0623  &  J0112+2244  \\
HD\,31305  & 2022-08-08 &  42 & 15--1302   &  J0510+1800  &  J0438+3004  \\
HD\,112532 & 2022-09-03 &  43 & 15--784    &  J1427-4206  &  J1315-5334  \\
HD\,131960 & 2022-08-13 &  47 & 15--1302   &  J1517-2422  &  J1457-3539  \\
HD\,141960 & 2022-08-29 &  45 & 15--784    &  J1517-2422  &  J1554-2704  \\
HD\,141960 & 2022-09-04 &  46 & 15--784    &  J1517-2422  &  J1553-2422  \\
HD\,144277 & 2022-09-07 &  44 & 15--784    &  J1517-2422  &  J1610-3958  \\
HD\,145101 & 2022-08-28 &  43 & 15--909    &  J1517-2422  &  J1553-2422  \\
HD\,152989 & 2022-08-23 &  43 & 15--1302   &  J1517-2422  &  J1700-2610  \\
HD\,155853 & 2022-08-29 &  45 & 15--784    &  J1617-5848  &  J1717-3342  \\
HD\,155853 & 2022-08-30 &  45 & 15--784    &  J1617-5848  &  J1717-3342  \\
HD\,159595 & 2022-06-03 &  42 & 15--784    &  J1924-2914  &  J1717-3342  \\
HD\,170116 & 2022-09-01 &  46 & 15--784    &  J1924-2914  &  J1851+0035  \\
HD\,170116 & 2022-09-01 &  46 & 15--784    &  J1924-2914  &  J1851+0035  \\
HD\,170116 & 2022-09-04 &  45 & 15--784    &  J1924-2914  &  J1851+0035  \\
HD\,176497 & 2022-08-13 &  46 & 15--1302   &  J1924-2914  &  J1937-3958  \\
HD\,176497 & 2022-08-13 &  46 & 15--1302   &  J1924-2914  &  J1937-3958  \\
\hline
\end{tabular}
\end{center}
\end{table}  


All data sets were calibrated and flagged using the Common Astronomy Software Applications package
\citep[\texttt{CASA};][]{mcmullin2007} and the standard pipeline scripts provided by the ALMA Observatory. 
For data processing, we used the default CASA versions specified in the pipeline:
for the observations of HD\,9985, HD\,144277, and HD\,170116, which were performed during the last period of the project, 
the \texttt{CASA} software version 6.4.1.12 was used, while for the earlier measurements the version 
6.2.1.7 was employed for data processing. No additional flagging was performed beyond the standard ALMA pipeline 
process. For objects that were observed at more than one epoch (Table~\ref{tab:obslog}), 
the resulting measurement sets were concatenated using the \textsc{concat} task. 
To ensure the relative weights between the different data sets are correct we recalculated the data weights with the 
\texttt{CASA} \textsc{statwt} task. For the analysis of line emission in the FDM spectral windows, we used the 
\textsc{uvcontsub} routine to fit a first-order polynomial to the line-free regions (Sect.~\ref{sec:co}) in the $uv$ space 
and subtracted the continuum from the obtained data.

\section{Analysis of continuum data} \label{sec:continuum}

\subsection{Imaging}
\begin{figure*}[h!]
\centering
\includegraphics[bb=20 80 490 450,width=0.32\textwidth]{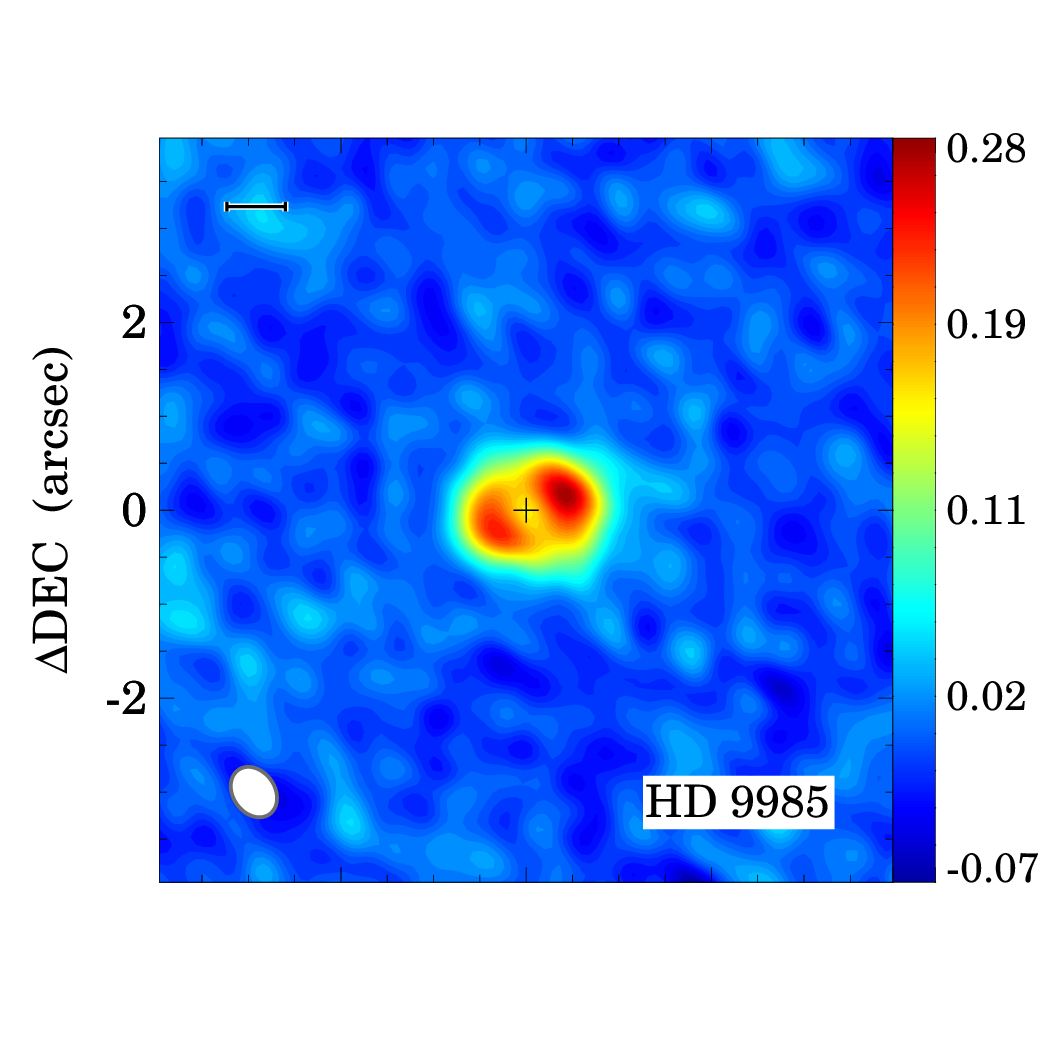}
\includegraphics[bb=60 80 490 450,width=0.292766\textwidth]{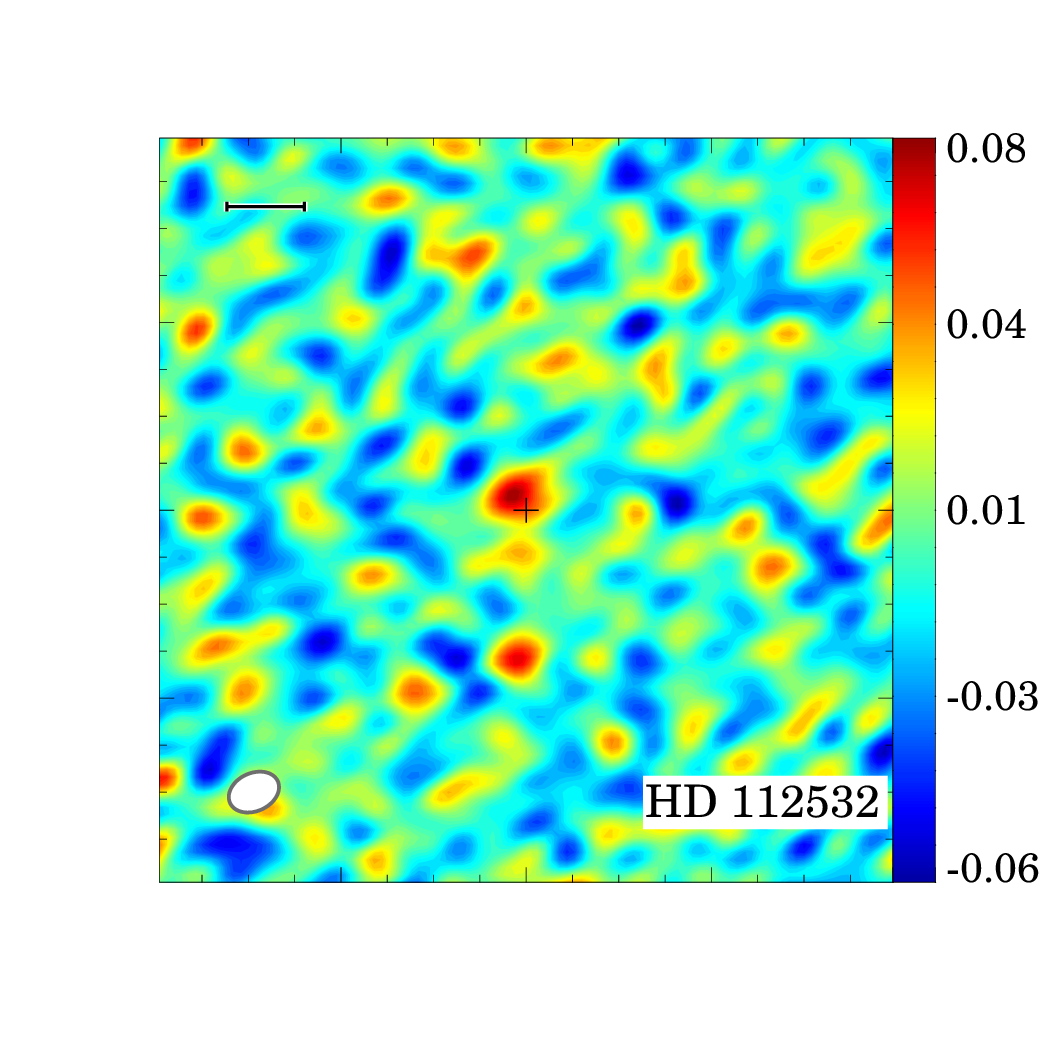}
\includegraphics[bb=60 80 490 450,width=0.292766\textwidth]{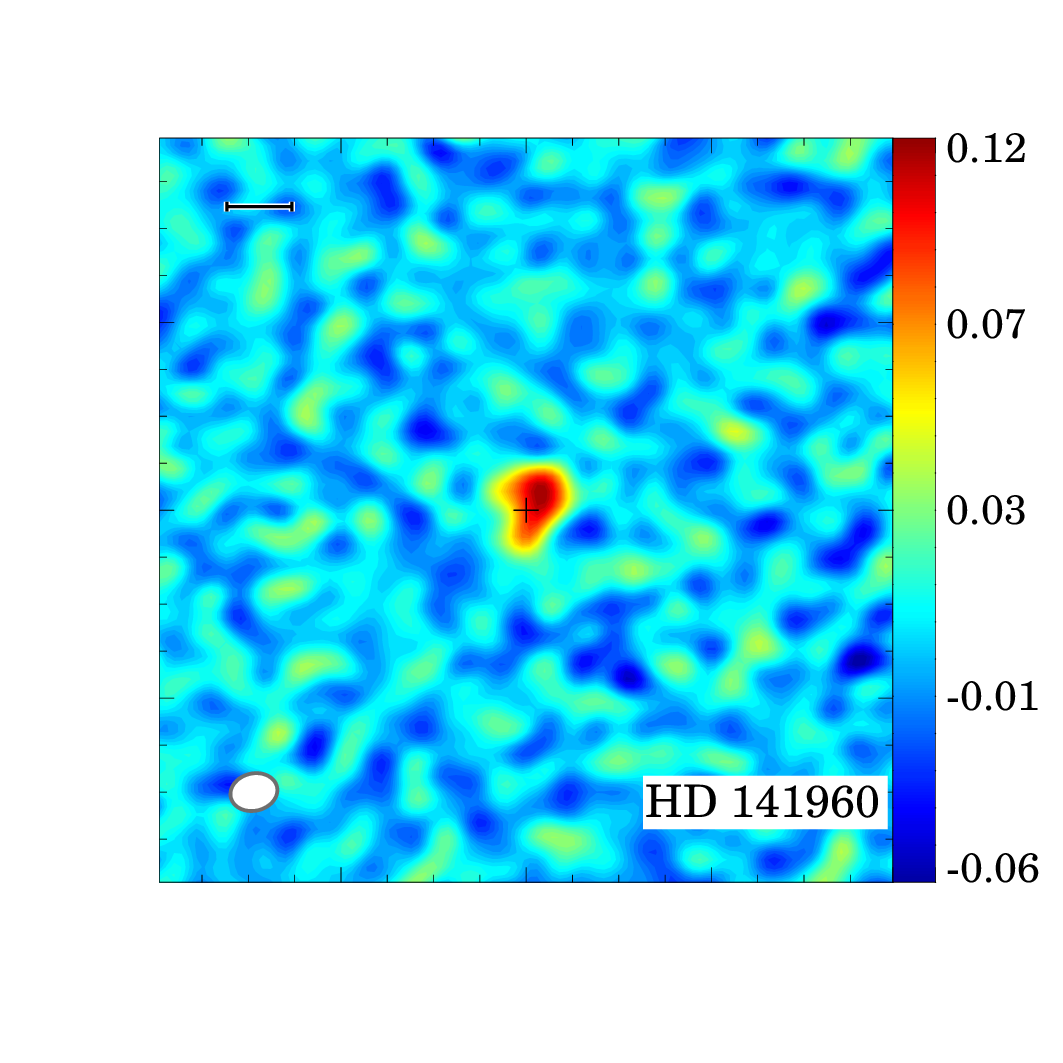}
\includegraphics[bb=20 80 490 450,width=0.32\textwidth]{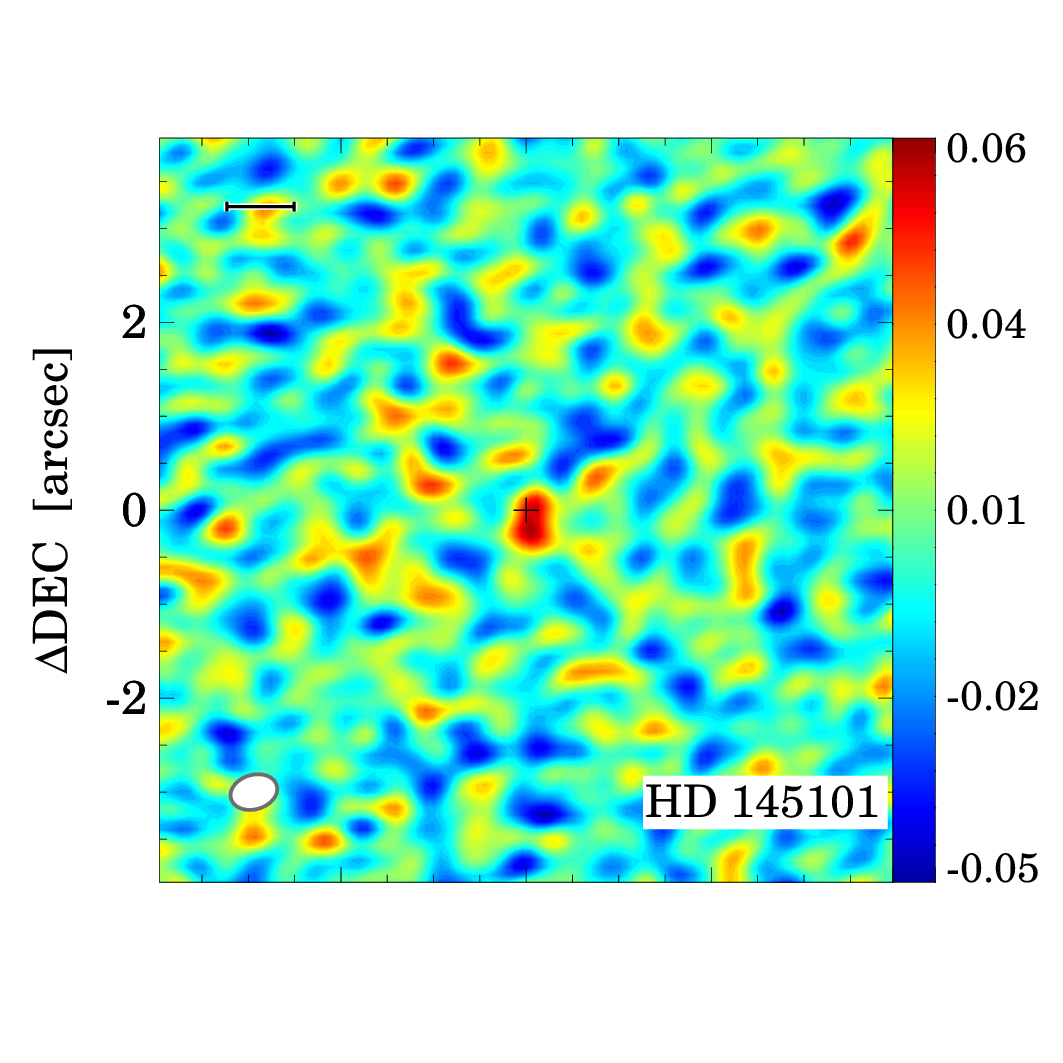}
\includegraphics[bb=60 80 490 450,width=0.292766\textwidth]{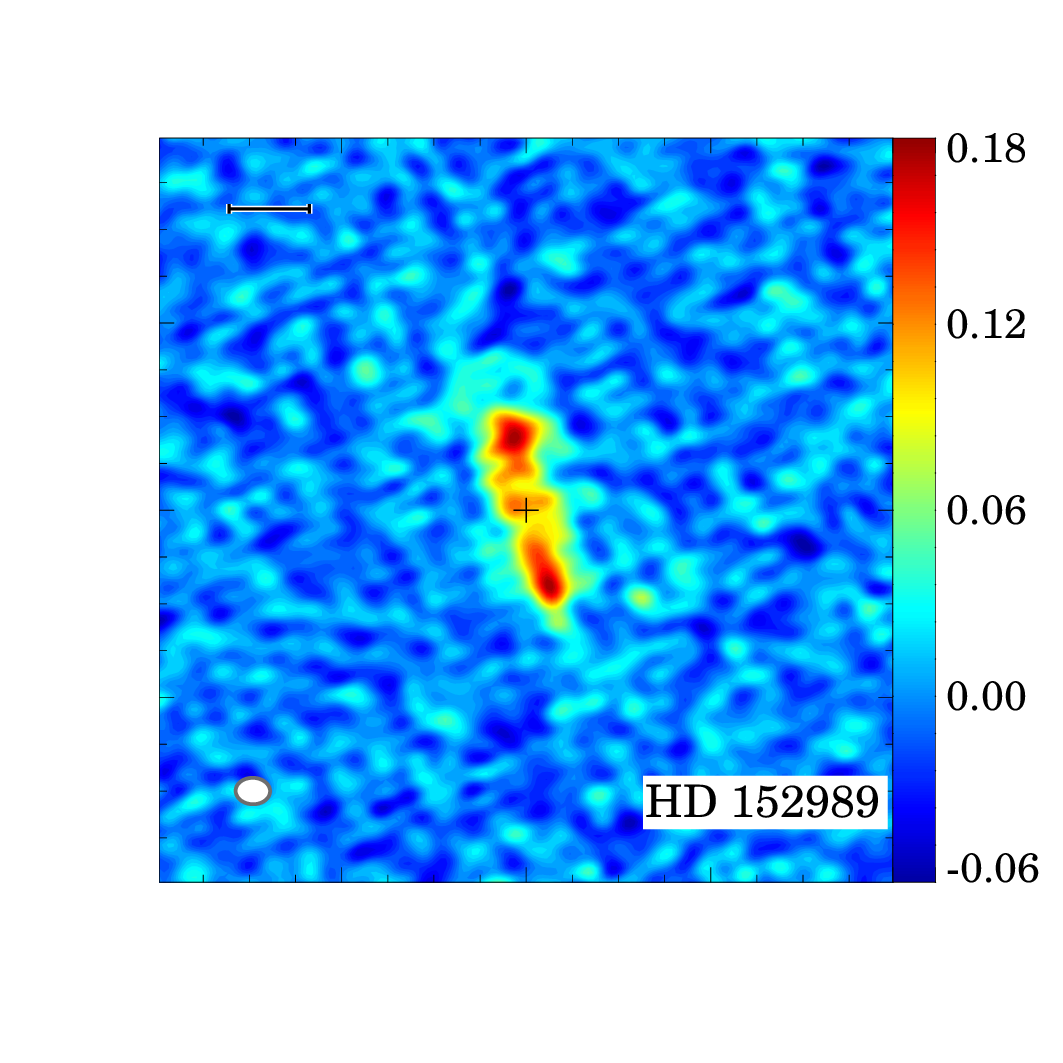}
\includegraphics[bb=60 80 490 450,width=0.292766\textwidth]{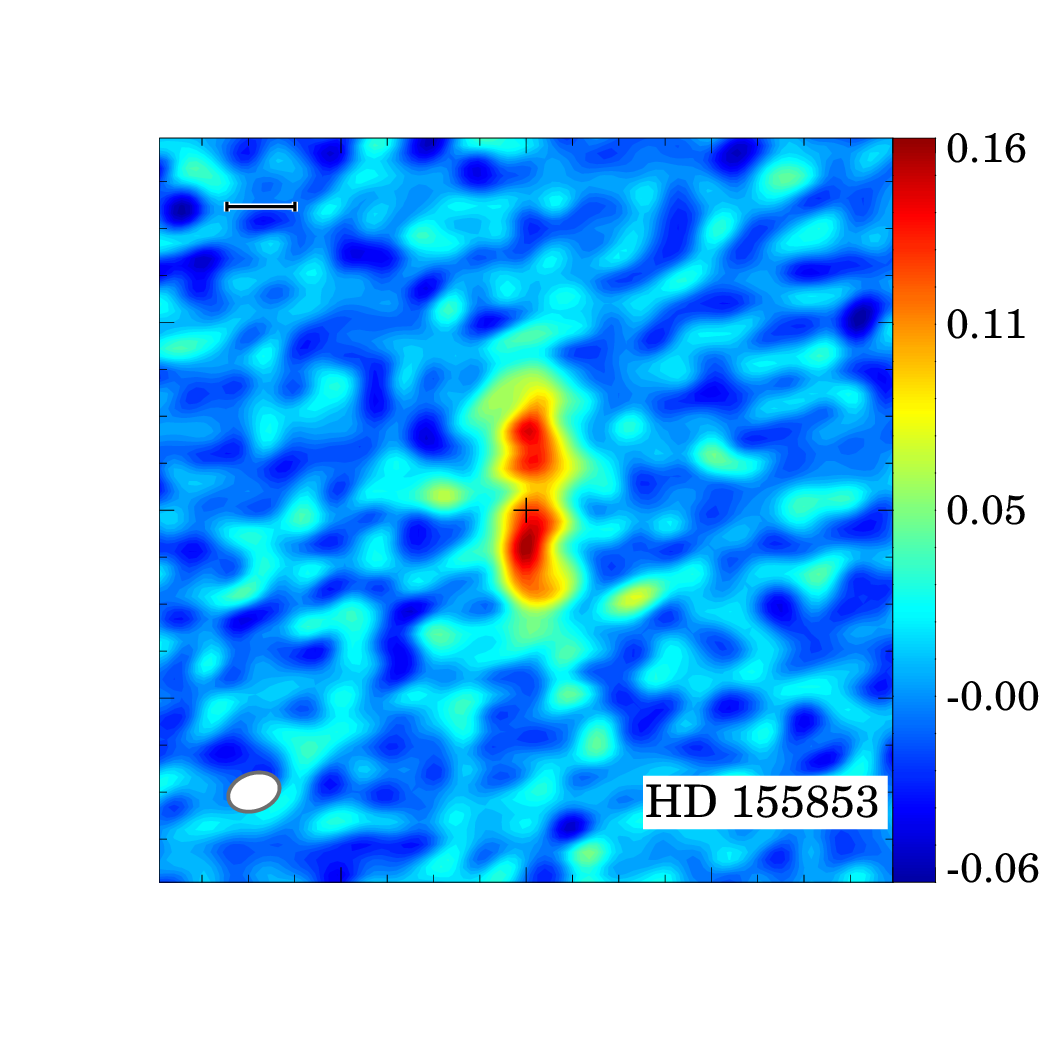}
\includegraphics[bb=20 80 490 450,width=0.32\textwidth]{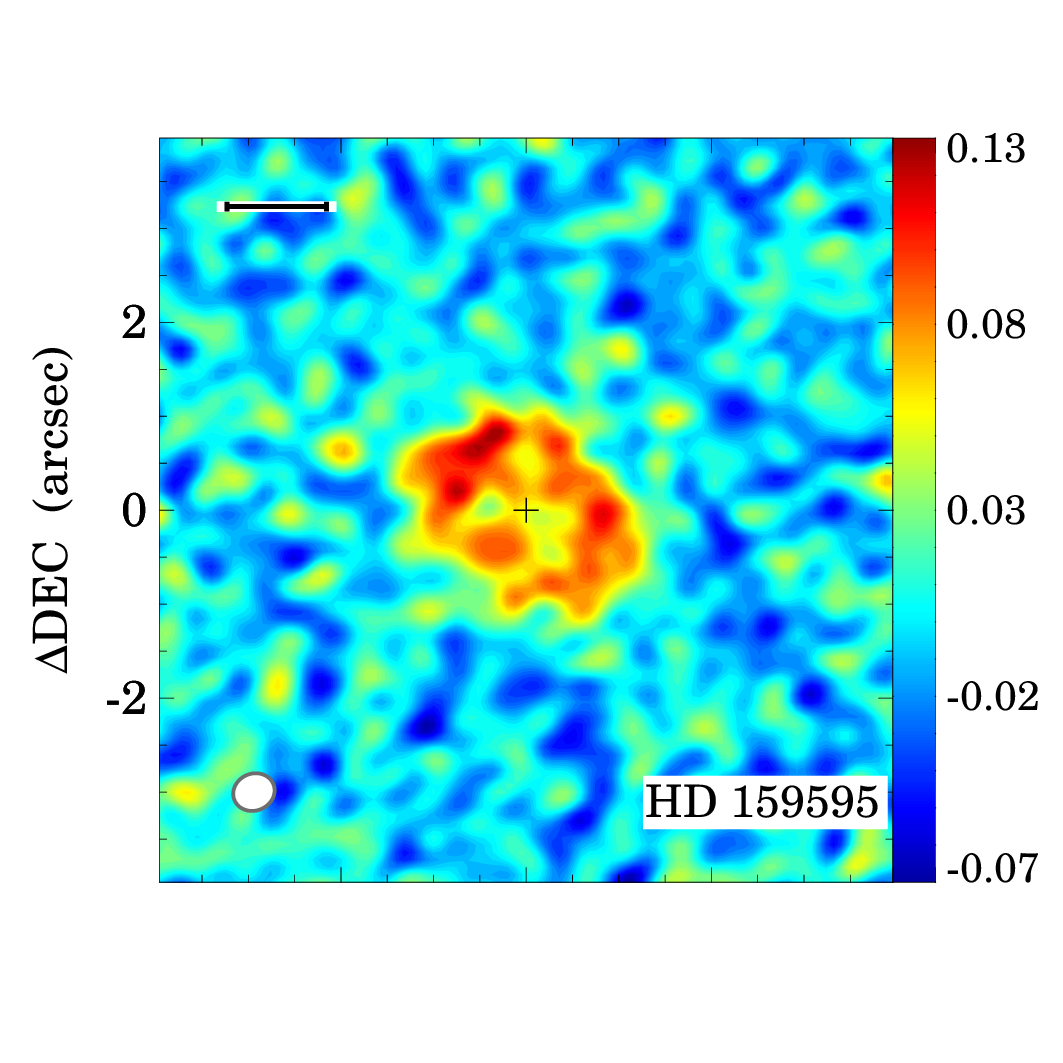}
\includegraphics[bb=60 80 490 450,width=0.292766\textwidth]{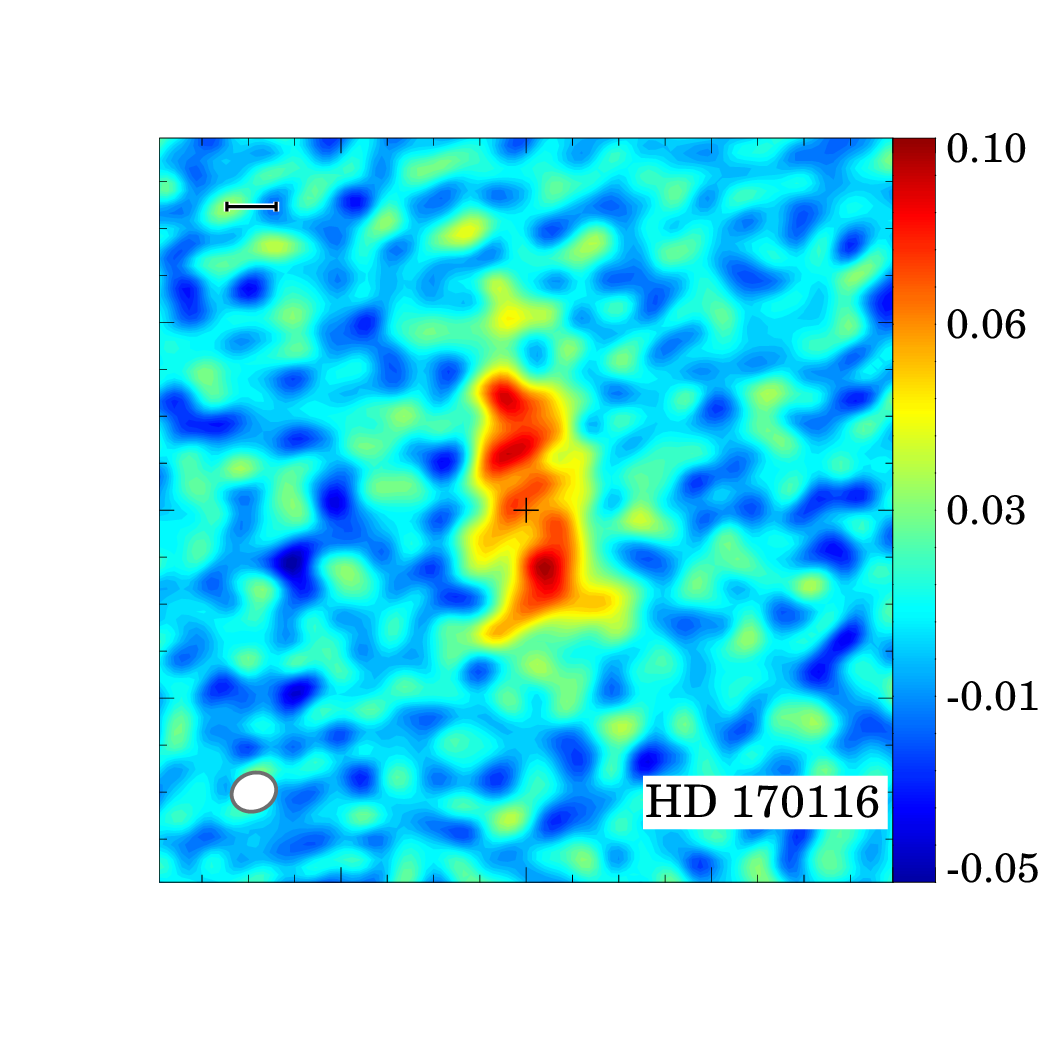}
\includegraphics[bb=60 80 490 450,width=0.292766\textwidth]{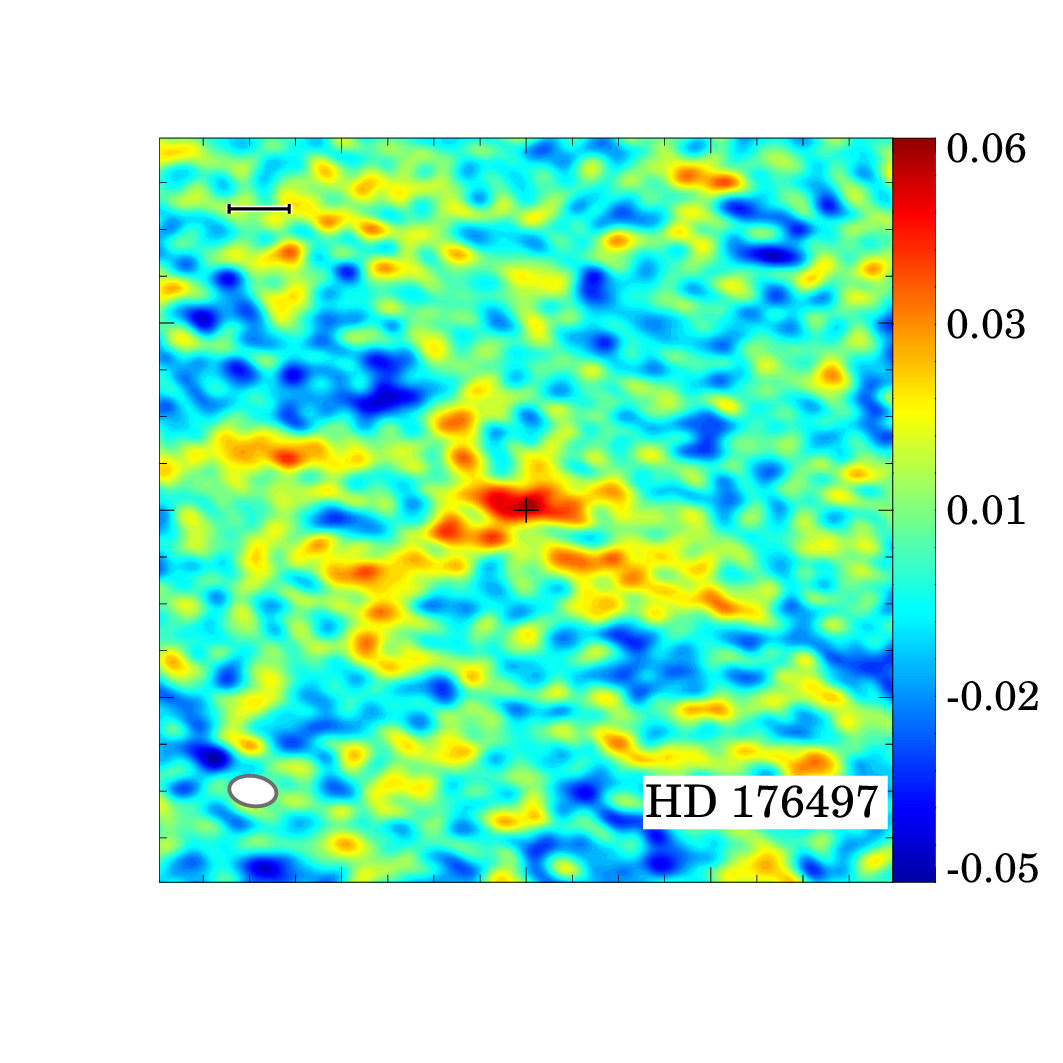}
\includegraphics[bb=20 30 490 475,width=0.32\textwidth]{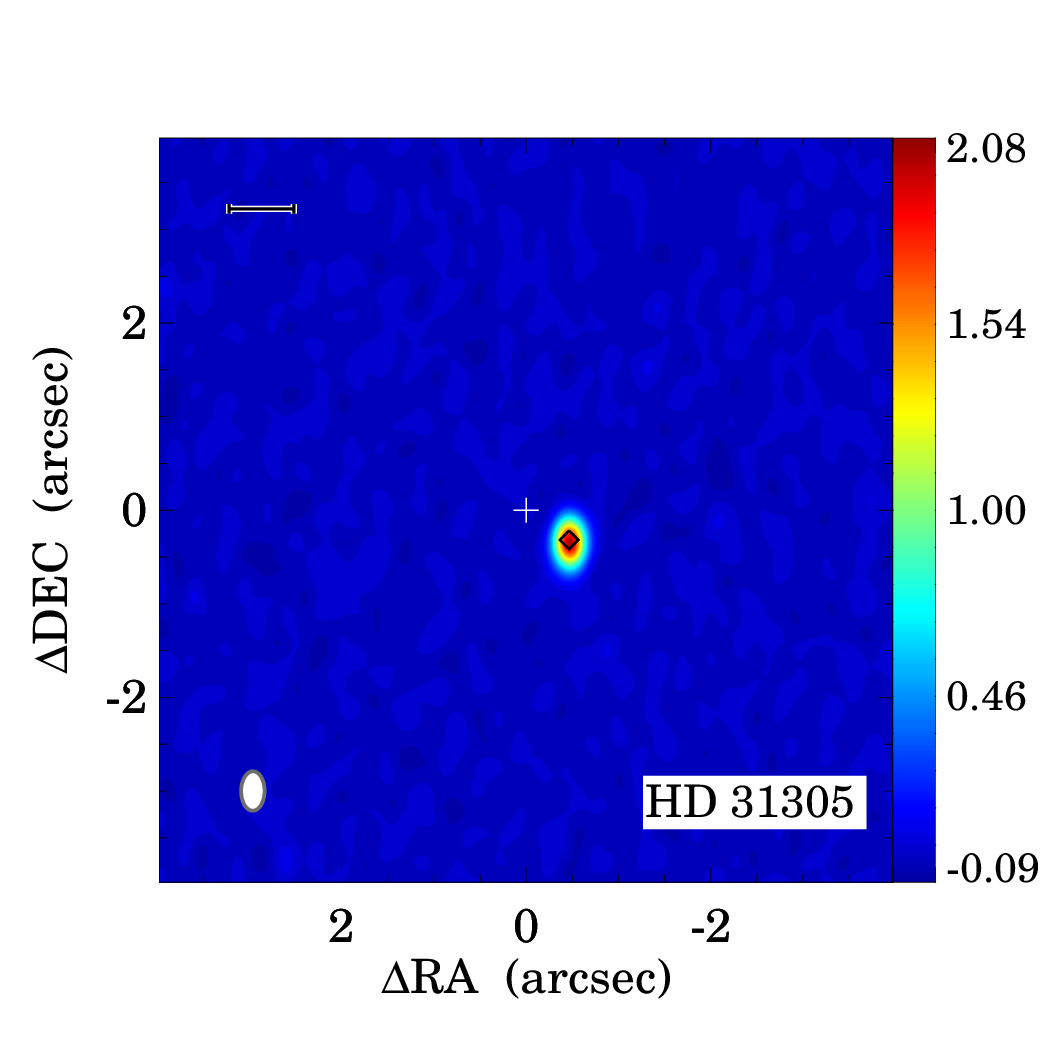}
\includegraphics[bb=60 30 490 475,width=0.292766\textwidth]{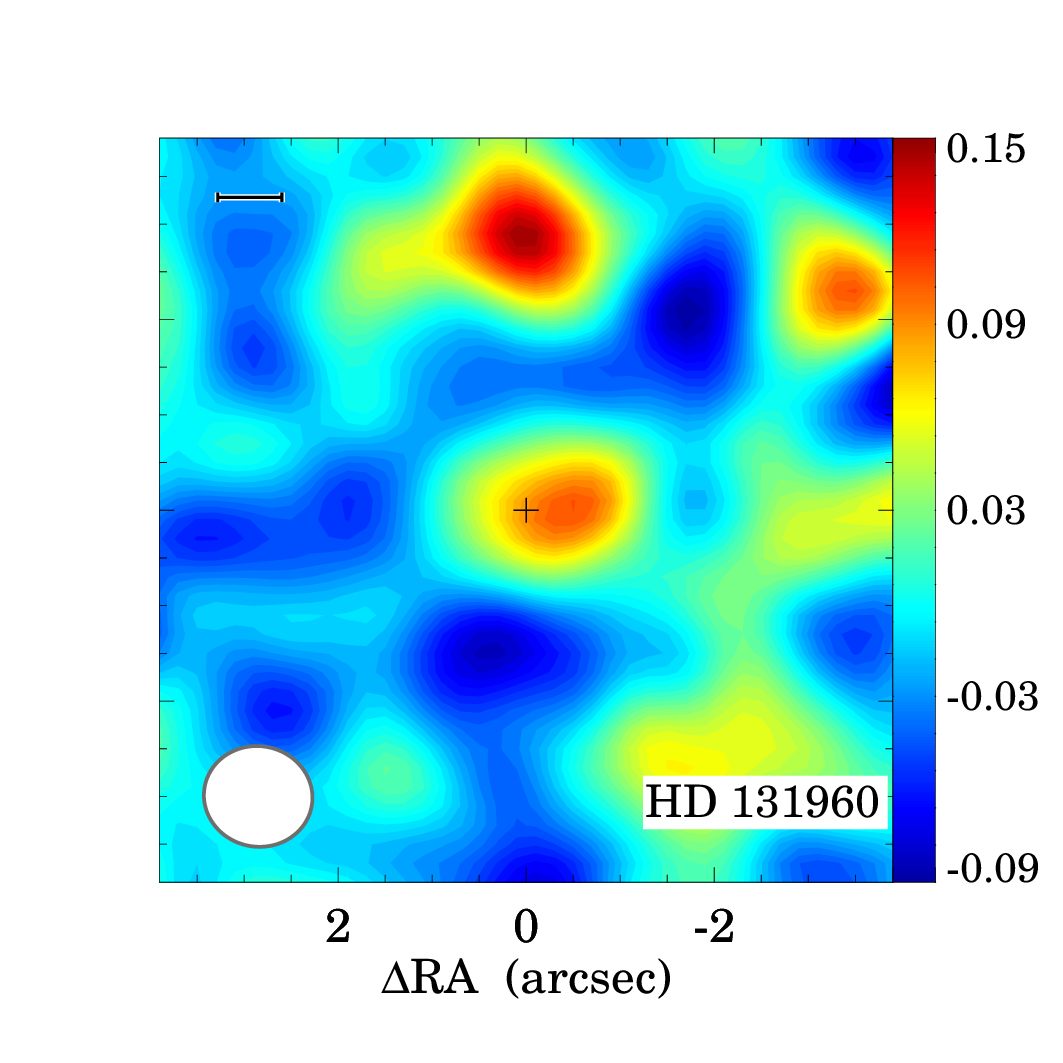}
\includegraphics[bb=60 30 490 475,width=0.292766\textwidth]{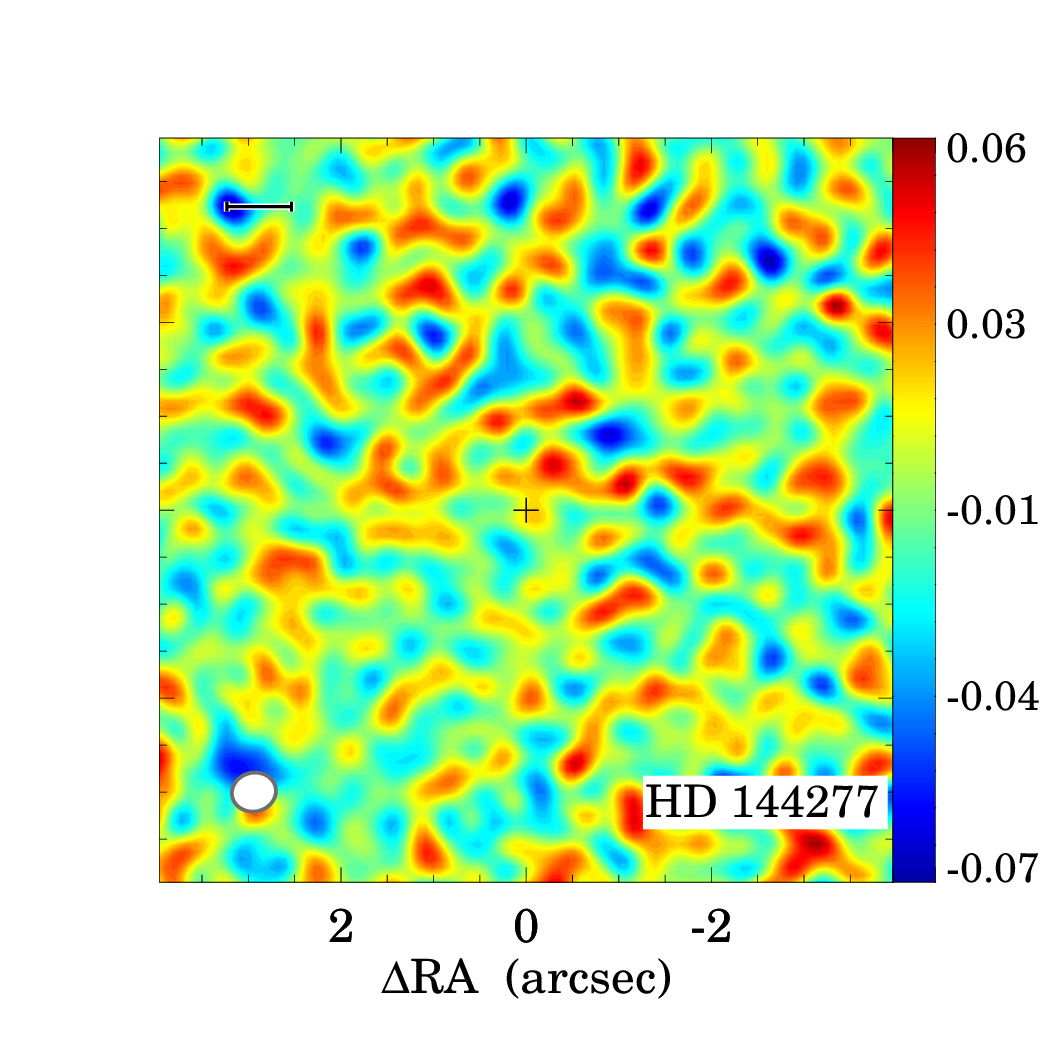}
\caption{ALMA Band~6 continuum emission for the observed targets. The stellar position is marked with a black plus sign. 
In the case of HD~31305, the position of the primary component (HD\,31305~A) is shown by a white plus sign, while 
the companion star (HD\,31305~B) is marked by a black diamond. 
For HD\,131960 the visibility data are tapered by a Gaussian with an FWHM of 1{\arcsec}.
At the bottom left of each panel a filled white ellipse shows the beam size. The length of the horizontal black 
bars corresponds to 100\,au. The color bar units are mJy~beam$^{-1}$.}
\label{fig:continuumplots}
\end{figure*}

To construct continuum images, we used the \textsc{tclean} task with the {\sl multiscale} deconvolution algorithm, setting a 
list of scale sizes of 0 (point source), 5, 15, and 25\,pixels (corresponding roughly to 1, 3, and 5 times the beam size). 
In this process, we used data from all of the spectral windows except for the channels where line emission was detected 
(Sect.~\ref{sec:co}). Figure~\ref{fig:continuumplots} presents the obtained 1.33\,mm continuum images showing the immediate 
(4\arcsec$\times$4\arcsec) environment of the targeted systems.
The beam sizes range between 0.25 to 1.16{\arcsec} and have a mean value of 0.5{\arcsec}. 
With the exception of HD\,144277 and HD\,176497, where natural weighting was used, all other images were compiled 
with  Briggs weighting where the robust parameter was set to 0.5. The panels in the top three rows of the figure show those nine 
cases where we detected significant (peak $>3\sigma$) signal at or near the position of the star. Six of them 
show clearly resolved disk structure, four with more edge-on (HD\,152989, HD\,155853, HD\,170116, and HD\,176497), and  
two with more face-on geometry (HD\,9985 and HD\,159595). The other three (HD\,112532, HD\,141960, and HD\,145101) are 
only marginally resolved or point-like. 

No significant continuum emission is found to be associated with HD\,144277. This conclusion is not changed either by 
using natural weighting or by applying Gaussian tapers with size of 1{\arcsec} or 2{\arcsec} to look for possible extended
emission. In the case of HD\,131960, neither the Briggs nor the natural weighted images show any source. However, by 
applying a taper of 1{\arcsec}, the resulting image (Fig.~\ref{fig:continuumplots}) shows a point source, located 
0\farcs2 from the position of the star, with a peak signal-to-noise ratio (S/N) of 2.6, just below the detection limit.
Although this observation is treated as a non-detection, this result suggests that a future deeper measurement may 
reveal a faint extended disk around this star.

HD\,31305 was not detected either, but at a separation of 0\farcs59 from the star there is a very bright source, 
whose position coincides well with that of HD\,31305~B, a K-type companion of our target (Appendix~\ref{sec:multiplicity}). 
The latter object, probably a protoplanetary disk around the companion star, is discussed separately in 
Appendix~\ref{sec:hd31305b}.

Table~\ref{tab:imagingpars} summarizes the relevant imaging properties, including the size and position angle 
of the synthesized beam and the achieved 1$\sigma$ rms noise levels.

\subsection{Modeling of the continuum emission}  \label{sec:uvmultifit} 
To model the spatial structure of the dust emission and to estimate the flux density of the disks at 1.33\,mm, 
we used the \textsc{uvmultifit} tool developed by \citet{martividal2014} to fit the observed visibilities. 
Only channels free of line emission were considered in the fits. For the six spatially well resolved
sources (HD\,9985, HD\,152989, HD\,155853, HD\,159595, HD\,170116, and HD\,176497), two surface brightness 
profile models were tested: an elliptical Gaussian and an azimuthally symmetric Gaussian ring model. 
The former source model has six free parameters: the offsets with respect to the phase center in right ascension 
and declination, the flux density, the diameter of the major axis, the ratio of the sizes of the minor and major 
axes, and the position angle. In the latter model, there is an additional parameter which gives the width of the 
Gaussian describing surface brightness profile of the ring. In the modeling of the faint disk around HD\,176497, 
this width parameter (the FWHM of the Gaussian) was fixed to be 0.7 of the ring radius 
\citep[i.e. $W_\mathrm{disk} = 0.7 R_\mathrm{disk}$, a typical value for debris disks][]{matra2025}.
For the other five objects, the ring width was also a free parameter.  
To evaluate which of the best-fitting models obtained for the two different profile types has better quality, we used 
the Bayesian Information Criterion (BIC), which favored the Gaussian ring approach in all cases. 
We note that, motivated by the system's CO measurements, for HD\,152989 we also developed a continuum disk 
model with two narrow rings (see Sect.~\ref{sec:cospatial}).
For the other three disks, we added a point source model to the profile types to be tested, and in the case of ring 
models the width parameter was always tied to the radius as $W_\mathrm{disk} = 0.7 R_\mathrm{disk}$. 
Using the BIC for HD\,112532, HD\,141960, and HD\,145101, we found the elliptical Gaussian, the Gaussian ring, 
and the point source models to be the best solution, respectively. The results are presented in 
Table~\ref{tab:contdiskprops}.

We note that the derived parameters refer to the radial profile of the surface brightness and those of the surface 
density profile are slightly different. Considering that the 1.3\,mm emission comes predominantly from large grains 
acting as blackbodies, the temperature dependence as a function of radius is $T(r) \propto 1/r^{0.5}$, and thus the 
peak of the surface density profile is found further away from the star. In addition, the width of the ring -- defined 
as the half-width of the maximum of the obtained profile -- also changes slightly compared to that inferred from the 
analysis of the surface brightness profile. However, in our cases these deviations are always smaller than the 
1$\sigma$ uncertainty of the estimated $R_\mathrm{disk}$ and $W_\mathrm{disk}$ parameters.


\begin{table*}                                                                  
\setlength{\tabcolsep}{1.7mm}                                             
\begin{center}                                                                                                                                       
\caption{Results from continuum observations.  
 \label{tab:contdiskprops} }
\begin{tabular}{lccccccccc}                                                     
\hline\hline
Target name &  $L_\mathrm{disk}/L_\mathrm{bol}$  & Model      & $F_\mathrm{1.33mm}$ & $R_\mathrm{disk}$ & $W_\mathrm{disk}$ &  $i$          & $PA$       & $M_\mathrm{dust}$ &  Label\\
            &       (10$^{-4}$)                  &            & (mJy)               &      (au)         &      (au)         & ({\degr})     & ({\degr})  &  ($M_\oplus$)     &       \\
\hline	     
\multicolumn{10}{c}{Detected disks}        \\
\hline
HD\,9985     & 12.6  & Gaussian ring & 1.26$\pm$0.14 (0.06)     & 84$\pm$9   & 57$\pm$25  & 48$\pm$3   & 104$\pm$11  & 2.6$\pm$0.3e-01  & 1  \\ 
HD\,112532   &  7.3  & Gaussian      & 0.119$\pm$0.038 (0.036)  & 36$\pm$17  &  \ldots    & \ldots     & 39$\pm$7    & 8.9$\pm$3.7e-03  & 2  \\  
HD\,141960   & 23.1  & Gaussian ring & 0.25$\pm$0.03 (0.02)     & 41$\pm$10  & 29         & 66$\pm$13  & 166$\pm$11  & 3.5$\pm$0.6e-02  & 3  \\ 
HD\,145101   &  6.1  & Point source  & 0.061$\pm$0.016 (0.015)  &  $<$33     & \ldots     & \ldots     &  \ldots     & 4.1$\pm$1.4e-03  & 4  \\ 
HD\,152989   & 75.8  & Gaussian ring & 1.42$\pm$0.15 (0.04)     & 103$\pm$7  & 44$\pm$9   & 79$\pm$2   & 14$\pm$1    & 2.1$\pm$0.2e-01  & 5  \\ 
             &       & Two narrow    &  0.56$\pm$0.20 (0.19)    &  83$\pm$13 & \ldots     & 77$\pm$2   & 15$\pm$1    & 8.3$\pm$3.0e-02  &    \\
	     &       & rings         &  0.91$\pm$0.19 (0.16)    & 120$\pm$15 & \ldots     & 77$\pm$2   & 15$\pm$1    & 1.3$\pm$0.3e-01  &    \\
HD\,155853   & 16.8  & Gaussian ring & 0.98$\pm$0.11 (0.04)     & 117$\pm$8  & 116$\pm$16 & 77$\pm$3   & 2$\pm$1     & 1.9$\pm$0.2e-01  & 6  \\ 
HD\,159595   & 21.2  & Gaussian ring & 1.79$\pm$0.21 (0.10)     & 83$\pm$7   & 72$\pm$18  & 46$\pm$10  & 60$\pm$10   & 1.4$\pm$0.2e-01  & 7  \\ 
HD\,170116   & 13.6  & Gaussian ring & 1.03$\pm$0.11 (0.04)     & 197$\pm$25 & 141$\pm$26 & 76$\pm$3   & 10$\pm$4    & 5.1$\pm$0.7e-01  & 8  \\ 
HD\,176497   & 17.0  & Gaussian ring & 0.135$\pm$0.032 (0.029)  &  72$\pm$17 & 50         & 78$\pm$7   & 83$\pm$7    & 2.7$\pm$0.7e-02  & 9  \\ 
\hline
\multicolumn{10}{c}{Non-detections}       \\
\hline
HD\,31305    & ?$^{a}$  & \ldots  & $<$0.063  & \ldots  & \ldots  & \ldots  & \ldots  & $<$5.5e-03 & \ldots \\ 
HD\,131960   & 13.0     & \ldots  & $<$0.120  & \ldots  & \ldots  & \ldots  & \ldots  & $<$2.1e-02 & \ldots \\ 
HD\,144277   &  5.9     & \ldots  & $<$0.072  & \ldots  & \ldots  & \ldots  & \ldots  & $<$1.2e-02 & \ldots \\ 
\hline
\end{tabular}
\tablefoot{The fractional luminosities ($L_\mathrm{disk}/L_\mathrm{bol}$) are taken from Appendix~\ref{sec:diskprops}. 
For detected disks, the estimates of the flux density at 1.33\,mm ($F_\mathrm{1.33mm}$), the disk radius ($R_\mathrm{disk}$), the ring width 
($W_\mathrm{disk}$), the inclination ($i$), 
and the position angle ($PA$) parameters are the results of model fitting with the \textsc{uvmultifit} tool. 
The quoted uncertainties of the flux densities are quadratic sums of the measurement errors (listed in brackets) and the
absolute calibration error (which were conservatively assumed to be 10\%).
With the exception of HD\,112532 and HD\,145101, where an elliptical Gaussian and a point source model were used, respectively, we applied Gaussian ring models to fit the visibility data of the detected disks (Sect.~\ref{sec:uvmultifit}). In the latter cases, the radius of the disk corresponds to the radius of the fitted ring, while for HD\,112532 to the semi-major axis 
(0.5\,$FWHM_\mathrm{maj}$) of the best-fitting Gaussian model. For HD\,152989, an alternative model consisting of two infinitesimally thin rings was also tested 
(Sect.~\ref{sec:cospatial}). To estimate the flux density upper limits for non-detections, we followed the method described in Sect.~\ref{sec:uvmultifit}. 
The 9$^\mathrm{th}$ column lists the derived dust masses (Sect.~\ref{sec:dustmasses}).
\tablefoottext{a}{In the case of HD\,31305, a significant fraction or even all of the observed excess of the system probably comes from the disk around the late-type companion, HD\,31305~B. 
Due to the low spatial resolution of the available IR data, the fractional luminosity of the disk around HD\,31305\,A (if it exists at all) cannot be constrained.}
The last column shows which label we use to mark the given object in Figs.~\ref{fig:rdiskls}--\ref{fig:primordial}.}
\end{center}
\end{table*}


Of the non-detected objects, for HD\,131960 and HD\,144277, their SEDs imply the presence of copious amount 
of warm (150--180\,K, Appendix~\ref{sec:diskprops}) circumstellar dust. However, the available measurements limited to 
wavelengths shortward of 25\,$\mu$m do not allow us to rule out that colder grains, located further away 
from the star, are also present in these systems. Taking this into account, for these objects 
we provide 3$\sigma$ upper limits for flux density derived from the noise measured in their 1{\arcsec} tapered
imaging, where the achieved beam size corresponds to 160--170\,au (we assumed that if they had a disk at all, then 
it would be a point source at this spatial resolution). Although the SED of HD\,31305 is much better sampled, with 
both good quality mid- and far-infrared data available, it is likely that a significant fraction or even all of the 
observed excess of the system actually comes from the disk around the late-type companion, HD\,31305~B, which is 
probably protoplanetary in nature (Appendix~\ref{sec:hd31305b}). 
Due to the presence of the nearby stellar companion (whose projected separation is $\sim$70\,au), if there is a disk 
around HD\,31305A at all, it may be strongly truncated, thus in this case the 3$\sigma$ upper limit on the flux 
density was derived from the noise of the Briggs weighted map that has a spatial resolution of $\sim$50\,au. 
The obtained flux density upper limits are also shown in Table~\ref{tab:contdiskprops}.

\subsection{Dust masses} \label{sec:dustmasses}
Using the obtained 1.33\,mm flux densities ($F_\mathrm{1.33mm}$) we estimated the mass of grains smaller than a 
few millimeters. Since the thermal emission of debris disks at millimeter wavelengths is optically thin, the dust mass 
can be estimated as 
\begin{equation}
M_\mathrm{dust} = \frac{{F_\mathrm{1.33mm}} d^2}{ B_{\nu}(T_\mathrm{BB}) \kappa_{\nu}}, \label{eq:mdust}
\end{equation}
where $d$ is the distance, $B_{\nu} (T_\mathrm{BB})$ is the Planck function at 1.33\,mm for a temperature of $T_\mathrm{BB}$, 
and $\kappa_{\nu}$ is the dust opacity at the given frequency. For the dust opacity we adopt 
$\kappa_{\nu}$=2.3\,cm$^2$\,g$^{-1}$ \citep{andrews2013}.
At such long wavelengths the radiation is dominated by large grains ($\gtrsim$100~{$\mu$}m)
which act as blackbodies. For spatially resolved disks, the characteristic blackbody temperature of large grains 
has been estimated using the derived disk radii (Table~\ref{tab:contdiskprops}) and the luminosity of the star 
(Table~\ref{tab:targets}), by applying the following formula: 
$T_\mathrm{BB} (\mathrm{K}) = 278~(L_*/L_\odot)^{0.25}~(R_\mathrm{disk}/\mathrm{au})^{-0.5}$. 
The $T_\mathrm{BB}$ values obtained this way range from 37\,K (for HD\,170116) to 90\,K (for HD\,112532).
For HD\,145101, we only have an upper limit on the disk size. From this, we can calculate a lower limit of $\sim$100\,K on 
$T_\mathrm{BB}$, while the temperature estimate from fitting its SED (163\,K, Table~\ref{tab:mbbprops}) provides an upper 
limit on this parameter. From this, two dust mass estimates can be calculated, the average of which is taken as the final result. 
For HD\,131960 and HD\,144277, we only have upper limits on the millimeter flux. As already mentioned above, although their 
SED analysis indicates only dust with temperatures of 150--180\,K, the presence of a colder disk component cannot 
be excluded, so we have assumed $T_\mathrm{BB} = 60$\,K in the calculations, which is the average of the 
dust temperature estimates we obtained for the 8 spatially resolved disks. If there is any disk around HD\,31305~A, it 
is probably limited in size to the gravitational influence of its nearby companion star. Therefore, in estimating a 
dust mass upper limit in this case, we have assumed a dust temperature of 110\,K, corresponding to a radial distance 
of $\sim$25\,au assuming blackbody grains (Sect.~\ref{sec:uvmultifit}). 

The dust mass estimates are listed in  Table~\ref{tab:contdiskprops}. We caution that the uncertainties given are formal 
errors only, taking into account the errors in the measured flux density, the distance of the system, and the derived dust 
temperature, but do not consider the possible systematic uncertainties associated with the value of the dust opacity, 
which can reach a factor of several \citep{krivov2021}, as well as the fact that the dust temperature is unlikely to be 
characterized by a single value.

\section{Analysis of line data} \label{sec:co}

\subsection{Imaging and CO line fluxes} \label{sec:coimaging}
To produce CO image cubes from the continuum subtracted visibility data (Sect.~\ref{sec:obsdanddatareduction}) we 
used the \textsc{tclean} task. Similar to the continuum imaging, the {\sl multiscale} algorithm was employed, with the 
same scale settings and Briggs weighting with the robust parameter set to 0.5 by default. LSR reference 
frame was used and the channel width was rebinned to 0.4\,km~s$^{-1}$ for $^{12}$CO line data and 1.6\,km~s$^{-1}$ 
for $^{13}$CO and C$^{18}$O observations. Basic properties of the synthesized beams and the 1$\sigma$ rms levels 
are given in Table~\ref{tab:imagingpars}.

Since $^{12}$CO is by far the most abundant of the three isotopologs studied, we expect the strongest signal 
from this molecule if circumstellar gas is present around a target. To search for $^{12}$CO (2--1) emission, we 
examined channels that are within $\pm$30\,km~s$^{-1}$ of the radial velocity of the target star in the LSR 
frame. The latter parameters were calculated from the heliocentric radial velocity data listed in 
Table~\ref{tab:targets}. We identified four systems, HD\,9985, HD\,145101, HD\,152989, and HD\,155853, with 
significant line emission ($>$3\,$\sigma$), spanning several adjacent channels, which spatially coincides well with 
the disk observed in the continuum. In the case of HD\,155853, which is located close to the galactic plane 
(galactic latitude $b = +2\fdg6$), in the velocity range where the disk is detected, some channels show 
bright extended emission close to, but not overlapping with the disk. Although it does not 
cause a direct contamination, it affects the background flux level of the specific channels. However, we 
found that this effect can be eliminated by omitting visibilities with baselines shorter than 60\,m 
in the cleaning process. For further analysis, the $^{12}$CO data cube obtained in this way was used.
In addition to these four discoveries, a faint but significant CO emission was identified in HD\,170116, 
which coincides well with the southern side of the continuum disk. We found no evidence of the presence of 
a gas disk at our other seven targets. In the case of HD\,159595, which is also located in the galactic plane ($b=-0\fdg11$), 
spatially very extended, bright line emission is present in a large fraction of the channels studied. However, the 
spatial distribution of the emitting regions bears no resemblance to the structure observed in the continuum, suggesting 
that they are associated with background molecular cloud material rather than the disk around the star. 
At HD\,31305, similar to the continuum observation, no emission is detectable at the primary companion, 
but HD\,31305B clearly hosts a bright gas disk. Since this is probably a protoplanetary disk, its line 
measurements are discussed in more detail separately in Appendix~\ref{sec:hd31305b}. 

By inspecting the channel maps of the rarer $^{13}$CO and C$^{18}$O isotopologs, we found significant emission 
at HD\,9985, HD\,145101 and HD\,155853 for the $^{13}$CO line, with similar spatial structure and velocity range 
as for $^{12}$CO. HD\,9985 and HD\,155853 show also C$^{18}$O line emission, but it can be detected 
only over a smaller area and in a narrower velocity interval than the $^{12}$CO and $^{13}$CO emissions. 
We note that the extensive background emissions observed in the $^{12}$CO data at HD\,155853 are absent 
from the $^{13}$CO and C$^{18}$O measurements. None of our other targets exhibit detectable $^{13}$CO or 
C$^{18}$O line emission. 

\begin{figure*}[h!]
\centering
\includegraphics[width=0.33\textwidth]{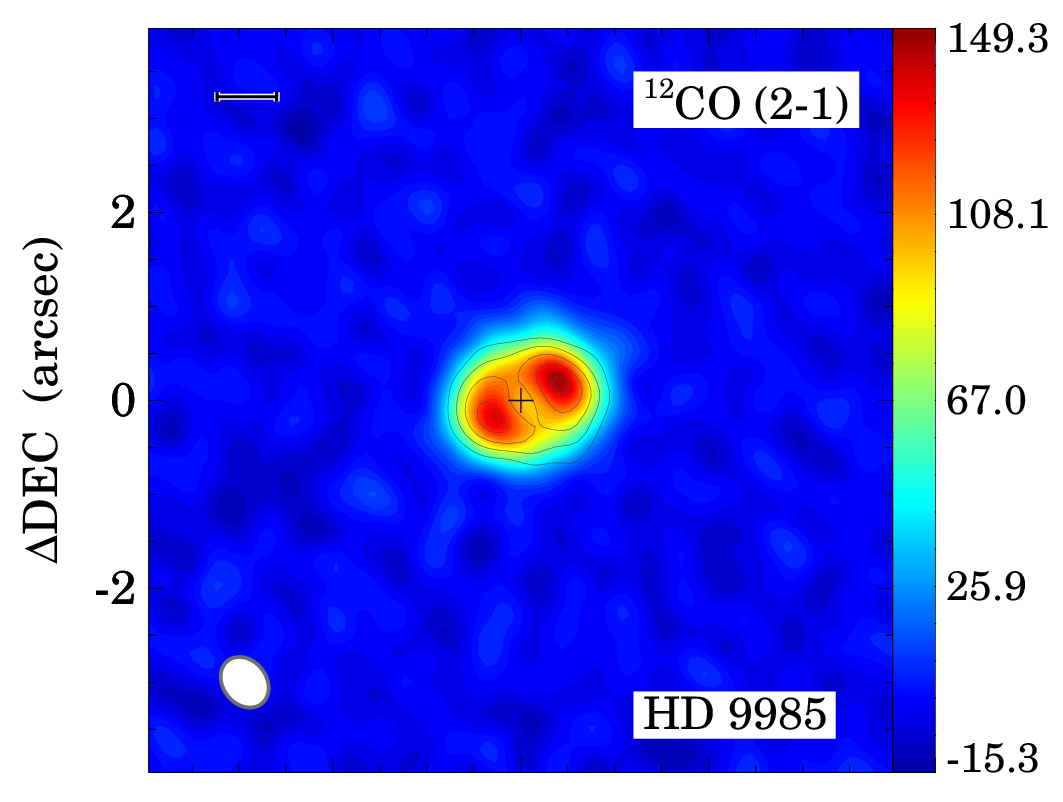}
\includegraphics[width=0.292967\textwidth]{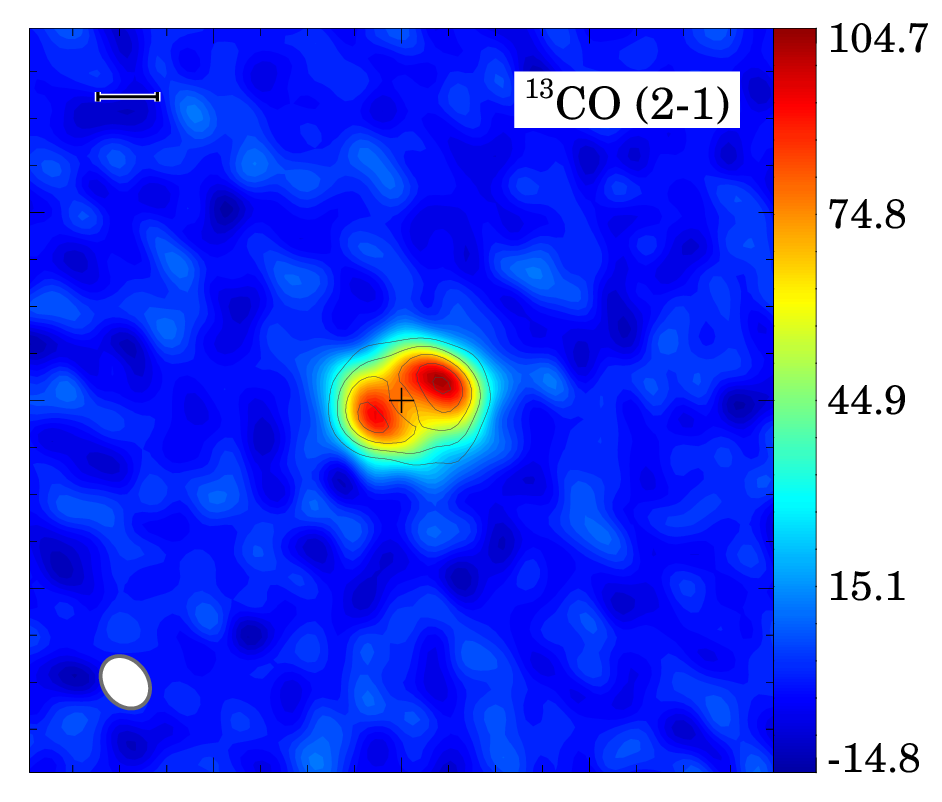}
\includegraphics[width=0.27\textwidth]{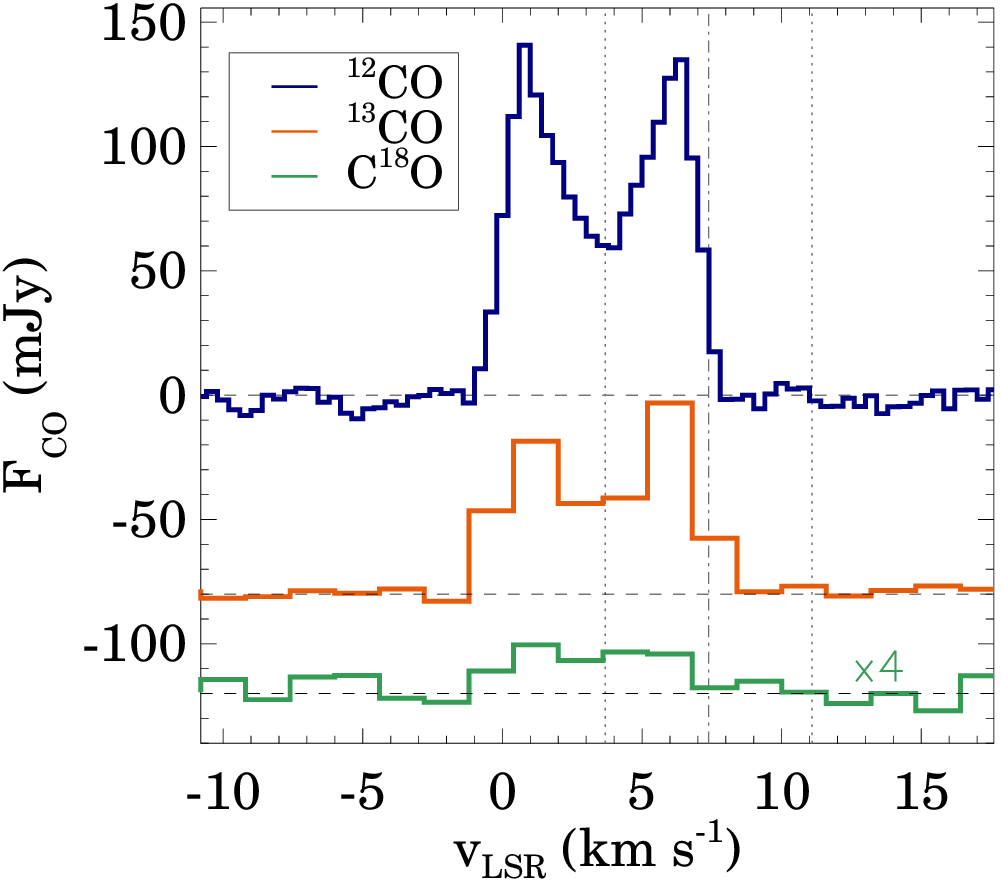}
\includegraphics[width=0.33\textwidth]{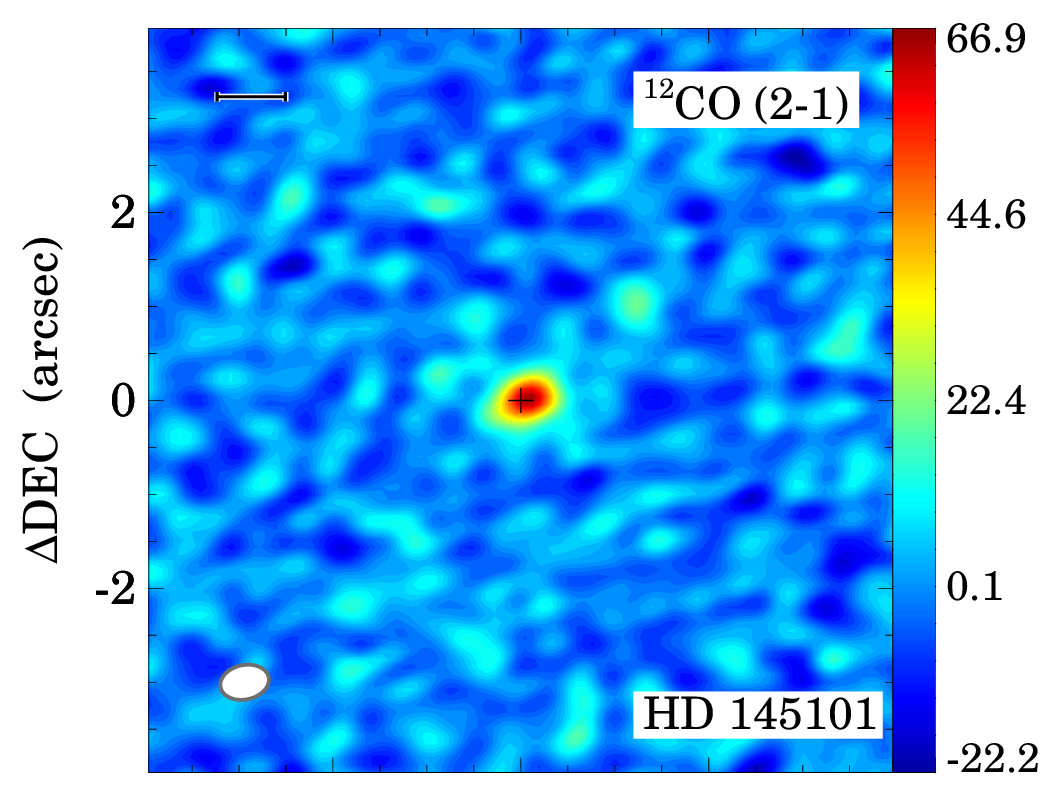}
\includegraphics[width=0.292967\textwidth]{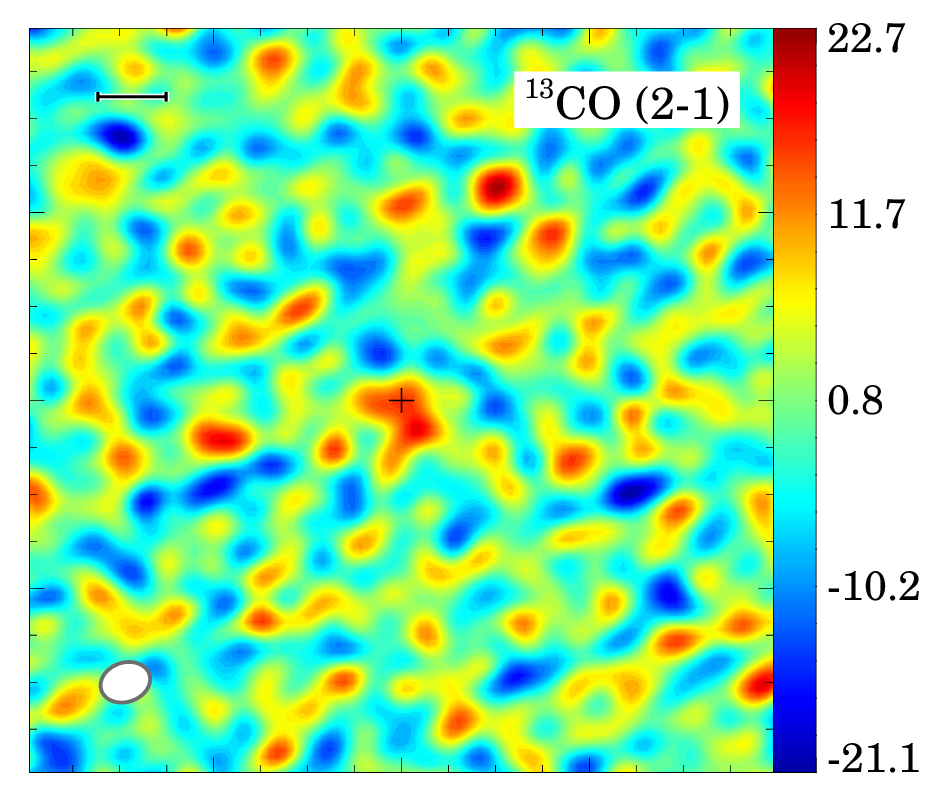}
\includegraphics[width=0.27\textwidth]{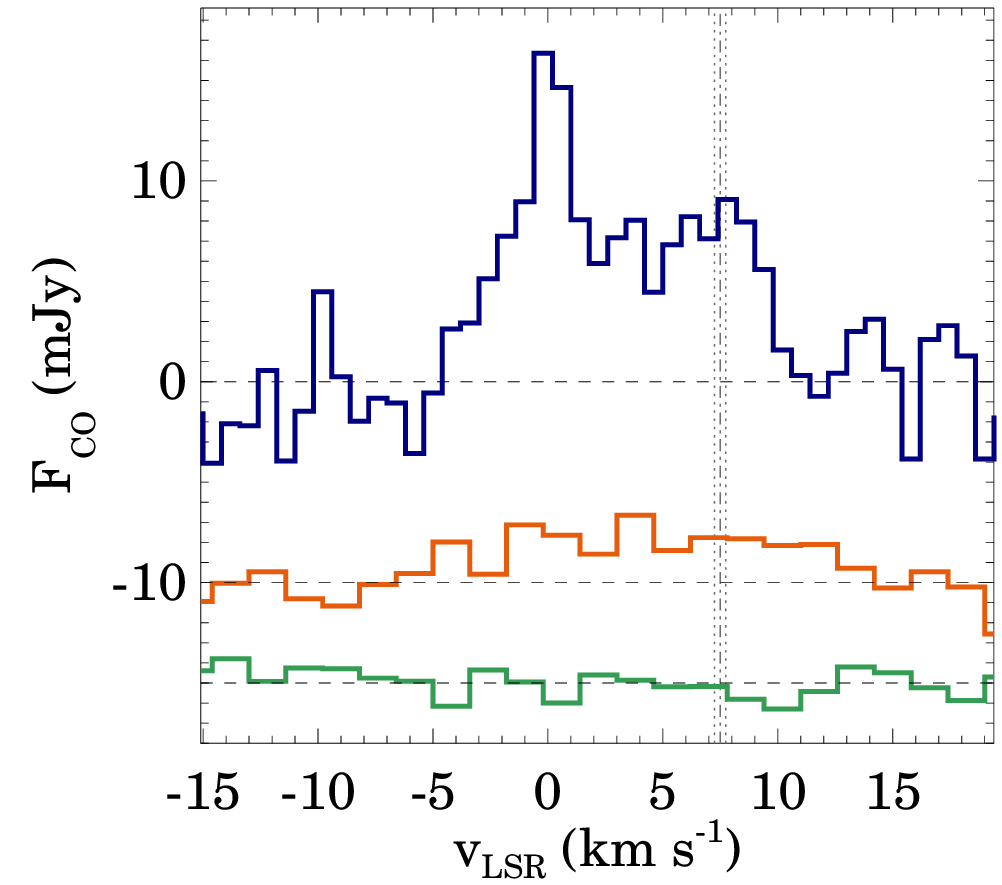}
\includegraphics[width=0.33\textwidth]{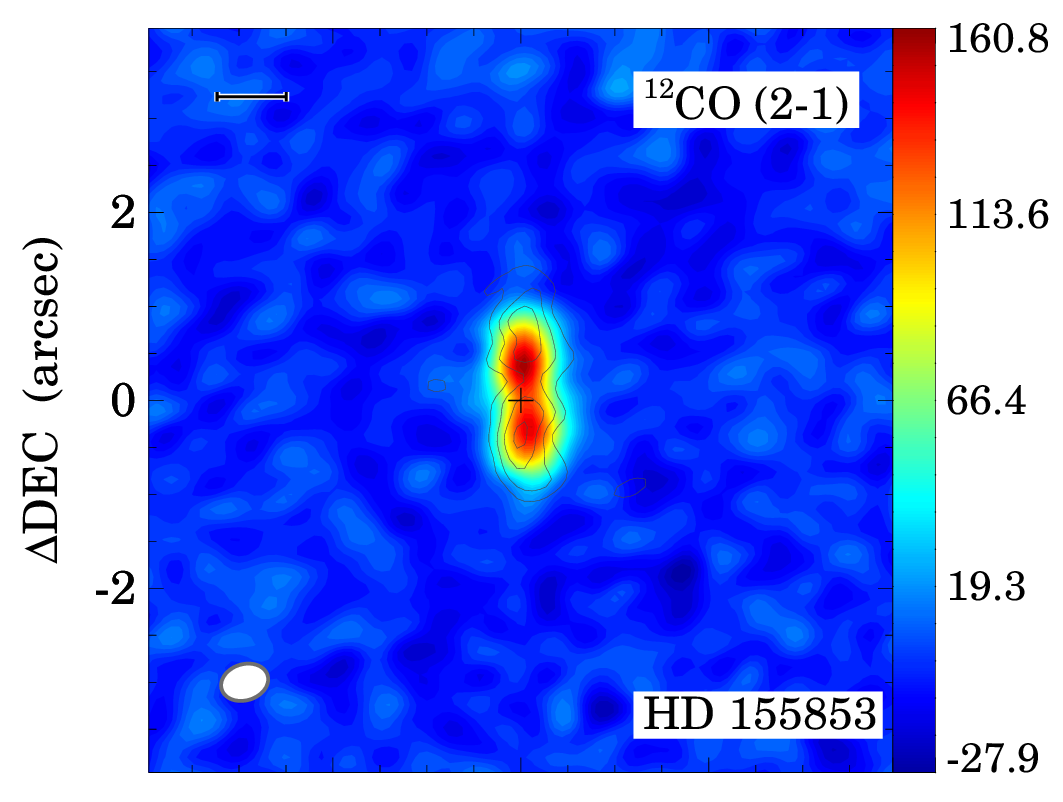}
\includegraphics[width=0.292967\textwidth]{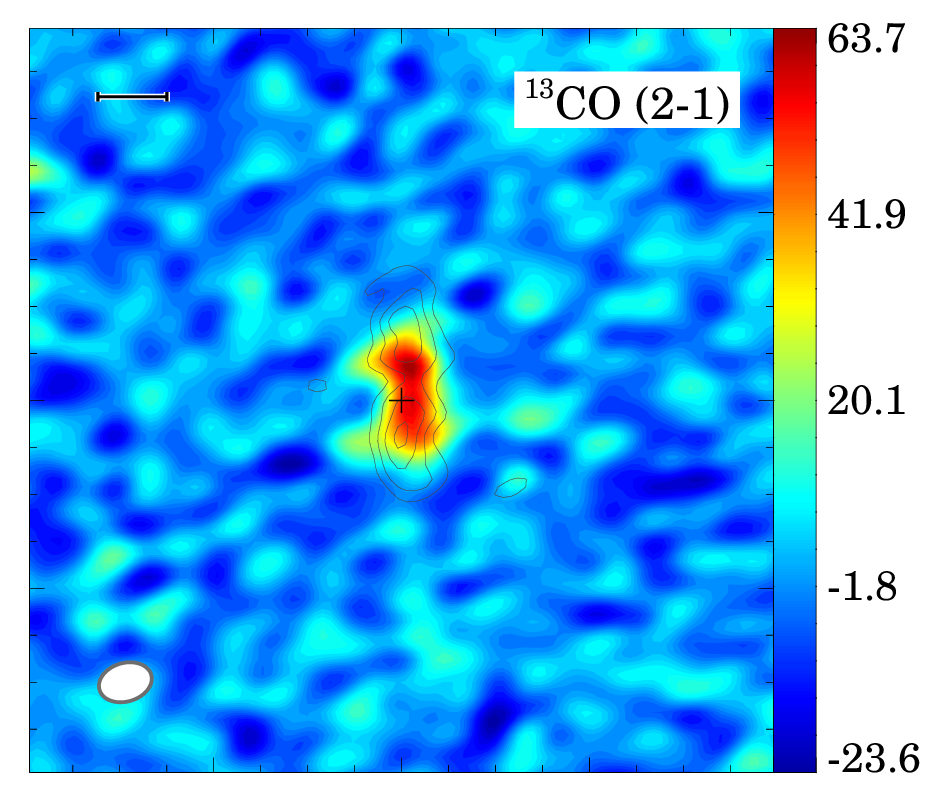}
\includegraphics[width=0.27\textwidth]{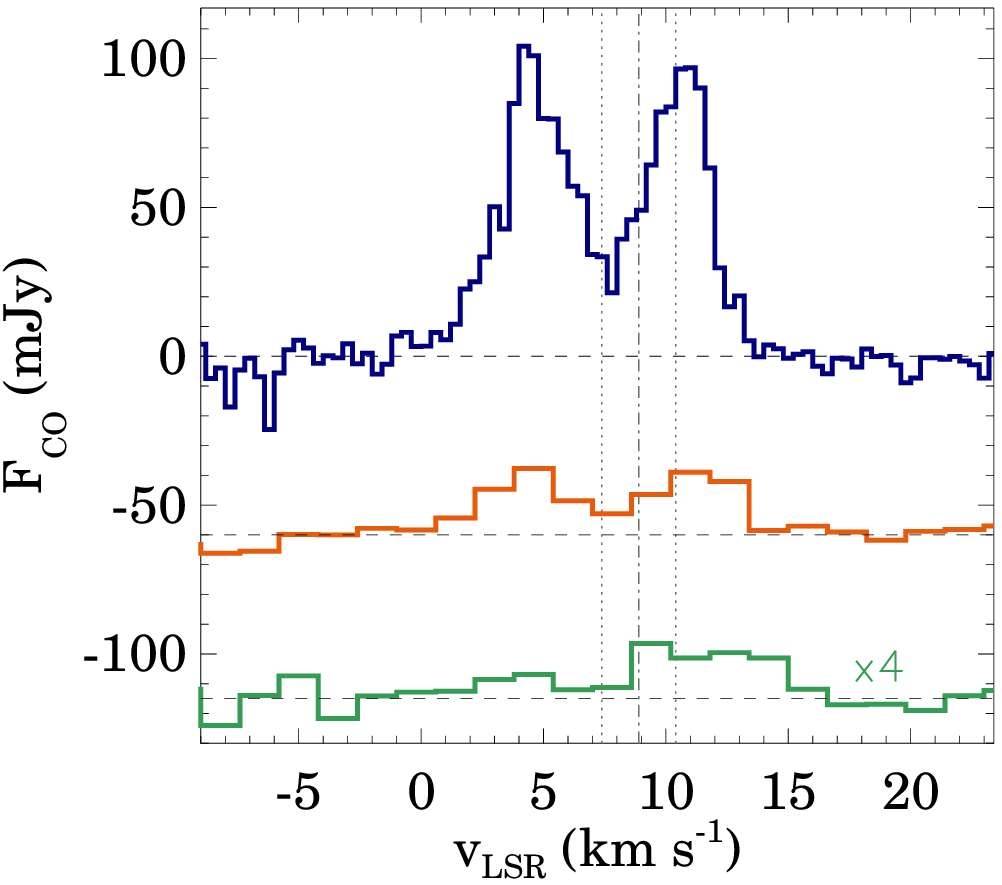}
\includegraphics[width=0.33\textwidth]{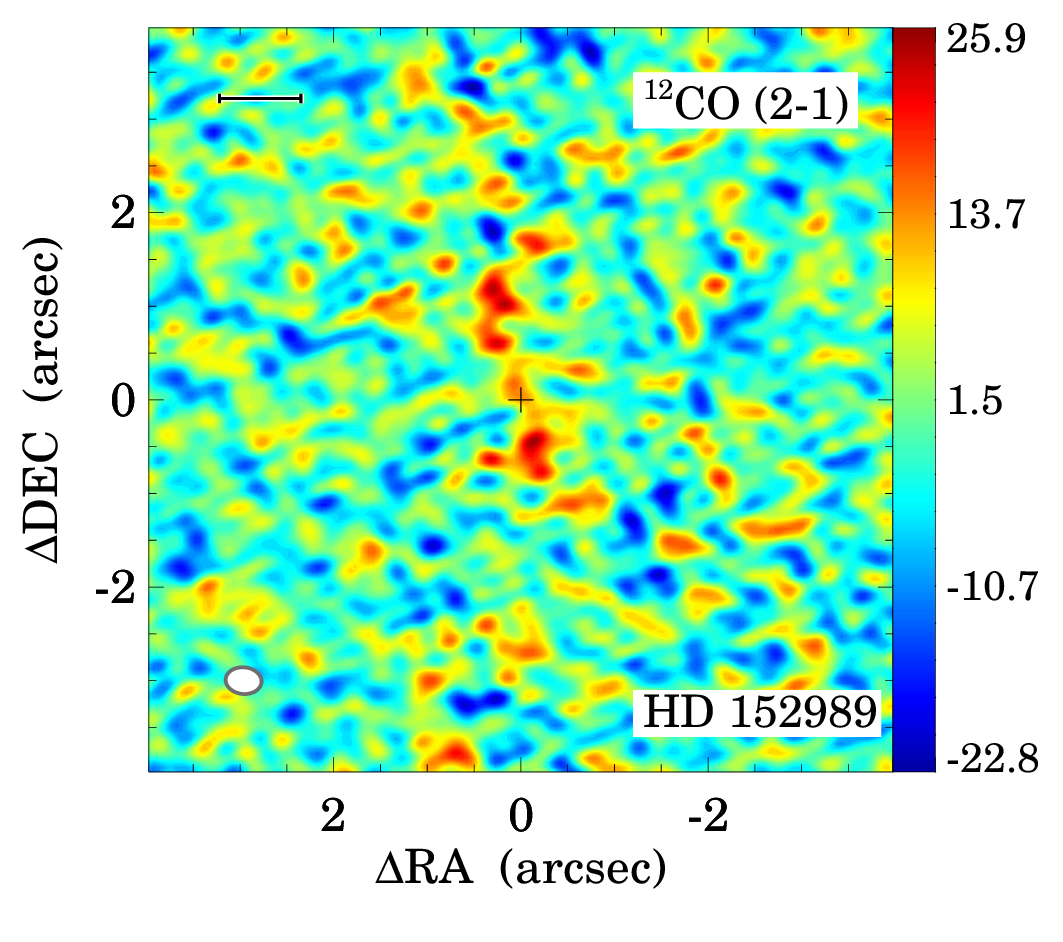}
\includegraphics[width=0.292967\textwidth]{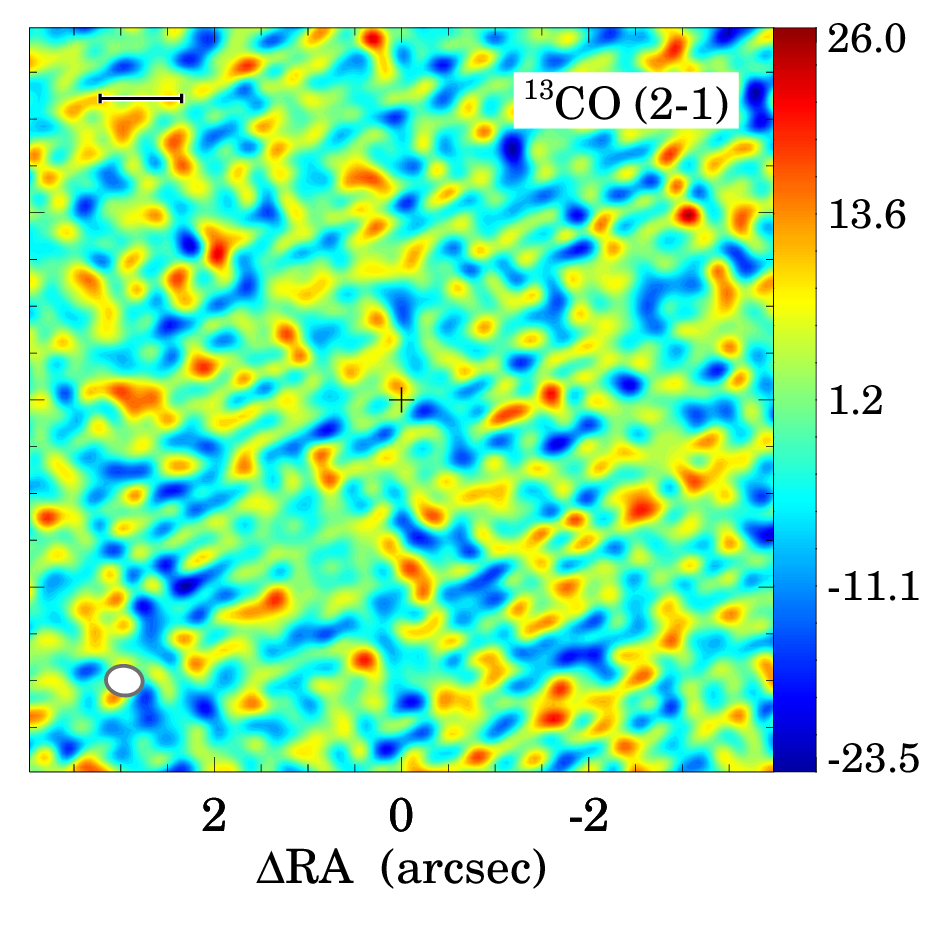}
\includegraphics[width=0.27\textwidth]{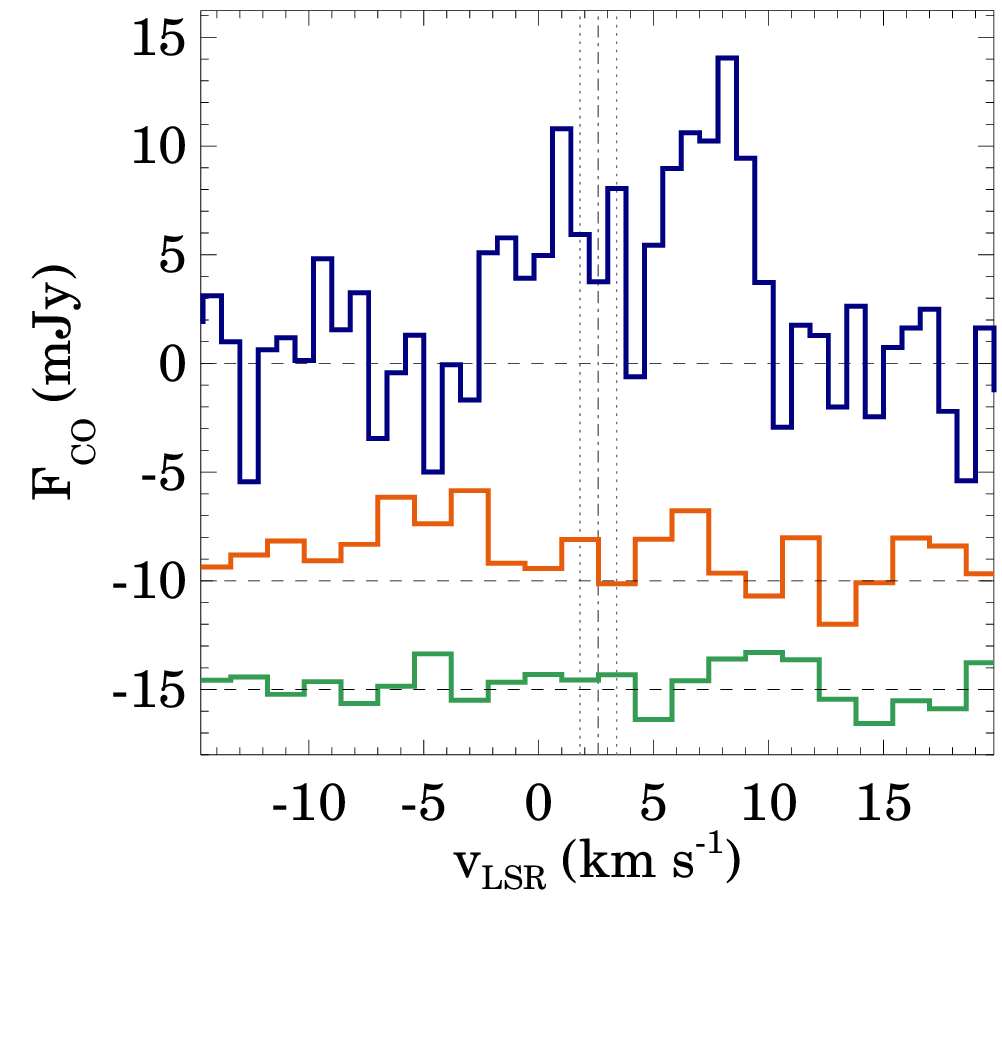}
\caption{$^{12}$CO (2--1) and $^{13}$CO (2--1) moment 0 maps for HD\,9985, HD\,145101, HD\,155853, and HD\,152989 
(left and middle columns). At the bottom left of each panel a filled white ellipse shows the beam size. The length of 
the horizontal black bars corresponds to 100\,au. The color bar units are mJy~beam$^{-1}$~km~s$^{-1}$. 
The contours plotted over the moment 0 maps of HD\,9985 and HD\,155853 show the continuum emission. The contour levels are in steps of [5,8,11,14,17]$\times$ rms noise of 15.9\,$\mu$Jy~beam$^{-1}$ for HD\,9985 and [4,6,8,10]$\times$ rms noise of 15.3\,$\mu$Jy~beam$^{-1}$ for HD\,155853.
The right column displays the obtained CO spectra in all three isotopologs. For better visibility, the $^{13}$CO and C$^{18}$O 
spectra have been shifted downward, and in the case of HD\,9985 and HD\,155853 the C$^{18}$O spectrum is multiplied 
by 4. The horizontal dashed lines show the zero flux levels of the spectra. The vertical dash-dotted and dotted lines 
mark the radial velocity of the star and its uncertainty in the LSR frame.}
\label{fig:coplots}
\end{figure*}

To construct $^{12}$CO moment 0 (velocity integrated) maps for HD\,9985, HD\,145101, HD\,152989 and HD\,155853, 
as a first step we identified those channels where significant, $>$3$\sigma$ emission is detected in the 
region corresponding to the continuum disk. The latter was defined by an ellipse fitted to the outer 
2-$\sigma$ contour of the structure detected in the continuum image. For the much fainter HD\,145101 
and HD\,152989, this analysis was done using data cubes binned in 2 velocity channels (resulting in 
a channel width of 0.8\,km~s$^{-1}$). The moment 0 maps, which are shown in the left column of Figure~\ref{fig:coplots}, 
were then obtained by integrating the data cube over the derived velocity range. 
For all of the above processing, we used self-developed routines implemented in \texttt{IDL} 
(Interactive Data Languages).
For HD\,9985 and HD\,155853 we repeated the above procedure to create their $^{13}$CO moment 0 maps (middle column 
of Fig.~\ref{fig:coplots}). We note that the velocity ranges we obtained are consistent with the ones found for $^{12}$CO.
For HD\,145101, which is very faint in this line, and for HD\,152989, which has no detection at all, 
however, it was not possible to define the velocity range of the integration based on the measured results.
Therefore, in these cases, we adopted the same velocity range extracted for the $^{12}$CO data to construct 
their $^{13}$CO moment 0 maps (middle column of Fig.~\ref{fig:coplots}). 
Finally, to produce the C$^{18}$O integrated maps, we adopted the velocity ranges used for the $^{13}$CO 
observations for all four sources.

To measure the integrated CO line flux densities we used the moment 0 maps. For disks with resolved structure, 
we first determined the integration aperture, i.e. the region that belongs to the disk. We considered pixels 
with intensities of $>$2$\sigma$, where $\sigma$ is the noise in the map, measured in regions outside the disk. 
We kept the area within 2$\sigma$ around the peak with disk emission and discarded any other $>$2$\sigma$ pixels 
disjoint from this region. Then we fitted an ellipse to the mask created this way and used it as an aperture to 
measure the line flux. By changing the size of the elliptical aperture, we obtained a growth curve of fluxes, 
which ended in a flat plateau in all cases. We used the aperture corresponding to the starting point of the 
plateau for the final flux extraction.
While this procedure worked well for the $^{12}$CO and $^{13}$CO maps of HD\,9985 and HD\,155853, where these 
disks form a coherent structure, to analyze the $^{12}$CO data of HD\,152989 we had to define three different 
elliptical aperture masks corresponding to the distinct, southern, central, and northern bright regions 
(see Fig.~\ref{fig:coplots}). To estimate the uncertainty of the integrated line fluxes, we placed 20 apertures 
identical to the source aperture at random positions outside the disk, and then derived the uncertainty as the 
standard deviation of the flux values measured in them. 

For the C$^{18}$O moment 0 map of HD\,9985 and HD\,155853, the above procedure was performed using the aperture mask 
obtained for the $^{13}$CO map. This resulted in measured line fluxes of (20.1$\pm$4.5)$\times$10$^{-23}$\,W~m$^{-2}$ 
and (18.0$\pm$5.8)$\times$10$^{-23}$\,W~m$^{-2}$, respectively, suggesting the presence of faint but detectable 
emission. Similarly, to estimate the line flux density of the disk around HD\,152989 in the two rarer isotopologs, 
we used the aperture obtained for the $^{12}$CO data. This approach yielded 3$\sigma$ upper limits in 
both lines. In the case of HD\,145101, the detected $^{12}$CO emission shows a compact structure, similar to its 
1.3\,mm continuum map. 
To estimate the integrated line flux, we used an elliptical aperture with the same aspect ratio and position angle 
as the corresponding beam, whose size, as in the above cases, was set using the curve of growth method. 
The resulting aperture was then employed to estimate the uncertainties as described above. To estimate the $^{13}$CO 
and C$^{18}$O line fluxes we used the same aperture defined based on the $^{12}$CO moment 0 map.

In the $^{12}$CO observation of HD\,155853, the background emission (see above) is only present in channels between
$v_\mathrm{LSR} = 2.2$ and 4.2\,km~s$^{-1}$, and outside this velocity range, the data are free from its influence.
This enables us to examine the impact of removing the visibility measurements with the smallest baseline on the 
line flux measurement. Using all baselines to construct the $^{12}$CO data cube and repeating the flux measurement 
procedure described above, we found that the flux in the velocity range unaffected by background emission was 9\% 
higher than in the short baseline-free data. Taking this into account, we scaled up the $^{12}$CO line flux, obtained 
from data with small baseline cut, by 9\%. The derived integrated line fluxes are listed in Table~\ref{tab:codiskprops}. 
We have also plotted the spectra extracted from the apertures for all three CO isotopologs in Figure~\ref{fig:coplots} 
(right column).

\begin{figure*}[h!]
\centering
\includegraphics[width=0.33\textwidth]{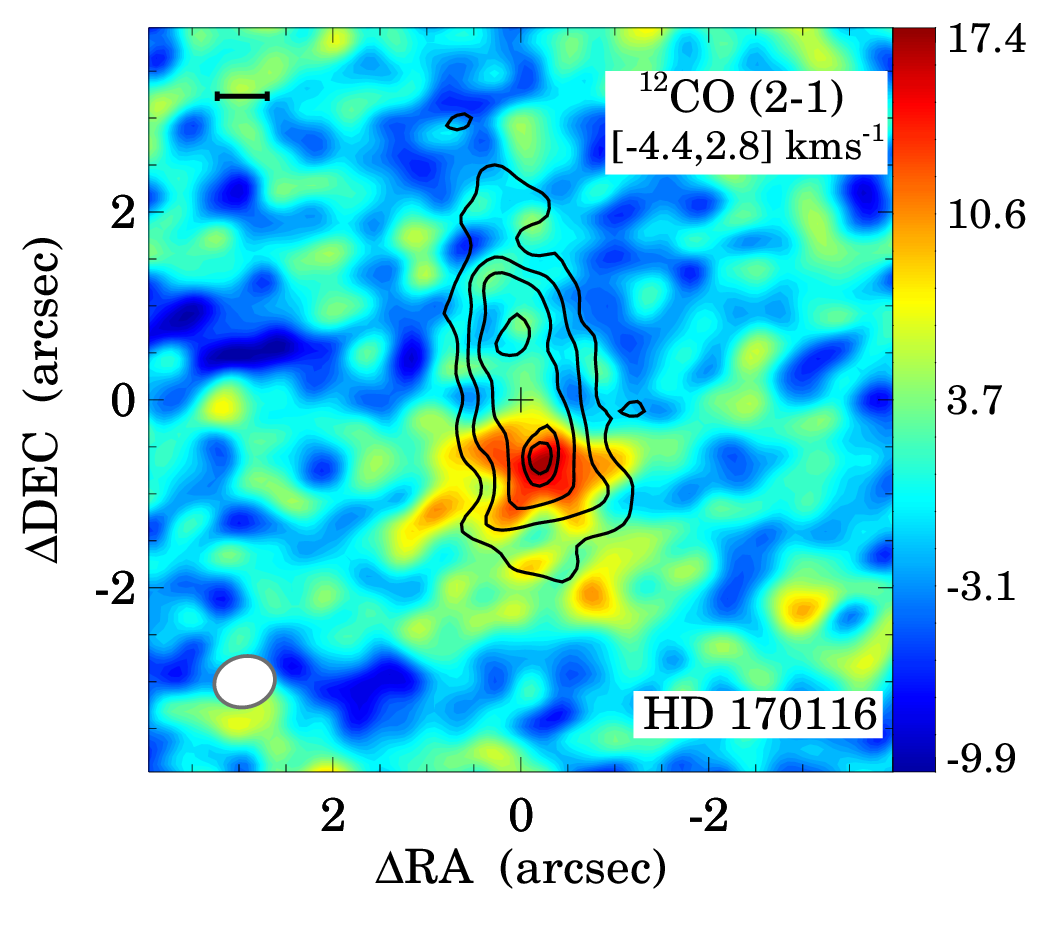}
\includegraphics[width=0.292967\textwidth]{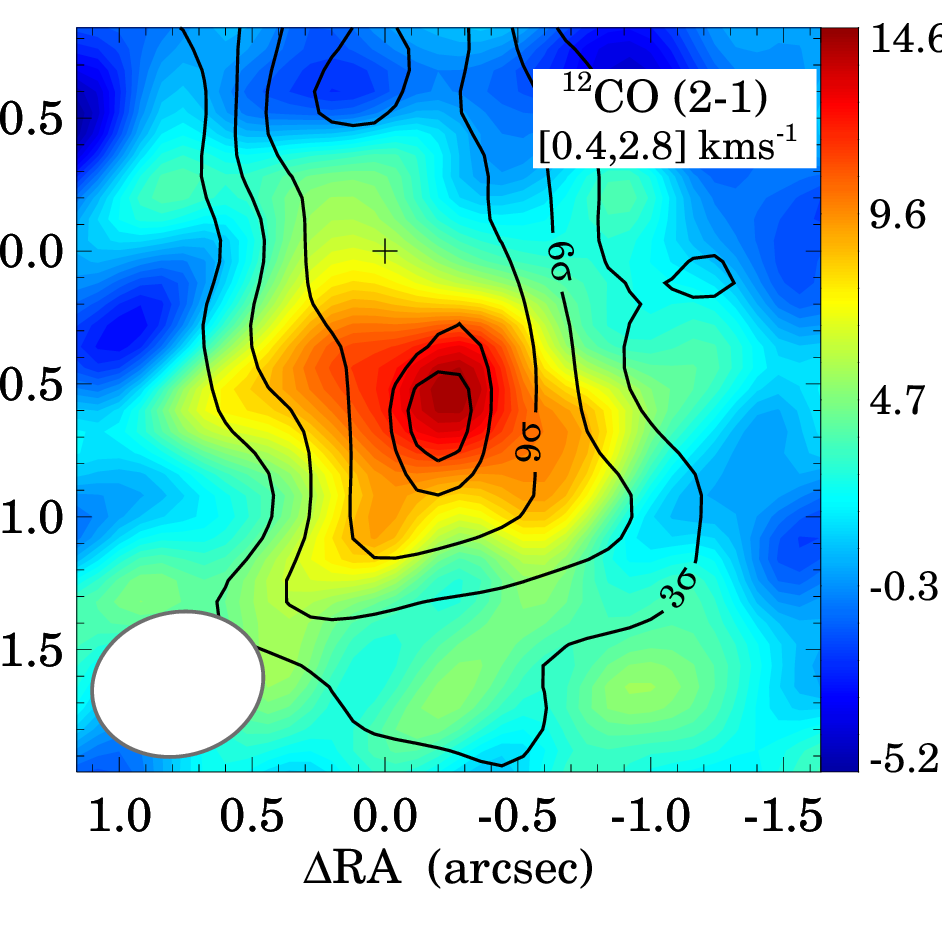}
\includegraphics[width=0.27\textwidth]{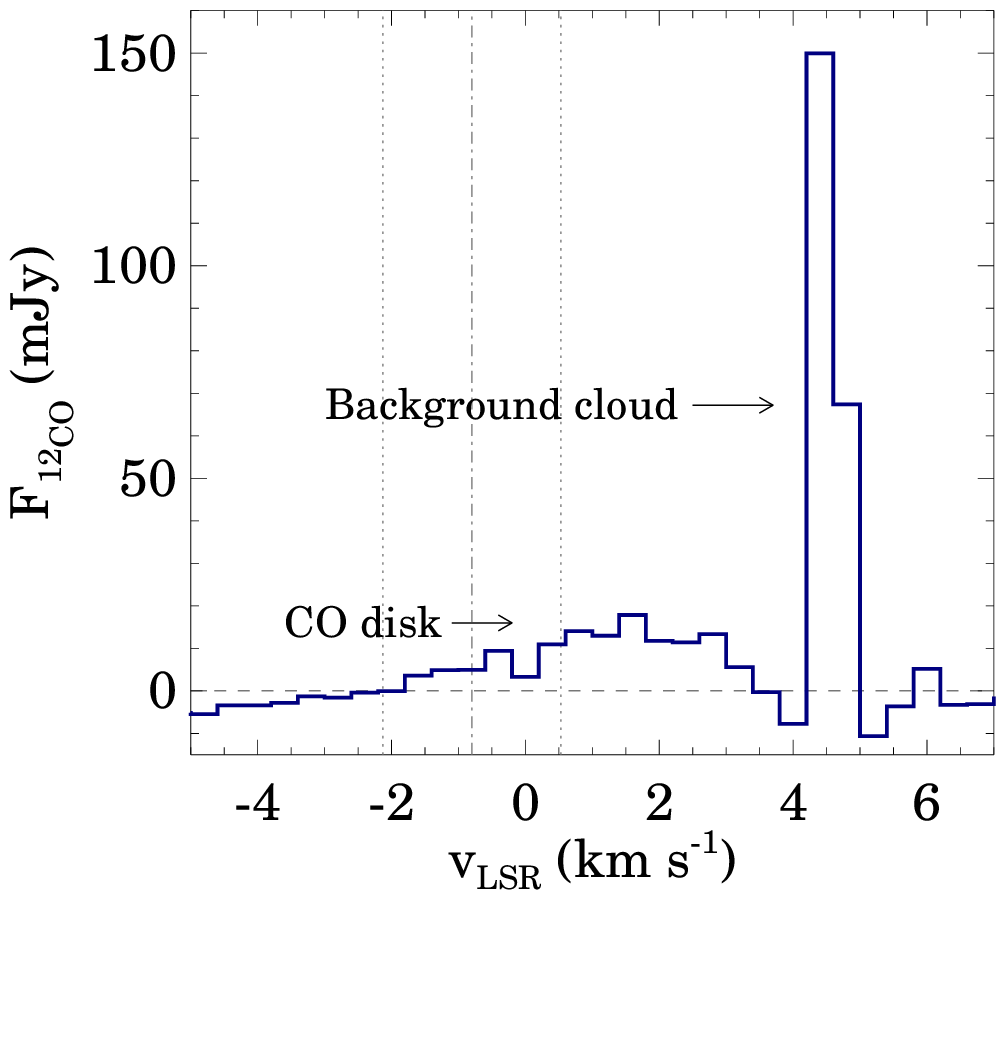}
\caption{Left panel: Integrated $^{12}$CO (2--1) intensity map of HD\,170116 between $v_\mathrm{LSR}$ of $-$4.4 and 
$+$2.8\,km~s$^{-1}$, using natural weighted data. The applied velocity interval was calculated under the assumption that 
the possible gas material is co-located with the detected dust (Sect.~\ref{sec:coimaging}). The color bar units are 
mJy~beam$^{-1}$~km~s$^{-1}$. The moment 0 map is overlaid by the contours of the  natural-weighted 
continuum image of the source. The contour levels are at 3, 6, 9, 12, and 13$\times$ rms of the continuum image. 
The white ellipse in the lower left corner shows the synthetic beam. The size of the horizontal black bar in the upper 
left corner corresponds to 100 au. The black plus sign indicates the position of the star. Center panel: same as the left 
panel, but zoomed in to the center of the observed localized CO brightness peak and using a narrower integration velocity range 
from $+$0.4 and $+$2.8\,km~s$^{-1}$, corresponding to the interval in which the emission is detected. Right panel: 
spectrum of the observed CO peak. The vertical dash-dotted and dotted lines mark the radial velocity of the star 
and its uncertainty in the LSR frame.
\label{fig:hd170116}}
\end{figure*}

To study the faint $^{12}$CO line emission found at HD\,170116, natural weighting was used to improve the 
signal-to-noise ratio. Assuming that the possible gas is co-located with the dust material we can calculate a velocity 
range in which the emission should be present: $v_\mathrm{sys} \pm \sqrt{{GM_*}/{R_\mathrm{in}}} \sin{i},$ 
where $v_\mathrm{sys}$ is the systemic velocity, $M_*$ is the stellar mass, $R_\mathrm{in}$ is the inner 
disk radius, while $i$ is the inclination of the disk. Using the radial velocity of the star in the LSR 
frame as the systemic velocity, taking $M_*$ and $i$ from Tables~\ref{tab:targets} and \ref{tab:contdiskprops} and 
assuming that $R_\mathrm{in} = R_\mathrm{disk} - W_\mathrm{disk}/2$ we obtained a $v_\mathrm{LSR}$ interval of 
[$-$4.4,$+$2.8]\,km~s$^{-1}$ for HD\,170116. Figure~\ref{fig:hd170116} (left) shows the naturally weighted moment 0 
map of the $^{12}$CO 2--1 transition for this velocity range overlaid by the contours of the
natural-weighted continuum image of the source. Significant CO emission is detected only in the southern side of 
the continuum structure, indicating a strong asymmetry in the CO brightness distribution. By applying an elliptical 
aperture fitted to the 3$\sigma$ contour of the continuum disk image, we obtain an integrated line flux of 
(43.6$\pm$10.3)$\times$10$^{-23}$\,W~m$^{-2}$. Analyzing separately the regions corresponding to the southern 
(south-southwest) and northern (north-northeast) parts of the continuum disk, they have
integrated CO fluxes of (40.5$\pm$7.3)$\times$10$^{-23}$\,W~m$^{-2}$ and (3.1$\pm$7.3)$\times$10$^{-23}$\,W~m$^{-2}$, 
respectively. By further examining the data cube, we found that the majority of detected line emission is indeed 
present at velocities between 0.4 and 2.8\,km~s$^{-1}$ (Fig.~\ref{fig:hd170116}, center panel). The observed 
CO structure has an integrated line flux of (28.5$\pm$3.6)$\times$10$^{-23}$\,W~m$^{-2}$ and its brightest part 
coincides well with a point-like clumpy structure detected in the continuum disk.
The spectrum of localized CO brightness peak is shown in Figure~\ref{fig:hd170116} (right). It is worth noting that, although 
in a different velocity range from the disk (and the brightness peak), between $+$4.4 and $+$4.8\,km~s$^{-1}$, additional, 
spatially extended CO emission can be detected in the HD\,170116 measurements. Given the close proximity of this star 
to the Galactic plane ($b = +4\fdg7$), it is likely that the latter, narrow CO line emission is related to a background molecular cloud. 
Although only the faintest parts of this cloud overlap with the disk, it shows up as a narrow peak in the spectrum of 
the CO peak (Fig.~\ref{fig:hd170116}, right). This background cloud has no detectable counterpart in the continuum 
measurement. 

For the analysis of possible $^{13}$CO or C$^{18}$O emission from the disk, moment 0 maps were generated for both the 
expected velocity range of the disk and the measured velocity range of the detected CO brightness peak. However, we found no 
signal above the noise in either line. The derived 3$\sigma$ line flux upper limits are listed in 
Table~\ref{tab:codiskprops}.

To look for gas toward the other seven sources (HD\,31305~A, HD\,112532, HD\,131960, HD\,141960, HD\,144277, HD\,159595, 
and HD\,176497), in addition to the Briggs weighted cubes, we also constructed naturally weighted and tapered (using 
Gaussian tapers of size 1{\arcsec} or 2{\arcsec}) data sets. With the exception of HD\,159595, no significant CO emissions 
were found at any other target. In the case of HD\,159595, however, as mentioned previously, the structure of the spatially 
very extended, bright emission that is present in almost all channels in the studied $v_\mathrm{LSR}$ range between $-$30 
and $+$30\,km~s$^{-1}$, is very different from that of the disk observed in the continuum and is likely to originate from 
background molecular clouds. To mitigate the contamination effect from this extended emission, in the following analysis of 
this source, we used a data cube where only visibilities with baselines of $>$\,100\,m were used in the \texttt{tclean} process. 
To constrain upper limits of the $^{12}$CO integrated line fluxes for these seven systems, we first derived their moment 0 
maps by using the expected velocity intervals. For HD\,112532, HD\,141960, HD\,159595, and HD\,176497 we assumed that the 
possible gas is co-located with the observed dust disk (Sect.~\ref{sec:uvmultifit}). In the case of HD\,131960, HD\,144277, 
and HD\,31305~A, no emission was detected around the stars in the mm continuum. While mid-IR measurements provide evidence 
of warm dust disks around the first two objects, it is unclear whether there is circumstellar dust around HD\,31305~A 
(Sect.~\ref{sec:uvmultifit}). Based on the dust temperatures obtained from SED fitting (Sect.~\ref{sec:diskprops}), 
we first estimated the blackbody radii of the disks around HD\,131960 and HD\,144277. Due to the presence of small dust 
particles that are inefficient emitters and thus warmer than blackbody grains present at the same location, the radii derived 
in this manner are lower limits \citep[e.g.,][]{pawellek2014}. To estimate the true disk radii, we applied an empirical 
relation based on spatially resolved debris disks to examine the ratio of the two radius types as a function of the central 
stars' luminosity \citep{pawellek2021}. Assuming that the disks are edge-on and the possible gas is co-located with the dust, 
we adopted a velocity interval of 20\,km~s$^{-1}$ and 16\,km~s$^{-1}$ centered on the radial velocity of the star in the LSR 
frame for HD\,131960 and HD\,144277, respectively. For HD\,31305~A we assumed a line width of 20\,km~s$^{-1}$ centered on the 
systemic velocity of the CO disk around HD\,31305~B (Appendix~\ref{sec:hd31305b}). To estimate the 3$\sigma$ upper limits of 
the integrated line fluxes, for objects where we detected the disk in the continuum (HD\,112532, HD\,141960, HD\,159595, and 
HD\,176497), we placed 20 apertures at random positions on the moment 0 maps and then computed the 1$\sigma$ uncertainty as 
the standard deviation of the flux values measured in them. The aperture was defined by fitting an ellipse to the detected 
continuum disk. For HD\,131960 and HD\,144277, the 3$\sigma$ upper limits were derived from the noise measured in their 
1{\arcsec} tapered moment 0 maps, while for HD\,31305~A we used the Briggs weighted map for this purpose. The obtained upper 
limits are listed in Table~\ref{tab:codiskprops}.


\begin{table*}                                                                                                             
\begin{center}                                                                                                                                       
\caption{Results from CO line observations.  
 \label{tab:codiskprops} }
\begin{tabular}{lccccc}                                                     
\hline\hline
Target name & $F_\mathrm{^{12}CO}$            &         $F_\mathrm{^{13}CO}$    &         $F_\mathrm{C^{18}O}$     &   $M_\mathrm{CO}$ & $\tau_\mathrm{^{12}CO}$ \\
            & (10$^{-23}$\,W~m$^{-2}$)        & (10$^{-23}$\,W~m$^{-2}$)        & (10$^{-23}$\,W~m$^{-2}$)         &  ($M_\oplus$)     &                         \\  
\hline	     
\multicolumn{6}{c}{Detections}        \\
\hline
HD\,9985              &  559.1$\pm$56.3 (6.9) & 316.9$\pm$32.4 (6.6) & 22.5$\pm$5.6 (5.2)  &  (2.2$\pm$0.2)$\times$10$^{-2}$  &  80$\pm$22 \\
HD\,145101            &  80.4$\pm$10.3 (6.5)  &  19.3$\pm$4.2 (3.7)  &  $<$10.1            &  (10.3$\pm$2.3)$\times$10$^{-4}$ &  25$\pm$8 \\
HD\,152989            &  67.7$\pm$10.7 (8.2)  &  $<$21.2             &  $<$16.7            &  (2.7$\pm$0.4)$\times$10$^{-5}$  &  \ldots  \\
HD\,155953            & 567.5$\pm$60.8 (21.7) & 134.5$\pm$15.9 (8.5) & 20.9$\pm$6.7 (6.4)  &  (7.0$\pm$0.8)$\times$10$^{-3}$  &  24$\pm$4 \\ 
HD\,170116            &  40.5$\pm$8.4 (7.3)   &   $<$21.5            &  $<$18.8            &  (4.4$\pm$0.9)$\times$10$^{-5}$  &  \ldots \\
HD\,170116 (CO peak)  &  28.5$\pm$4.0 (2.8)   &   $<$6.8             &  $<$5.5             &  (3.1$\pm$0.4)$\times$10$^{-5}$  & \ldots \\
\hline
\multicolumn{6}{c}{Non-detections}       \\
\hline
HD\,31305  &  $<$25.1  & \ldots & \ldots  &  $<$1.6$\times$10$^{-5}$  & \ldots \\
HD\,112532 &  $<$17.5  & \ldots & \ldots  &  $<$7.7$\times$10$^{-6}$  & \ldots \\
HD\,131960 &  $<$31.8  & \ldots & \ldots  &  $<$2.1$\times$10$^{-5}$  & \ldots \\
HD\,141960 &  $<$13.6  & \ldots & \ldots  &  $<$8.5$\times$10$^{-6}$  & \ldots \\
HD\,144277 &  $<$22.1  & \ldots & \ldots  &  $<$1.4$\times$10$^{-5}$  & \ldots \\
HD\,159595 &  $<$107.0 & \ldots & \ldots  &  $<$2.8$\times$10$^{-5}$  & \ldots \\
HD\,176497 &  $<$25.4  & \ldots & \ldots  &  $<$1.8$\times$10$^{-5}$  & \ldots \\
\hline
\end{tabular}
\tablefoot{Integrated CO line fluxes ($^{12}$CO, $^{13}$CO, and C$^{18}$O (2--1) fluxes for disks with detected $^{12}$CO emission and 
only the $^{12}$CO flux upper limits for non-detections) and CO mass estimates ($M_\mathrm{CO}$) for our twelve targets. 
The quoted uncertainties of the line fluxes are quadratic sums of the measurement errors (given in brackets) and the
absolute calibration error (which were conservatively assumed to be 10\%). 
For the three disks detected in both the $^{12}$CO and the $^{13}$CO emission lines, the last column provides the estimated $^{12}$CO 
optical depth. The quoted uncertainties of the $\tau_\mathrm{^{12}CO}$ estimates are formal and only consider the uncertainties of the 
line flux measurements, but not account for the uncertainties of the adopted isotopolog abundance ratio.
}
\end{center}
\end{table*}


\subsection{CO mass estimates}
Taking an optically thin $J=2-1$ rotational transition line of the observed isotopologs, $^{12}$CO, $^{13}$CO, or C$^{18}$O, 
the total mass of $^{12}$CO gas can be estimated as:
\begin{equation} 
M_{CO} = {4 \pi m d^2} \frac{F_\mathrm{21}}{x_2 h \nu_\mathrm{21} A_\mathrm{21}} f_{^{12}C^{16}O/^{x}C^{y}O},    
\end{equation}
where $m$ is the mass of a $^{12}$CO molecule, $d$ is the distance of the object, $F_\mathrm{21}$ is the measured $J=2-1$ 
line flux of the given $\mathrm{^{x}C^{y}O}$ isotopolog, $h$ is the Planck constant, $\nu_{21}$ and $A_{21}$ are the rest 
frequency and the Einstein coefficient of the transition, while $x_2$ is the fractional population of the upper, $J=2$ level. 

The $f_{^{12}C^{16}O/^{x}C^{y}O}$ is the abundance ratio of the $^{12}$CO molecule relative to the specific 
isotopolog. In the local interstellar medium, the [$^{12}$C]/[$^{13}$C] and [$^{16}$O]/[$^{18}$O] ratios are estimated to be 
$\sim$77 and $\sim$560, respectively \citep{wilson1994}. In the following analysis, we assume corresponding 
CO isotope ratios in the observed gas disks. Using this assumption, the line opacities can be estimated from the obtained flux 
ratios \citep[][eq.~1]{difolco2020}. For HD\,9985, HD\,145101, and HD\,155853, the measured $^{12}$CO to $^{13}$CO flux ratios 
are between 1.7 and 4.5, indicating highly optically thick $^{12}$CO emission in all cases with $\tau_\mathrm{^{12}CO}$ of 
$\sim$80, $\sim$25, and $\sim$24, respectively. However, these results also suggest that their $^{13}$CO line 
may be optically thin. This statement is further strengthened in the case of HD\,9985 and HD\,155853 by their measured
$F_\mathrm{^{12}CO}/F_\mathrm{C^{18}O}$ ratios. For HD\,9985, HD\,145101, and HD\,155853, we therefore use the 
$^{13}$CO line measurements to estimate the CO gas mass, while for HD\,152989 and HD\,170116, where only $^{12}$CO 
emission was detected, we assume that the given line is optically thin.     

The fractional population of the $J=2$ level ($x_2$) depends on how often the CO molecules collide with other particles, the 
nature of the collision partners, and the radiation field to which they are exposed. In dense media, the collisions can be 
frequent enough to dominate the excitation of the rotational levels, leading to local thermodynamic equilibrium (LTE) in which 
the level populations follow the Boltzmann distribution. If we assume this situation in our newly discovered gas disks, i.e. 
that LTE holds, then we need the gas excitation temperature to estimate $x_2$. Using the measured peak flux values of the 
optically thick $^{12}$CO data we obtained minimum gas brightness temperatures of 11\,K for HD\,9985 and HD\,155853 and 5\,K 
for HD\,145101. In previously identified CO-bearing debris disks, where the gas excitation temperature could be estimated, 
similarly low values between 8 and 30\,K were typically obtained \citep{kospal2013,flaherty2016,matra2017b,higuchi2020,difolco2020}.
In our calculations we have finally adopted an excitation temperature of 20\,K for all gaseous disks. The obtained CO gas mass 
estimates and their formal uncertainties, which take into account the errors in the distance and flux values, are summarized 
in Table~\ref{tab:codiskprops}.

Several simplifications have been made in the above calculations, and the uncertainties in the CO mass estimates are in fact 
likely larger than the calculated formal errors. To assess how feasible the hypothesized LTE approximation is, we should know 
what types of collision partners are present and what their densities are. However, no measurements are available for other 
possible important constituents of the observed gas disk, such as H$_\mathrm{2}$ molecules if the gas is primordial, or H, C, 
O atoms or electrons in the case of second generation gas material. We also know little about the real gas temperatures. For 
our adopted excitation temperature of 20\,K and LTE, the $J=2$ rotational level is very efficiently populated, but a significantly 
different temperature and/or non-LTE environment may result in a lower fractional population than assumed and, accordingly, a 
higher CO mass.

For three disks, the $^{12}$CO mass was estimated from the measured $^{13}$CO line fluxes, assuming a $^{12}$CO/$^{13}$CO 
abundance ratio that is typical of the local interstellar material. However, isotope selective photodissociation 
or isotopic fractionation can significantly affect the abundances of these molecules leading to very different 
ratios from disk to disk \citep{visser2009}. While the former process tends to result in a higher $^{12}$CO/$^{13}$CO abundance 
ratio, the fractionation can even reduce it if the gas is sufficiently cold. 

Taking all these effects into account, the actual mass of CO may differ by as much as several factors from what we have 
estimated, and is more likely to be higher.

For those targets where no gas emission was detected, we derived an upper limit for the CO mass. To do this, we used the 
obtained CO flux upper limits and assumed LTE conditions and a gas excitation temperature of 20\,K. 
The results are given in Table~\ref{tab:codiskprops}. 
As a caveat, we note that in these presumably gas-poor systems, it is likely that radiative processes 
rather than the collisions govern the population of the rotational levels. Such subthermal excitation 
can lead to a lower population of the J=2 rotational level ($x_2$) than we assumed, resulting in 
higher upper limits for CO mass \citep{matra2015,cataldi2023}.

\subsection{Spatial distribution of CO gas} \label{sec:cospatial}
\begin{figure*}[h!]
\centering
\includegraphics[width=0.98\textwidth]{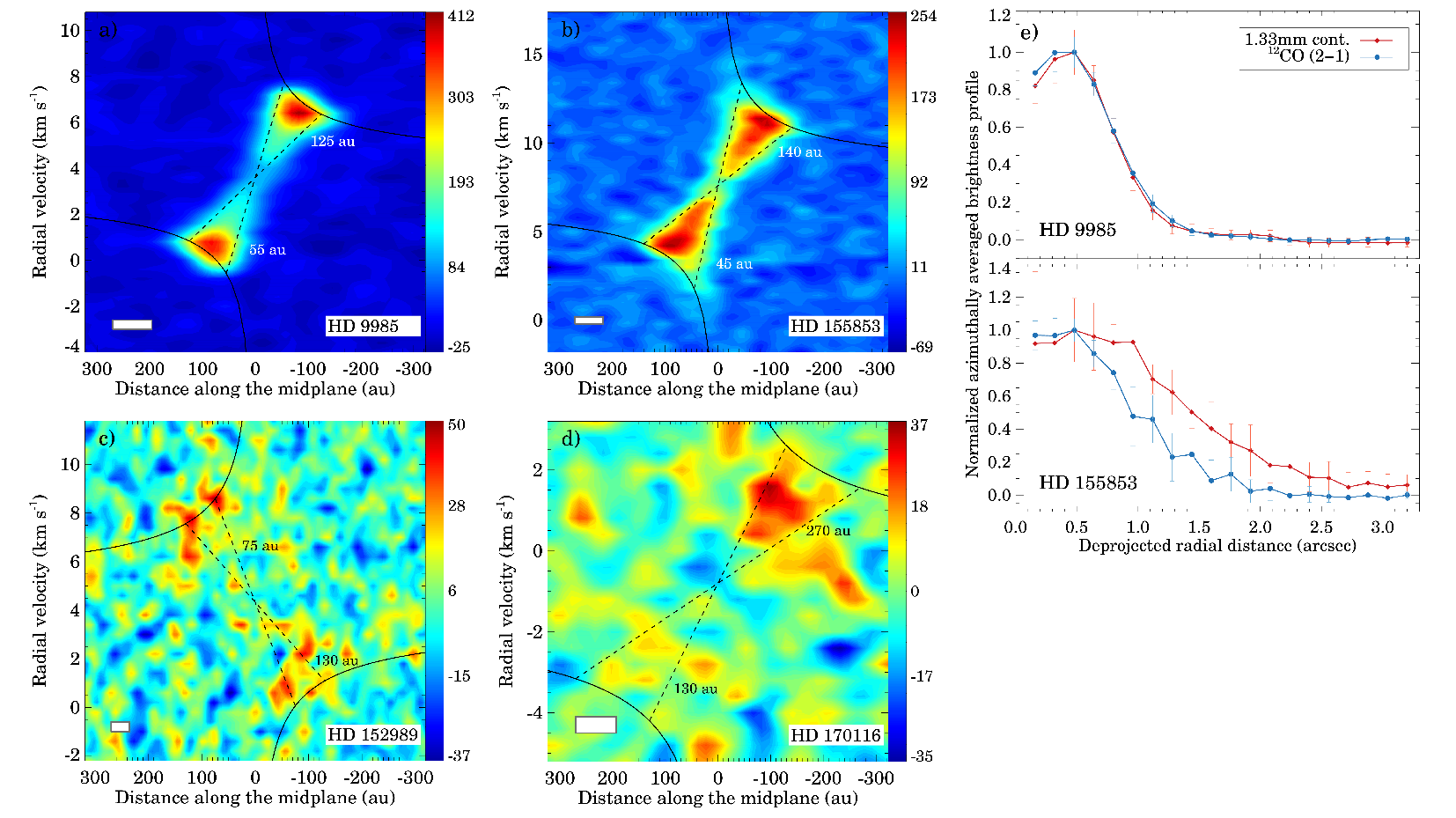}
\caption{a--d): Position-velocity diagrams of $^{12}$CO (2-1) emission for HD\,9985 (a), HD\,155853 (b), 
HD\,152989 (c), and HD\,170116 (d). The color bar units are mJy~beam$^{-1}$~channel$^{-1}$.
The black solid curves show the tangential velocity of the gas, 
while the black dashed diagonal lines display the line-of-sight velocity of the gas at a fixed orbital radius 
as a function of projected separation. To derive these, we assume a Keplerian rotation profile for the gas, 
the distance and mass of the star are taken from from Table~\ref{tab:targets}, and the position 
angle and inclination of the gas disk are from Sect.~\ref{sec:cospatial}. The white rectangle at the bottom left 
represents the spectro-spatial resolution. e): Azimuthally averaged radial profile for the continuum and 
$^{12}$CO (2-1) data for HD\,9985 (top) and HD\,155853 (bottom). For comparison, the profiles are 
normalized to unity.
\label{fig:pvd}}
\end{figure*}

To investigate whether the detected CO gas is co-located with the dust in HD\,9985, HD\,155853, HD\,152989, and 
HD\,170116, we constructed their $^{12}$CO position-velocity (PV) diagrams along the major axis of the disks 
(Figs.~\ref{fig:pvd}a-d). 
The overplotted black solid curves show the tangential velocity of the gas as a function of projected distance 
from the star, while the diagonal black dashed lines mark the radial velocities and positions of the gas orbiting 
at two orbital radii, assuming Keplerian rotation. To calculate the diagrams, we need as input parameters the 
position angles and inclinations of the disks as well as the distances and stellar masses \citep{matra2017b}. 
For HD\,152989 and HD\,170116, where the CO emission is relatively weak, the disk position angles and inclinations
obtained for the continuum data were adopted from Table~\ref{tab:contdiskprops}. In the case of HD\,9985 and HD\,155853 
we used their moment 0 maps to estimate these disk parameters. To this end we made cuts from the stellar position 
toward position angles between 0 and 360{\degr} in 5{\degr} increments, determined the location of maximum intensity 
along these cuts, and finally the derived positions were fitted with an ellipse. In this way, we obtained 116$\pm$7{\degr} 
and 49$\pm$3{\degr} for the position angle and inclination of the CO disk around HD\,9985, and 2.5$\pm$1{\degr} and 
77$\pm$2{\degr} for HD\,155853. We note that these disk parameters are in very good agreement within the uncertainties 
with the $PA$ and $i$ estimates obtained from their continuum observations (Table~\ref{tab:contdiskprops}). The distances 
and stellar masses were taken from Table~\ref{tab:targets} for all four systems.

 The PV diagram shows that the bulk of the CO gas in HD\,9985 lies between $\sim$55 and $\sim$125 au (Fig.~\ref{fig:pvd}a). 
By modeling its continuum data we found that the peak of the dust radial distribution is at $\sim$84\,au and the FWHM of 
the fitted Gaussian profile is $\sim$57\,au (Table~\ref{tab:contdiskprops}), indicating that the densest part of the 
dust ring is confined between $\sim$55\,au and $\sim$115\,au, i.e. the dust and gas in this system are probably 
quite well co-located. The good agreement of the images and normalized azimuthally averaged radial flux profiles for the continuum 
and $^{12}$CO moment 0 maps in Figures~\ref{fig:coplots} and \ref{fig:pvd}e (top) further supports this finding.
As Figures~\ref{fig:continuumplots} and \ref{fig:coplots} show, for HD\,9985, the brightness distributions 
of the $^{12}$CO and $^{13}$CO integrated maps, and the continuum emission all peak in a similar location at the northwestern 
ansa of the ring. The maximum brightness of the northwestern part of the ring exceeds that of the southwestern part by 2.4, 
2.4, and 2.6$\sigma$ in the continuum, $^{12}$CO, and $^{13}$CO maps, respectively. Although the differences measured in 
individual maps are not significant in themselves, considering them together, they hint at a brightness asymmetry, given the 
positional coincidence. Detailed analysis of this brightness asymmetry will be the topic of a future study. 

In the case of HD\,155853, we found that the CO gas is mostly situated between 45 and 140\,au (Fig.~\ref{fig:pvd}b), 
closer to the star than large dust particles whose emission comes mainly from radial distances between 60 and 175\,au 
(Table~\ref{tab:contdiskprops}). Figures~\ref{fig:coplots} and \ref{fig:pvd}e (bottom) also clearly illustrate this 
discrepancy in the radial spatial distributions of gas and dust. This property of HD\,155853 makes it most similar to 
HD\,21997 and HD\,131488 among the previously known CO-bearing disks, in which the gas and dust components also show 
significant spatial differences. In these disks the inner boundary of the gas appears substantially closer to the star 
than that of the large dust particles \citep{kospal2013,pawellek2024}. Based on the $^{12}$CO and $^{13}$CO moment 0 maps 
(Fig.~\ref{fig:coplots}), the northern part of the disk is brighter than the southern one. However, the difference in peak 
emissions between the northern and southern regions is less than 1.7\,$\sigma$ in both maps. Further higher sensitivity 
observations are needed to confirm and study the possible brightness asymmetry. We note that in the continuum image, 
contrary to the CO moment 0 maps, the southern part of the disk is brighter, but the peak is not significant, since the 
brightness difference compared to the northern part is only $\sim$1$\sigma$.

Assuming a dust ring with a Gaussian radial brightness profile, the analysis of the continuum data for HD\,152989 indicated that 
the majority of the large grains are located between 80 and 125\,au (Table~\ref{tab:contdiskprops}). In contrast, the PV diagram 
of the object suggests that there is little gas between these two radial distances, and that the CO is instead mostly concentrated 
in two rings at $\sim$75 and $\sim$130\,au (Fig.~\ref{fig:pvd}c). Motivated by this result, we examined whether the dust 
distribution could be consistent with such a morphology. To do this, we adopted a \textsc{uvmultifit} model in which the emission 
is associated with two infinitesimally thin rings with the same position angle and inclination. The model fit yielded a position 
angle of 15$\pm$1{\degr} and an inclination of 77$\pm$2{\degr}, both are consistent with the results of the Gaussian ring model 
within the uncertainties (Table~\ref{tab:contdiskprops}). For the radius and flux density of the inner ring we obtained 
83$\pm$13\,au and 0.56$\pm$0.20\,mJy, while the best estimate for the radius and flux density of the outer ring are 120$\pm$15\,au 
and 0.90$\pm$0.19\,mJy. Using BIC to assess the goodness-of-fit of the two assumed models -- the single Gaussian ring and the double 
narrow ring, which have 7 and 8 free parameters, respectively -- we found no significant difference between them.
The best-fit radii of the two continuum dust rings agree reasonably well with the estimated radii for the possible two gas rings. 
It is therefore conceivable that the dust and gas are in fact quite well co-located and exist in two rings separated by 
a zone depleted in both materials. This would make this system unique as we do not know other gas-bearing debris disks 
where the gas is located in multiple distinct rings.

According to the continuum model of HD\,170116, the bulk of the dust observed at 1.33\,mm lies between 
130 and 270\,au (Fig.~\ref{fig:pvd}d). The PV diagram of the detected $^{12}$CO gas, which is concentrated in the southern wing 
of the disk, also shows that the gas appears to reside roughly within this radial interval. The bright CO clump is located close 
to the inner edge of the disk. Note that while for the other three objects we used the $^{12}$CO data to estimate the systematic 
velocity of the gas, this was not possible for the HD\,170116, where we therefore adopted the measured radial velocity of the star.

By fitting a two-dimensional elliptical Gaussian model to the $^{12}$CO moment 0 map of HD\,145101, we found that its 
gas disk is marginally resolved and has FWHMs of 0\farcs39$\pm$0\farcs11$\times$0\farcs30$\pm$0\farcs10, implying 
a characteristic disk radius of 27$\pm$8\,au. This disk size is consistent with the upper limit obtained from 
the analysis of the continuum data ($<$33\,au, Table~\ref{tab:contdiskprops}).

\section{Discussion} \label{sec:discussion}
In our mini-survey we observed 12 young intermediate mass stars, 9 of which are found to have detectable millimeter continuum 
emission from surrounding dust-rich debris disks, and 5 of which also exhibit CO line emission. 
Over the past decade, dozens of young, dust-rich debris disks have been targeted in various ALMA observing programs to study 
their fundamental properties and assess their CO content. In the following discussion, we combine our results with data from 
the literature to investigate the occurrence of CO gas in dust-rich young debris disks, as well as the similarities or differences 
of the newly discovered CO-bearing debris disks to previously identified ones. The unified sample also allows us to improve our 
knowledge of the general characteristics of young CO-bearing systems, to examine the environmental conditions under which they 
can form, and to confront the observational results with the current theories of the origin of the gas. The dust continuum  
has been spatially resolved for most of our targets allowing us to investigate possible mechanisms for the dynamic excitation 
of these disks and whether the result of this analysis could indicate the presence of larger planet(s) in the system.

\subsection{Sample of young dust-rich debris disks observed by the ALMA}
For one of our targets, HD\,31305, the millimeter continuum and line emission has been found to originate from a disk around its 
companion star, presumably of protoplanetary nature. This discovery raises the possibility that the observed  infrared excess of 
the system is also more likely associated with this young gas-rich disk, and that HD\,31305 may have no circumstellar material 
at all, or that its possible disk is much fainter than previously thought. We have therefore excluded this system from our further 
analysis. The other 11 disks -- 10 of which are hosted by A-type stars and 1 possessed by an early F-type star -- are all younger 
than 50\,Myr and have fractional luminosities $>$5$\times$10$^{-4}$. In the literature, we identified 46 additional debris disks 
around stars with spectral types from A to K that are similarly young and dust-rich and for which ALMA millimeter CO line 
measurements are available 
\citep{kospal2013,dent2014,lieman-sifry2016,marino2016,moor2017,moor2019,difolco2020,kral2020,schneiderman2021,lovell2021,rebollido2022}. 
The relevant properties of the {46} systems are summarized in Table~\ref{tab:codisksample}. Of the total sample of {57} disks, 
{30} are around A-type stars, while the other {27} host stars have F--K spectral types 
\footnote{Our original target list included 28 dust-rich debris disks, out of which 12 
were observed in the project (Sect.~\ref{sec:sampleselection}).  We note that of the 16 
targets that were finally not measured, eight have since been observed with the ALMA 
12-m Array as part of two other projects (2022.1.00968.S, PI: James Miley; 2024.1.00852.S, 
PI: Joshua Lovell). 
The latter measurements were also performed in Band\,6 and similarly to our project they also 
targeted the $^{12}$CO, $^{13}$CO, and C$^{18}$O rotational lines in addition to the continuum. 
Following their analysis, the sample can be expanded further in the future.}.

\subsection{Detection rate of CO gas} \label{sec:detrates}
In the combined sample, the presence of CO gas has been detected in {19} disks around A-type stars and in {4} disks surrounding 
later-type stars. So, in line with previous results \citep{lieman-sifry2016,moor2017}, the majority of CO-bearing disks are found 
around A-type stars. However, the new, larger sample now allows a finer analysis of the detection rate as a function of the 
luminosity of the host stars. All but two gaseous disks have $^{12}$CO luminosities 
$L_\mathrm{{{12}{CO(2-1)}}}>$9.5$\times$10$^{16}$\,W. The gas disks around HD\,172555 and 
HD\,181327 have substantially lower $^{12}$CO (2-1) luminosities of 1.3x10$^{16}$\,W and 6.5x10$^{15}$\,W, respectively. They are 
among the closest stars in the sample, and similarly faint gas disks such as they host would be undetectable for most of the other 
objects in the sample, which are typically located $>$100\,pc away. 
The sample also includes three objects whose 3$\sigma_{L_\mathrm{{12}{CO(2-1)}}}$ detection limits are higher than 
9.5$\times$10$^{16}$\,W (HD\,145263, HD\,145880, and HD\,159595). To avoid detection bias when examining the dependence of detection 
rates on stellar luminosity, the above five objects were not taken into account.
We then divided the remaining {52} systems into four subgroups of {13} objects each, according to their stellar luminosity.
In the case of HD\,106906, which is a nearly equal mass binary with a projected separation of 0.36--0.58\,au 
\citep{rodet2017}, $\sim$3.5\,L$_\odot$, half of the total luminosity of the system was taken 
into account in the course of grouping. From the lowest to the highest luminosities, the subgroups L1 (0.18\,L$_\odot<$ L$_*<$ 3.52\,L$_\odot$), 
L2 (3.53\,L$_\odot<$ L$_*<$ 6.47\,L$_\odot$), L3 (6.54\,L$_\odot<$ L$_*<$ 11.7\,L$_\odot$), and 
L4 (11.8\,L$_\odot<$ L$_*<$ 21.9\,L$_\odot$) typically include stars with K7--F5, F5--A9, A8--A4 and A4--A0 spectra. 

\begin{figure}[h!]
\centering
\includegraphics[width=0.47\textwidth]{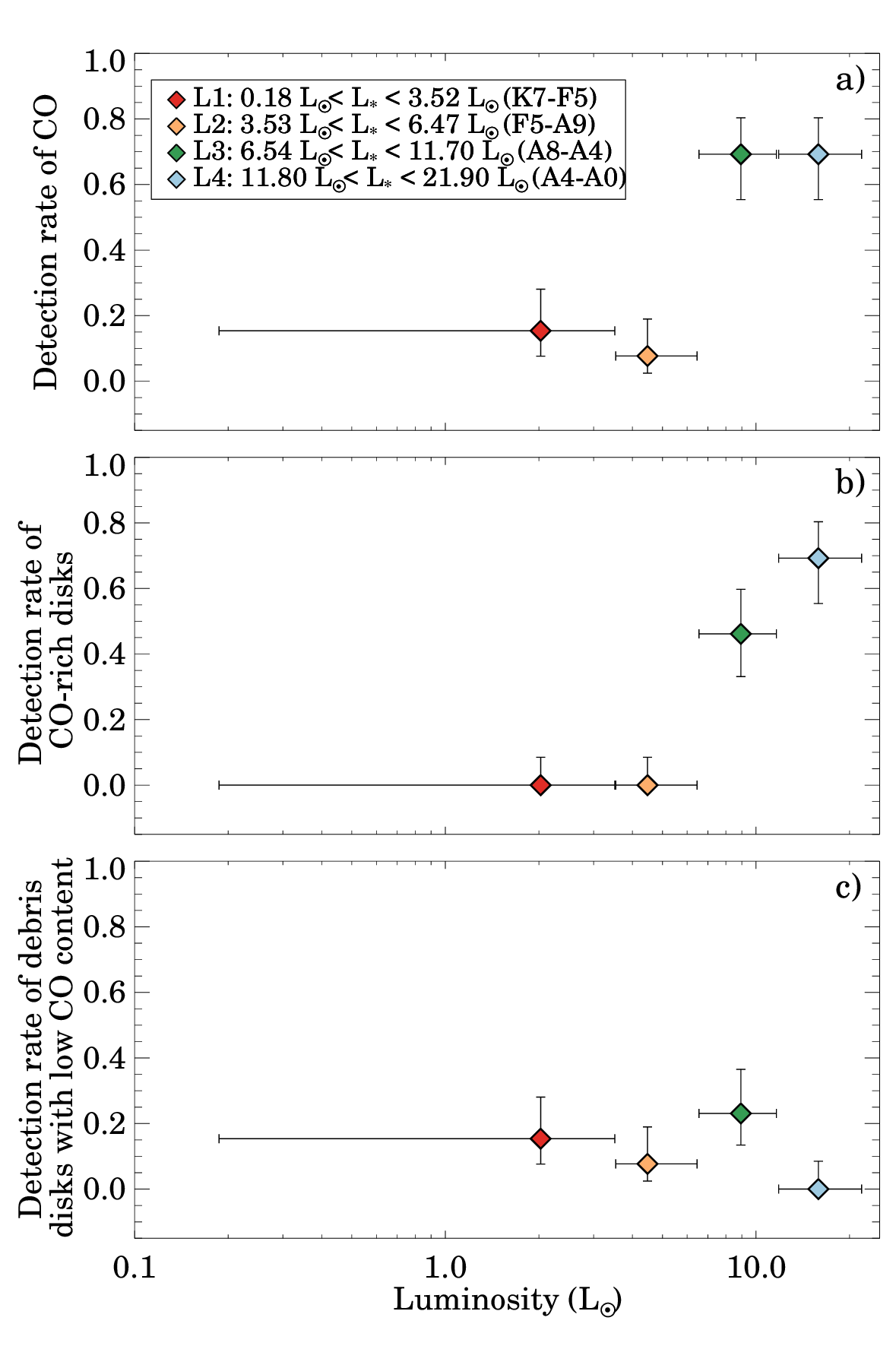}
\caption{a) Detection rates of CO gas in debris disks as a function of host star luminosity. Within the {52}-member sample, 
four subgroups (L1-L4) were defined according to luminosities, with {13} disks in each subgroup. 
b) Detection rates of CO-rich debris disks as a function of host star luminosity. 
c) Detection rates of debris disks with low CO content as a function of host star luminosity.}
\label{fig:detrate}
\end{figure}

Figure~\ref{fig:detrate}a shows the proportions of debris disks showing detected $^{12}$CO (2--1) emission with a luminosity 
of $>$9.5$\times$10$^{16}$\,W in the four stellar luminosity intervals. While in the two lower luminosity subgroups (L1 and L2) 
the detection rates of CO gas are only 15.4$^{+12.7}_{-7.8}$\% and 7.7$^{+11.3}_{-5.2}$\%, respectively, for disks around 
stars belonging to the higher luminosity L3 and L4 subgroups, the detection rates are significantly higher,
at 69.2$^{+11.1}_{-13.8}$\% in both cases\footnote{Because of the small samples, the lower and upper bounds of the 68.2\% 
confidence interval were computed using the approach proposed by \citet{agresti1998} and available in the \texttt{binom} 
library of the \texttt{R} statistical package.}. If we consider only CO-rich debris disks (Fig.~\ref{fig:detrate}b), i.e. 
systems with $M_\mathrm{CO} > 10^{-3}$\,M$_\oplus$, we find that there are no such disks in L1 and L2 subgroups, while in 
L3 we obtain a slightly lower detection rate (46.2$^{+13.6}_{-13.0}$\%) than when we consider also objects with lower CO mass 
(see above). Interestingly, the CO-bearing debris disks around stars with luminosity $>$13.2\,L$_\odot$ are all CO-rich. When 
we compare the combined L1 and L2 and the combined L3 and L4 subgroups, the Fisher's exact test\footnote{We employed the 
\texttt{fisher.test} routine implemented in the R statistical package.} we applied shows a significant relationship between 
stellar luminosity and the occurrence of CO-bearing disks, regardless of whether all CO detections or only 
CO-rich objects are considered. Finally, we note that if only the low CO content ($M_\mathrm{CO}<10^{-4}$\,M$_\oplus$) debris 
disks are considered, their detection rates were similar in all four luminosity groups: 15.4$^{+12.7}_{-7.8}$\%, 
7.7$^{+11.3}_{-5.2}$\%, 23.1$^{+13.5}_{-9.6}$\%, and 0$^{+8.5}_{-1.4}$\% in the L1, L2, L3, and L4, respectively 
(Fig.~\ref{fig:detrate}c).  

There are a total of 26 disks in the L3 and L4 subgroups, of which 8 have fractional luminosities between 5$\times10^{-4}$ 
and 10$^{-3}$, and 18 with $f_\mathrm{d}>10^{-3}$. In the "low" $f_\mathrm{d}$ subset there are 3 disks with detected CO gas (37.5$^{+17.6}_{-14.8}$\%), 
whereas in the "high" $f_\mathrm{d}$ subset there are {15} (83.33$^{+7.1}_{-10.6}$\%, only HD\,95086, HR\,4796, and HD\,176497 
have no CO detection). However, using again Fisher's exact test to compare the rates, we obtained a probability of 0.06, 
implying that the difference between the occurrences of CO-bearing disks in the "low" and "high" $f_\mathrm{d}$ groups is 
statistically not significant. 

Systems belonging to the L3--L4 subgroups have ages between 6 and 48\,Myr; however, their age distribution -- 
due to most of them being members of the Sco-Cen association (15 out of {26}) -- is far from uniform. The sample contains 
{19} objects younger than {20}\,Myr and 7 with ages between 20 and 48\,Myr, while for HD\,32297 only an upper limit of 30\,Myr 
is available for the age. By assigning the latter object to the "older" systems, we obtain a CO gas detection rate 
of 6/7 (85.7$^{+9.3}_{-18.2}$\%) in this group, while in the "younger" group this rate is 12/19 (63.2$^{+10.2}_{-11.5}$\%). 
When comparing the proportion of CO-bearing debris disks in the two different age groups, we found no statistically significant 
difference between them by applying Fisher's exact test. This remains true even if we assign HD\,32297 to the younger subsample. 
It is safe to say that within the studied age range of 6 to 48\,Myr, we found no evidence for an evolutionary trend in the 
detection rates of CO-bearing disks with age. However,the sample size of older disks is still quite small. 

\begin{figure}[h!]
\centering
\includegraphics[width=0.48\textwidth]{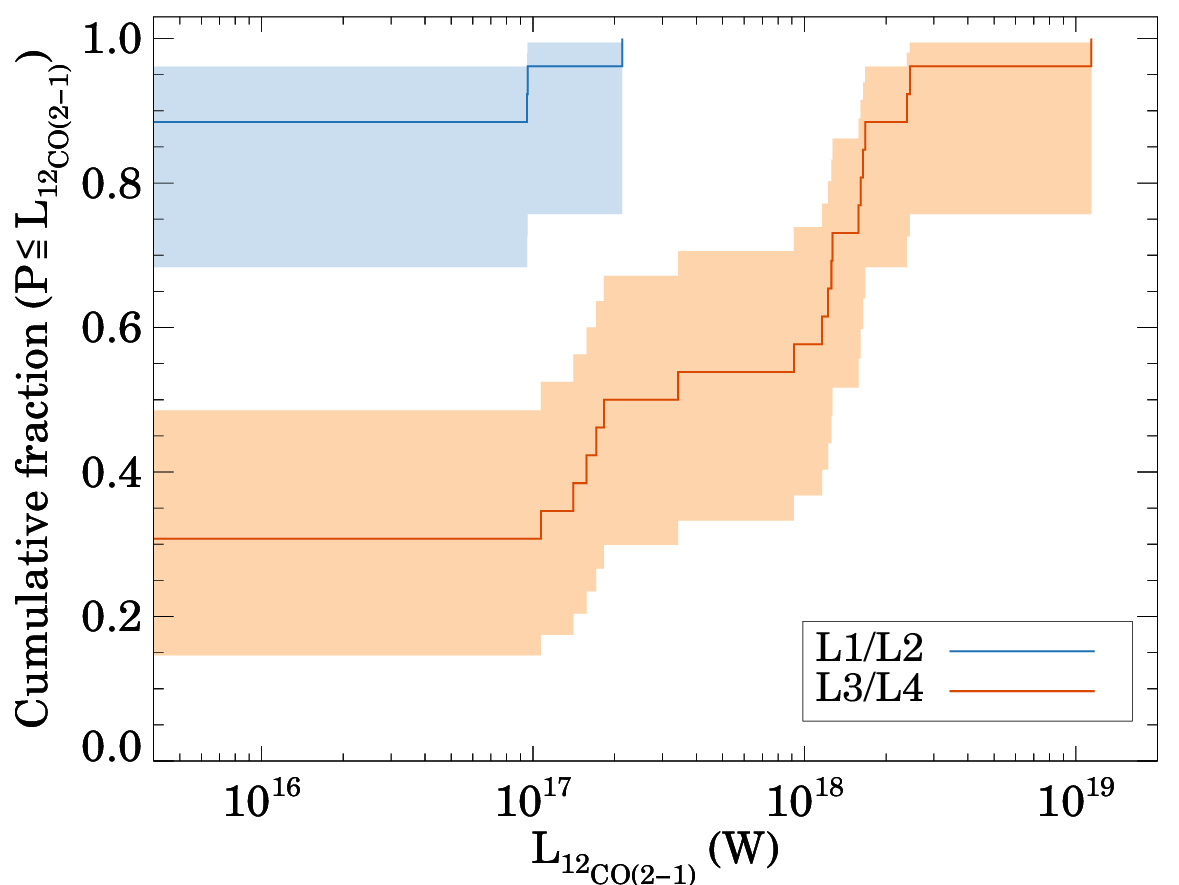}
\caption{Cumulative distribution functions of the $^{12}$CO line luminosities of debris disks belonging to the 
L1/L2 (blue curve) and L3/L4 groups (orange curve). The shaded regions show 95\% confidence intervals.}
\label{fig:kmplot}
\end{figure}

Finally, we compared the cumulative distributions of the $^{12}$CO (2$-$1) line luminosities of the disks in the 
combined L1/L2 ($L_* < 3.52 L_\odot$) and L3/L4 ($L_* > 3.53 L_\odot$) subgroups. With the exception of NO\,Lup, for which 
the J=3--2 rotational transition of $^{12}$CO  was measured, observations of the J=2--1 line are available for all objects 
in the sample. For NO\,Lup, the obtained $L_\mathrm{12CO(3-2)}$ luminosity was converted to $L_\mathrm{12CO(2-1)}$ by 
assuming LTE for gas at a temperature of 20\,K. To infer the cumulative distribution functions of the given groups, we used 
the Kaplan-Meier estimator implemented in the \texttt{lifelines} package \citep{Davidson-Pilon2019}, which considers not only 
detections, but also the measured CO line luminosity upper limits. A logrank test showed that the observed difference between 
the distributions of $^{12}$CO line luminosities of the two subgroups (Fig.~\ref{fig:kmplot}) is statistically significant. 
In line with our previous results based on detection rates (see above), this test also confirms that the CO component of debris 
disks around early type stars with higher luminosity differs significantly from that of disks around late type, low luminosity 
stars.

\subsection{Radial location of planetesimal belts in CO bearing debris systems} \label{sec:diskradii}
\begin{figure}[h!]
\centering
\includegraphics[width=0.48\textwidth]{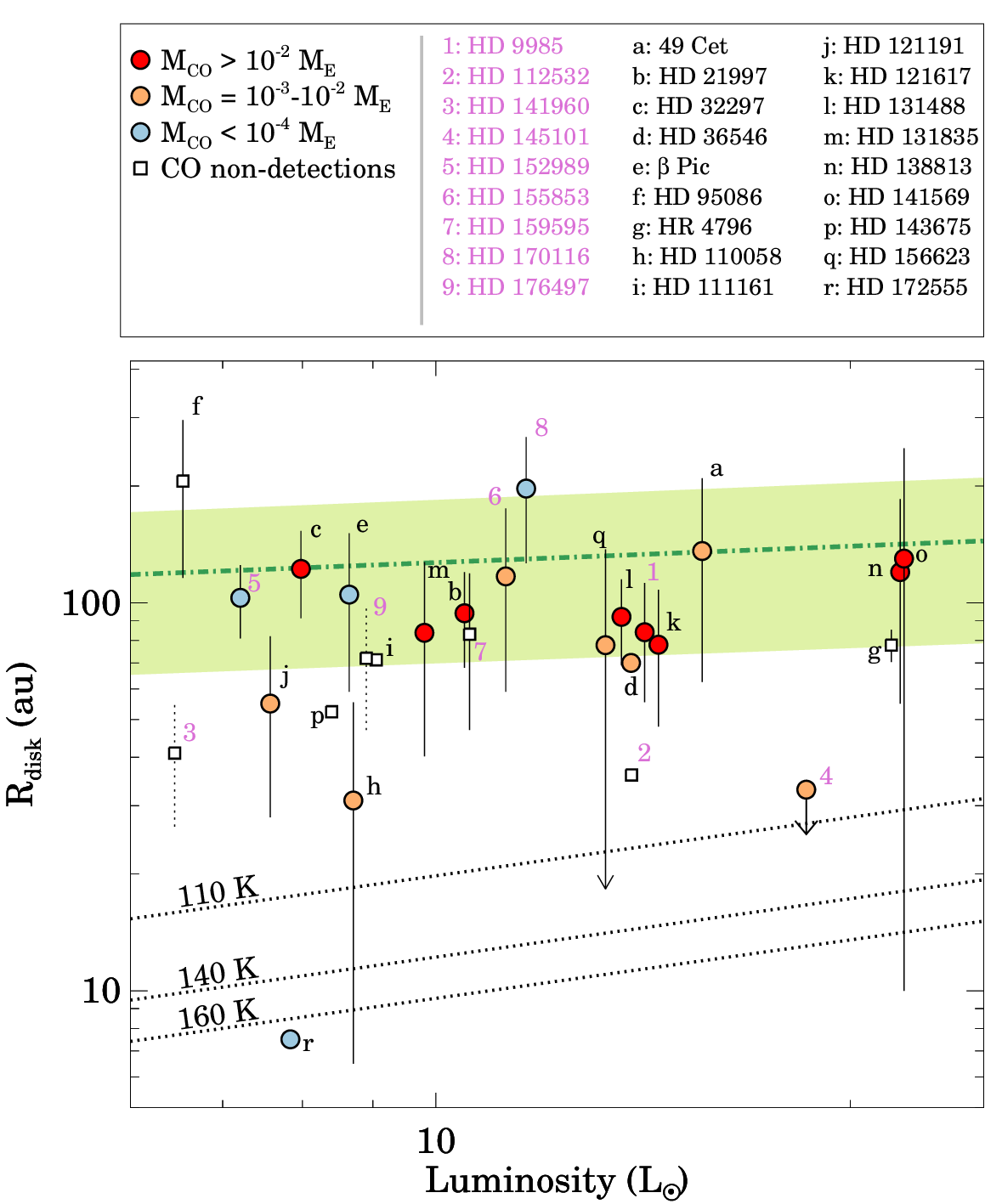}
\caption{Radii of dust disks as a function of stellar luminosities. In addition to the 9 disks we detected (indicated by numbers) 
17 additional, previously identified debris disk systems (indicated by letters) are also displayed. The data for the latter objects 
are taken from Table~\ref{tab:codisksample}. The vertical bars represent the radial extent of the disks. For disks where the width 
parameter was fixed in the modeling the bars are shown by dotted lines. In the case of HD\,156623 (marked by q) only an upper 
limit is available for the position of the inner edge of the disk. With the exception of HD\,111161 (i), HD\,143675 (p), and 
HD\,172555 (r), the size estimates are derived from millimeter continuum observations. Different symbols indicate those systems 
where CO gas is detected and those where it is not, and within the first group three categories are further distinguished by 
CO mass estimates. The green dashed-dotted line represents the best fit of the empirical planetesimal belt radius--stellar luminosity 
relationship \citep{matra2025}, while the green shaded region shows the $\pm$68\% confidence taking into account 
both the intrinsic scatter as well as the uncertainties of the obtained parameters.}
\label{fig:rdiskls}
\end{figure}

Large dust particles probed at 1.3\,mm are little influenced by stellar 
radiative/wind forces and thus can serve as a proxy for the spatial distribution of unseen parent planetesimals, the possible 
source of gas. We detected continuum millimeter emission from nine of the debris disks we observed with ALMA. In 
Figure~\ref{fig:rdiskls} we plotted the obtained radii and, where available, the widths of these disks as a function of the 
stellar luminosity. In cases where the widths of the belts were fixed to be $0.7 R_\mathrm{disk}$, the vertical bars indicating 
these parameters are marked with dotted lines. For HD\,152989, we used the parameters of the single Gaussian ring model for 
simplicity, but note that the double ring model also gives very similar results for the inner and outer edges of the debris disk 
(Sect.~\ref{sec:cospatial}). Systems where no CO emission is detected are shown by empty squares, while CO-bearing debris disks 
are marked by filled circles. For the latter, different colors were used according to the derived CO mass, distinguishing three 
categories: disks with low $M_\mathrm{CO}$ of $<$10$^{-4}$\,M$_\oplus$, and CO-rich debris disks with $M_\mathrm{CO}$ between 
10$^{-3}$ and 10$^{-2}$\,M$_\oplus$, and with $M_\mathrm{CO}$ of $>$10$^{-2}$\,M$_\oplus$. 
In the combined sample, radii derived from spatially resolved data are available for {17} additional disks around A-type stars 
having luminosities similar to our 9 objects ($>6\,L_\odot$). These systems are also plotted in Fig.~\ref{fig:rdiskls}.
For {15} of them, the results are also based on measurements at millimeter 
wavelength, while in the case of HD\,111161 and HD\,143675 the disk has only been detected in scattered light. Since small 
particles observed in scattered light are significantly affected by the radiation pressure, for the latter two objects we only 
assumed that the peak of the radial distribution of the small dusts roughly coincides with that of the larger particles measured 
at mm wavelengths, but did not consider the inferred radial extent, which could be much wider than the planetesimal belt due 
to interactions with the radiation forces. Finally, we also added the disk of HD\,172555 to the plotted sample, whose radius 
estimate was taken from \citet{schneiderman2021}. Thus we plotted 27 disks in total. While the nine disks from our program are 
marked with numbers, the other objects from the literature are marked with letters. Using spatially resolved millimeter 
observations of debris disks, \citet{matra2018} derived an empirical relationship between the radius of the belts and the 
luminosity of the central star. Figure~\ref{fig:rdiskls} shows the updated version of this relationship based on a larger 
sample \citep{matra2025}.

As Fig.~\ref{fig:rdiskls} demonstrates, the radii of the planetesimal belts in CO-rich systems are not different from those in 
which only small amounts of CO gas have been found, or even no gas was detected at all. In the most CO-rich objects 
($M_\mathrm{CO}>$10$^{-2}$\,M$_\oplus$) the peak radii of the planetesimal belts lie in a relatively narrow range
from 78 to {130}\,au. The relative width ($W_\mathrm{disk}/R_\mathrm{disk}$) of these belts, in most cases, are between 0.5 and 
1.1. Three of these particularly CO-rich systems, HD\,9985, HD\,121617, and HD\,131488, are very similar to each other, both in 
terms of the size of their planetary belts and the luminosity of their host stars. However, if the radial extent of the gas disk 
is taken into account in the comparison, the strong similarity holds only for the first two systems since while in HD\,9985 and 
HD\,121617 the gas and dust are quite well co-located \citep[Sect.~\ref{sec:cospatial} and][]{cataldi2023}, in HD\,131488 the gas 
disk extends much further inwards than the dust disk \citep{pawellek2024}.   

With the exception of HD\,172555, where the observed warm dust and CO gas may have been released in a giant collisional 
event \citep{lisse2009,schneiderman2021}, the planetesimal belts in the other systems are situated at tens of au from 
their star. Of the CO-rich debris disks, the one around the newly discovered HD\,145101 has the smallest radial extent. 
In a sufficiently high temperature environment CO and CO$_2$ ices trapped in an amorphous water ice matrix can desorb. 
Laboratory experiments indicate significant gas releases at the phase change from amorphous to crystalline water at $\sim$140\,K 
and also when the desorption of water ice occurs at $\sim$160\,K \citep{collings2003}. In real astrophysical environments, however, 
the heating of ice is much slower than in the laboratory, resulting in lower desorption temperatures 
\citep{collings2003,martin-domenech2014}. Considering realistic heating rates of comets, water ice desorption is more likely to 
occur at temperatures around 110--120\,K \citep{collings2004}. In the figure we indicated regions corresponding to the relevant 
temperatures mentioned above. These were calculated assuming planetesimals with an albedo of 5\%. This suggests that desorption 
processes in the inner regions of the disks around HD\,110058, HD\,141569, and HD\,156623 and in the disk of HD\,145101 may also 
contribute to CO gas production. In the case of HD\,110058, this possibility has already been raised by \citet{hales2022} in their 
analysis of the gas disk. While in the case of HD\,145101 we only have an upper limit for the size of the continuum disk, by 
analyzing its $^{12}$CO moment 0 map we obtained an estimate of 27$\pm$8\,au for the characteristic radius of the gas disk 
(Sect.~\ref{sec:cospatial}), which corresponds to a temperature of 110$\pm$16\,K for planetesimals with an albedo of 5\% using the 
luminosity of the host star. This suggests that CO and CO$_2$ gas can be released from the surface layers of larger icy bodies, 
even as a result of desorption. The temperature of smaller dust particles, as they are inefficient emitters, could be even higher. 
From the available SED data, we inferred a characteristic dust temperature of 163$\pm$15\,K for 
this disk (Sect.~\ref{sec:diskprops}), suggesting that the ice mantles of small dust particles, if any, 
may sublimate and contribute to gas production. 

Of the eight disks successfully resolved in our study, those around HD\,112532 and HD\,141960 are among the smallest, with 
sizes well below those predicted by the empirical relation (Fig.~\ref{fig:rdiskls}). The radius of the belt around HD\,170116, 
however, is one of the largest among the resolved disks around young A-type stars. This conclusion regarding the disk's 
outstanding size remains valid even when compared to the sample of 74 debris belts studied in the REASONS millimeter survey, 
which also includes disks around older stars with different spectral types \citep{matra2025}.

\subsection{Second generation gas produced in a steady-state collisional cascade?} \label{sec:secondary}
The observed dust content of most known debris disks can be explained by the slow, steady state collisional grinding of a 
planetesimal belt that is more or less co-located with the dust ring \citep{wyatt2008}. If the minor bodies involved are rich 
in volatiles, collisions can result not only in smaller and smaller fragments and eventually dust particles, but also in the 
release of second generation gas material \citep{zuckerman2012,matra2015,kral2017}. Assuming that the icy exo-planetesimals are 
similar to comets in the Solar System \citep{mumma2011,bockelee-morvan2017}, the release of H$_2$O, CO, and CO$_2$ molecules is 
expected to be predominant. If there is no effective shielding mechanism in operation, these molecules dissociate relatively 
quickly \citep[$t_\mathrm{phd}\lesssim$120\,yr even for CO molecules, the most resistant among them,][]{visser2009,heays2017} 
under the influence of interstellar and stellar UV photons. 

According to \citet{matra2017a} the mass loss rate of solids at the bottom of the ideal collisional cascade can be estimated as 
\begin{equation} 
\dot{M}_\mathrm{D_{min}} ({M_\oplus} yr^{-1}) = 1.2\times10^{-3} \bigg( \frac{R_\mathrm{disk}}{\rm au} \bigg)^{1.5}
		     \bigg(\frac{W_\mathrm{disk}}{\rm au} \bigg)^{-1} f_\mathrm{d}^2 
		    \bigg(\frac{L_*}{L_\odot}\bigg) 
		    \bigg(\frac{M_*}{M_\odot} \bigg)^{-0.5}, 
\label{eq:mdustprod}		      
\end{equation}
Assuming that the gas release is mediated by collisions that expose new icy surfaces or release trapped gas, 
a simple estimate of its production rate and finally the mass of CO can then be derived \citep{matra2019}: 
\begin{equation}
M_\mathrm{CO,pred} = \dot{M}_\mathrm{D_{min}} t_\mathrm{phd} \frac{f_\mathrm{CO+CO_2}}{1-f_\mathrm{CO+CO_2}}, 
\label{eq:mco}
\end{equation}
where $f_\mathrm{CO+CO_2}$ is the mass fraction of CO+CO$_2$ in the planetesimals involved in the cascade.
In reality, the estimate of $M_\mathrm{CO,pred}$ is subject to large uncertainties.  
First, the typical formal uncertainty of the $\dot{M}_\mathrm{D_{min}}$ estimate is 40--60\%, and the real 
uncertainty could be even substantially higher as there are a number of simplifying assumptions behind
Eq.~\ref{eq:mdustprod}.  
For example, the luminosity is included because brighter stars have a stronger radiation pressure and can therefore blow out 
larger dust particles, meaning that the same fractional luminosity corresponds to a larger mass in small dust particles. 
However, there are debris disks where the size of the smallest particle does not necessarily correspond to the blowout grain 
size, it can either be larger than that \citep{pawellek2014}, or smaller in disks with a very high $f_\mathrm{d}$ and/or 
having significant gas content \citep{thebault2019,bhowmik2019,singh2021}. 
Moreover, even if we consider only comets in the Solar System, we see significant differences in their CO+CO$_2$ abundance 
($f_\mathrm{CO+CO_2}$), and it is reasonable to assume that for comets in different exoplanetary systems 
this parameter may show even larger diversity. The photodissociation lifetime of CO gas also has its own uncertainty 
e.g. depending on the possible shielding processes. 

\begin{figure}[h!]
\centering
\includegraphics[width=0.48\textwidth]{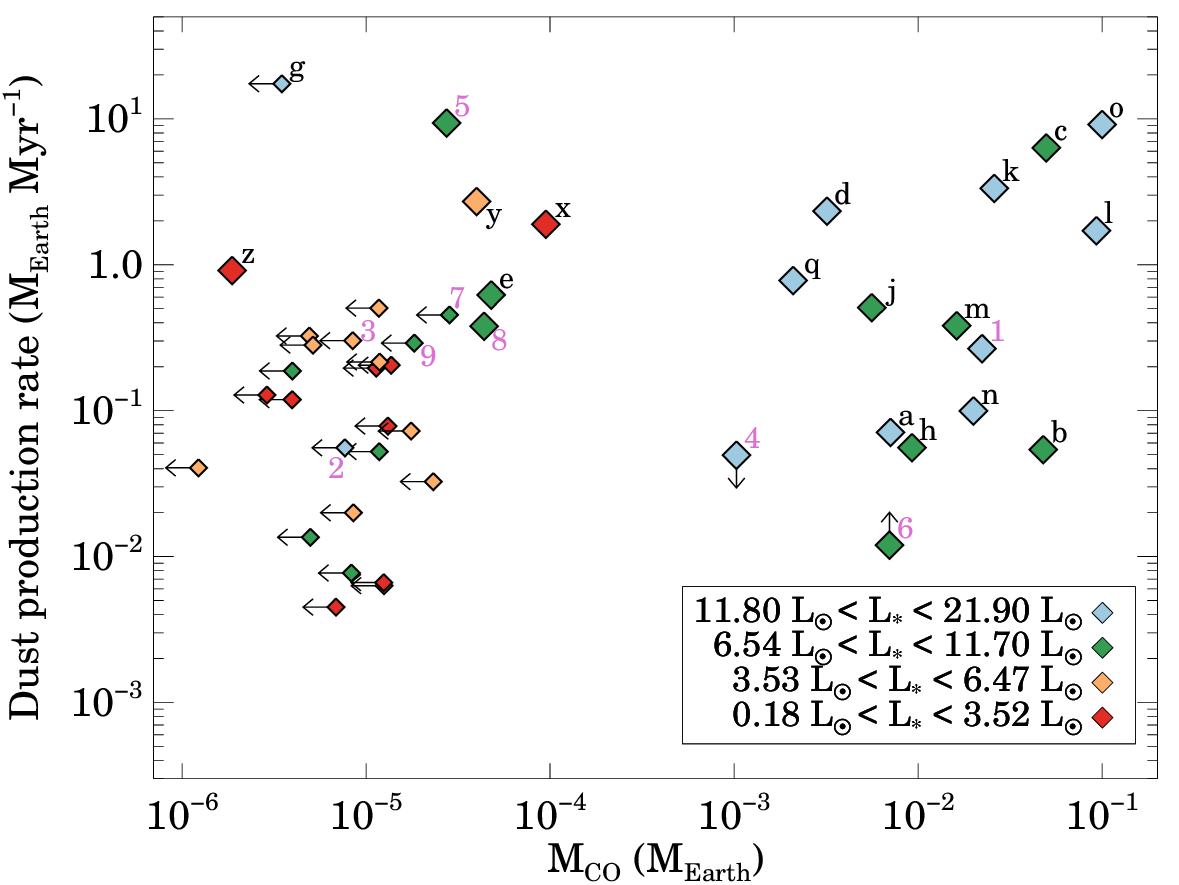}
\caption{Dust production rate as a function of the obtained CO masses. The different colors of the symbols 
correspond to the luminosity groups defined earlier in Sect.~\ref{sec:detrates}. The number and letter designations of 
the CO-bearing debris disks are the same as in Fig.~\ref{fig:rdiskls}. The three objects that were not included in 
that figure, HD\,129590, HD\,146897, and HD\,181327, are labeled with the letters x, y, and z, respectively. 
For the HD\,155853, we have only a lower limit for the $f_\mathrm{d}$ of the cold component (Appendix~\ref{sec:diskprops}), 
and hence also for the $\dot{M}_\mathrm{D_{min}}$.}
\label{fig:mdustprod_mco_ls}
\end{figure}

Figure~\ref{fig:mdustprod_mco_ls} shows the mass loss rates ($\dot{M}_\mathrm{D_{min}}$) computed from Eq.~\ref{eq:mdustprod} 
as a function of the observationally estimated CO mass for {46} debris disks including 9 from our sample. Only disks with radii 
derived from spatially resolved data were considered. Additionally, we excluded systems whose circumstellar material is believed 
to be of transient origin rather than related to a steady-state collisional cascade. Following these principles, two CO-bearing 
debris disks were excluded from this analysis. For HD\,172555, it was suggested that the observed warm dust and CO gas 
may have transient nature, presumably resulting from a recent giant collision \citep{lisse2009,su2020,schneiderman2021}. 
In the case of NO\,Lup, only an upper limit of the disk size is available, and the fractional luminosity of the cold dust 
component is subject to large uncertainty \citep{lovell2021}, making the estimate of the dust production rate to be very 
uncertain as well. Therefore, this system was also discarded.

The data for the calculations were taken from 
Tables~\ref{tab:targets}, \ref{tab:contdiskprops}, \ref{tab:codiskprops}, and \ref{tab:codisksample}. As before, in cases where 
no estimate of the width of the dust ring is available, or where the disk has so far only been resolved in scattered light, 
we assumed that $W_\mathrm{disk} = 0.7 R_\mathrm{disk}$. For debris disks whose excess spectrum requires two different temperature 
components to be properly modeled, we used the fractional luminosity of the cold components in the calculations. In the case of 
HD\,145101, we used the derived upper limit of $R_\mathrm{disk}<33$\,au in the calculations, thereby also obtaining an upper limit 
for $\dot{M}_\mathrm{D_{min}}$. For the sake of simplicity, for HD\,152989 we adopted the $R_\mathrm{disk}$ and $W_\mathrm{disk}$ 
parameters derived from its Gaussian ring model.

As shown in the figure, the estimated CO masses exhibit a large spread of at least five orders of magnitude with a gap 
separating debris disks with low ($M_\mathrm{CO}\lesssim10^{-4}$\,{M$_\oplus$}) and high (CO-rich, 
$M_\mathrm{CO}\gtrsim10^{-3}$\,{M$_\oplus$}) CO content. It is worth noting here that for the latter group of disks the 
$^{12}$CO line is probably optically thick and therefore their mass estimates are based on isotopolog measurements, which 
introduces a further uncertainty in the estimates due to the uncertain abundance ratios of the CO isotopologs. The large 
spread is evident for all $\dot{M}_\mathrm{D_{min}}$.
In debris disks with low CO content, in the absence of efficient shielding mechanism, the photodissociation lifetimes 
of CO molecules is limited to a few hundred years \citep{kral2017,matra2017b,marino2022}, so the long-term persistence of the gas disk requires continuous replenishment. Therefore, there is a general consensus that the gas in these systems, if present at all, is of 
secondary origin \citep{kral2017,matra2017b,matra2017a,marino2020}.
There are six disks in this subsample in which CO gas has already been detected: $\beta$\,Pic, HD\,129590, HD\,146897, HD\,181327, 
HD\,152989, and HD\,170116 (marked by e, x, y, z, 5, and 8 in Fig.~\ref{fig:mdustprod_mco_ls}). 
With the exception of HD\,181327, their estimated $M_\mathrm{CO}$ masses are between 10$^{-5}$ and 10$^{-4}$\,M$_\oplus$. 
All of these CO-bearing debris disks have $\dot{M}_\mathrm{D_{min}}\geq$0.38\,M$_\oplus$Myr$^{-1}$.
Even if we discard those disks (including HD\,181327 and HD\,159595) that do not meet our detection limit criterion 
(Sect.~\ref{sec:detrates}), the CO detection rate for disks having $\dot{M}_\mathrm{D_{min}}\geq$0.38\,M$_\oplus$Myr$^{-1}$ is 
5/7, while for disks with lower production rates it is 0/22.
Using these results, the null hypothesis that the CO detection rates in the high and low dust production groups would be the same 
can be rejected on the basis of a Fisher's exact test which yielded a low probability of 1.8$\times$10$^{-4}$.
This suggests that collisions may play some role in CO gas production in the low CO content disks.

Both newly discovered low CO content gaseous debris disks show interesting structural features. For HD\,152989, the ALMA observations 
hint at the presence of two gas-bearing dust rings with a gap between them, while the disk around HD\,170116 is similar 
to the one around $\beta$\,Pic, not only in terms of CO mass and dust production rate, but also in the fact that 
the CO brightness distribution shows a significant asymmetry \citep[Sect.~\ref{sec:cospatial} and ][]{dent2014}. 
Assuming that the $^{12}$CO line is optically thin, which seems plausible, this implies that the mass distribution of the gas 
is also highly asymmetric. The nature of the bright CO clump observed in the disk of $\beta$\,Pic is still debated,
it could either originate from a recent giant collision between Mars-sized bodies, from a tidal disruption event, 
or from collisions of planetesimals trapped in mean-motion resonance with a planet \citep{dent2014,matra2017b,cataldi2018}. 
The asymmetry is even more pronounced in HD\,170116, where only one side of the disk shows detectable CO emission 
(Fig.~\ref{fig:hd170116}). This CO disk also has a brighter clumpy part, which coincides well with the brightest part of 
the continuum dust emission. The possible scenarios to explain the observations here may be similar to those proposed for 
$\beta$\,Pic, but further detailed investigation of these is beyond the scope of this work. In any case, this suggests that in both 
disks there is a mechanism that produces CO gas on top of the collisional cascade, thereby increasing the steady state gas 
(and dust) content, albeit possibly only temporarily.

The presence and persistence of the observed  large amounts of CO gas in CO-rich debris disks, which sometimes reach levels 
similar to those in protoplanetary disks, is only possible if some kind of shielding gas material exists in the system. The 
nature of this material and the observed gas component is still debated. It may be either second generation (see below) or 
residual primordial (Sect.~\ref{sec:hybrid}).
According to the shielded secondary gas disk model \citep{kral2019,marino2020}, the high gas content of CO-rich 
systems can be explained if at some point during their evolution, due to a sufficiently high gas production rate, the emerging 
neutral carbon gas -- mainly derived from the photodissociation of CO and CO$_2$ -- becomes optically thick and can shield CO 
from UV photons. Together with the subsequent self-shielding of CO molecules, this process can prolong the photodissociation 
lifetime of CO significantly, allowing large gas masses to accumulate in such debris disks. 

Assuming that both gas and dust originate from the collisions of icy solids, one might naively expect on the basis of 
this picture that CO-rich disks should tend to have a higher dust production rate than debris systems that do not 
harbor such massive gas disks. However, the observational results, as demonstrated by Fig.~\ref{fig:mdustprod_mco_ls}, 
do not support this expectation. 
The derived dust production rates for the CO-rich debris disks range between 0.01 and 9.2\,{M$_\oplus$}{Myr$^{-1}$}, while 
for the six gas-bearing disks with a low CO content are between 0.4 and 9.4\,{M$_\oplus$}{Myr$^{-1}$}. 
Thus, interestingly the two ranges significantly overlap, and the production rates for the former group is even slightly 
smaller on average (1.7\,{M$_\oplus$}{Myr$^{-1}$} versus 2.6\,{M$_\oplus$}{Myr$^{-1}$}). Even if we focus only on the CO-rich 
subsample, we see no correlation between $\dot{M}_\mathrm{D_{min}}$ and $M_\mathrm{CO}$. 
We note that, based on a smaller sample of gas-rich debris disks, \citet{cataldi2023} also found no correlation between the 
predicted CO+CO$_2$ production rates and the measured CO masses.
\citet{marino2020} found that the timescale of the viscous evolution of the gas can be longer than that of the evolution of the 
collisional cascade. This means that the massive gas component can persist for a long time after the dust (and gas) production 
rate of the cascade has dropped, i.e. it is possible that the decline of the gas disk follows the decline of the dust disk with 
a significant delay. So if we are interested in the formation and evolution of CO-rich disks, then the past collisional mass loss 
should be taken into account, which could be much higher than the present one. This could even open up the possibility that there 
was a period in the early evolution of CO-rich disks when, for some reasons (e.g. a dynamic instability), the rate of dust and gas 
production increased peculiarly, allowing the formation of a CO-rich environment. The question is, whether this possible decoupling 
of the evolution of the dust and gas components is sufficient in itself to explain why the observed amount of CO does not depend 
on the mass loss rate of solids. 

Another possible solution to this issue is that the ratio of the gas to dust production rates tends to be higher in the CO-rich 
systems, e.g. because the abundances of CO/CO$_2$ ice in their planetesimals are higher, and/or because other, 
possibly more efficient, outgassing mechanisms are at work in addition to collisions. The heating of icy planetesimals, 
either by stellar irradiation or by the decay of radioactive nuclides, can result in the release of gas 
\citep{davidsson2021,kral2021}. Examples where sublimation induced by stellar irradiation may contribute to the CO production, at least in 
the inner parts of the disks, are HD\,110058 and HD\,145101 (see above), which are among the CO-rich debris systems 
with the lowest $\dot{M}_\mathrm{D_{min}}$. The release of gas by radiogenic heating can be significant in the first $\sim$30\,Myr, 
but mostly only if large bodies of hundreds of kilometers in size are present in the planetesimal population 
\citep{davidsson2021,bonsor2023}. This could suggest that CO-rich disks are systems where the planetesimals have 
become larger for some reason. However, these are among the brightest debris disks, where, if the maximum planetesimal 
size were so large, the solid matter content would significantly exceed the amount of material observed to be available 
in protoplanetary disks, suggesting that in these systems, especially in the outer zones, the planetesimals are 
instead small \citep{krivov2021,bonsor2023}. Thus, despite the possible scenarios outlined above, questions remain as 
to why CO-rich disks can form in some systems but not in others at the same dust production rate, and why this phenomenon 
is restricted to disks around higher luminosity A-type stars.

\subsection{Residual gas from the primordial disk phase?} \label{sec:hybrid}
\begin{figure}[h!]
\centering
\includegraphics[width=0.48\textwidth]{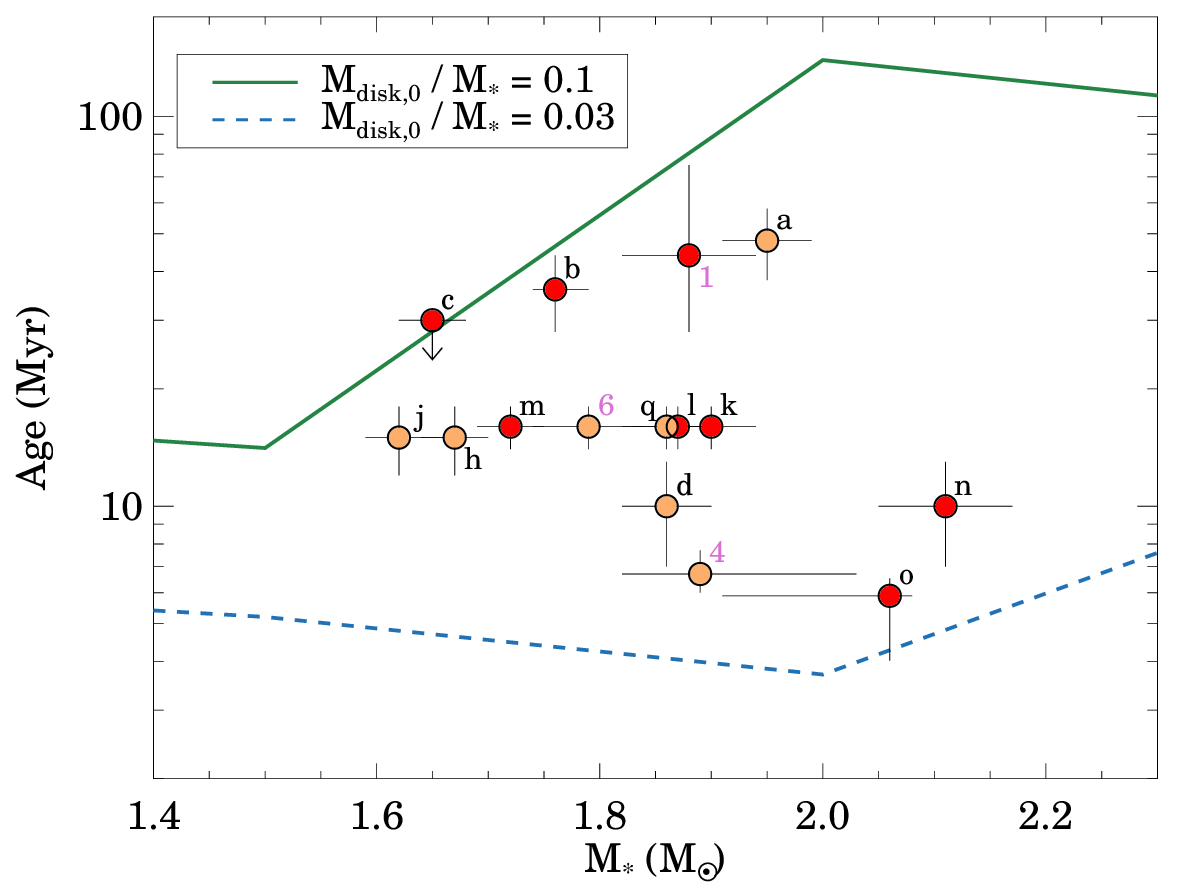}
\caption{Age as a function of stellar mass for known CO-rich debris systems. As in Fig.\ref{fig:rdiskls}, the red symbols 
indicate disks with estimated CO masses $>$0.01\,M$_\oplus$, while the orange ones represent disks with $M_\mathrm{CO}$ between 
0.001 and 0.01\,M$_\oplus$. The labels of the disks (numbers or letters) are also the same as in Fig.~\ref{fig:rdiskls}. 
The solid green and dashed blue curves show the lifetimes of primordial gas in disks depleted in submicron size dust particles 
assuming initial disk masses of $M_\mathrm{disk,0} = 0.1M_*$ and $M_\mathrm{disk,0} = 0.03M_*$, respectively. The curves are 
taken from \citet{ooyama2024}.
}
\label{fig:primordial}
\end{figure}

A possible alternative explanation is that the gas in CO-rich debris disks, or at least some of them, is residual material 
from the previous protoplanetary phase.
In fact, already after the discovery of the first CO-rich disks -- motivated by their young age -- it was proposed that the 
observed CO is only a small fraction of the gas material, which is predominantly composed of primordial H$_2$ gas that could 
significantly extend the photodissociation lifetime of CO molecules by efficiently shielding them from UV radiation 
\citep{kospal2013,pericaud2017}.
Since the dust component of these systems is of secondary origin, this scenario assumes a disk of hybrid nature \citep{kospal2013}. 
It should be noted that, although H$_2$ is primordial in this model, the observed CO gas is not necessarily, and secondary 
processes may play a role or even dominate its formation. Although most of the CO-rich disks have ages between 10 and 20\,Myr 
(Fig.~\ref{fig:primordial}), at least three objects, including the newly discovered HD\,9985, have been identified that are 
probably older than 30\,Myr. A key question for the viability of the hybrid disk scenario is whether a primordial gas disk 
around an A-type star can survive to such old ages.

We note that although Herbig Ae disks typically disperse on a timescale of a few million years \citep{brittain2023,pfalzner2022}, 
we know of some that are older than 10\,Myr \citep[e.g.,][]{arun2019,wichittanakom2020} so their age distribution overlaps with 
that of the youngest CO-rich debris disks. These disks, however, are very different from the ones we have studied: though there 
is also an overlap in the CO content between the least gas-rich Herbig Ae disks and the most gas-rich debris disks, the dust 
content is orders of magnitude higher in the former systems providing additional very efficient shielding to not only CO but 
other molecules as well. Because of this latter effect, there may be a significant difference in the composition of the gas, 
and thus in the molecules that can be detected in the two types of disks \citep{klusmeyer2021,smirnov-pinchukov2022}.

To assess the potential lifetime of primordial disks around stars of different masses, \citet{nakatani2023} 
used a semi-analytic 0D disk evolution model and found that protoplanetary disks depleted in submicron dust 
particles and PAHs can persist for tens of millions of years around intermediate-mass stars, as the lack of 
small grains leads to much slower far-UV photoevaporation. By extending this model using 1D disk evolution simulations, 
\citet{ooyama2024} further investigated this question. In Fig.~\ref{fig:primordial} the solid green curve shows 
the lifetimes of primordial gas disks as a function of stellar masses in their fiducial model where they 
assumed initial disk masses of $M_\mathrm{disk,0} = 0.1M_*$ 
\citep[for further default parameters of the fiducial model, see][]{ooyama2024}. 
As the figure shows, according to this model, primordial gas can persist even up to $\sim$140\,Myr around a 
2\,M$_\odot$ star, and the predicted maximum disk lifetimes exceed the age of the known CO-rich debris systems. 
\citet{ooyama2024} also found that by assuming smaller initial disk masses ($M_\mathrm{disk,0} = 0.03M_*$), the predicted 
lifetime of the primordial gas is dramatically reduced (blue dashed curve), well below the age of the known 
systems. These results suggest that the gas in CO-rich systems could be of primordial origin, and if so, 
they may be the descendants of the most massive protoplanetary disks.

\subsection{Dynamical excitation of the disks} \label{sec:stirring}
For collisions between planetesimals to result in their erosion and eventual dust production through a collision cascade, 
the relative velocity of the bodies must be sufficiently high, which requires dynamical excitation of the disk \citep{wyatt2008}. 
Assuming that the disk has not been prestirred in some way since its birth, some mechanism is required to initiate and maintain 
an excited population of planetesimals. This may be the gravitational influence of the largest planetesimals embedded in the 
disk \citep[self-stirring,][]{kb2004,krivov2018} or perturbation from a planet or stellar companion present in the system 
\citep{mustill2009}, and in some cases stellar flybys may also play a role \citep{kb2002}. Knowing some basic properties of the 
central star and the planetesimal belt, it is possible, under certain assumptions, to estimate a minimum disk mass that would be 
required to excite the disk by self-stirring over the lifetime of the system \citep{pearce2022}. We used the 
\texttt{MinSelfStirringDiscMass\footnote{\url{https://www.tdpearce.uk/public-code/}}} python script developed by \citet{pearce2022} 
to compute this parameter for all disks from our sample whose continuum emission was spatially resolved. The required input 
parameters are: the mass and age of the star and the radii of the inner and outer edges of the planetesimal belt (assuming that 
the disk is axisymmetric) and the measured dust mass (from which the total disk mass is estimated). The stellar parameters and their 
uncertainties are taken from Table~\ref{tab:targets}. Since those large particles that dominate the observed millimeter-wavelength 
emission are negligibly affected by the radiative forces, their spatial distribution is supposed to follow that of the parent 
bodies. Therefore, in the calculations we have assumed that the planetesimal belts' inner and outer edge radii equal to 
$R_\mathrm{disk}-0.5W_\mathrm{disk}$ and $R_\mathrm{disk}+0.5W_\mathrm{disk}$, respectively (Table~\ref{tab:contdiskprops}). 
For HD\,112532, HD\,141960, and HD\,176497, where we had no estimate for the disk width from the measurements, we adopted 
$W_\mathrm{disk} = 0.7R_\mathrm{disk}$. For HD\,152989, both the Gaussian ring and the double ring models were considered, in 
the latter case utilizing the radii of the inner and outer ring as the inner and outer edges of the disk.
The dust mass estimates of the disks were taken from Table~\ref{tab:contdiskprops}. 


\begin{table}                                                                                                           
\begin{center}                                                                                               
\caption{Stirring of spatially resolved debris disks in our sample. 
 \label{tab:stirring} }
\begin{tabular}{lcc}                                                     
\hline\hline
Target name & $M_\mathrm{disk,self-stir}$ & $M_\mathrm{pl,stir}$ \\
            & ($M_\oplus$)                & ($M_\mathrm{Jup}$)   \\ 
\hline
HD\,9985        &  600$^{+200}_{-300}$    &  0.10$^{+0.11}_{-0.13}$ \\ 
HD\,112532      &  40$^{+40}_{-40}$       &  0.04$^{+0.13}_{-0.13}$ \\
HD\,141960      &  100$^{+100}_{-100}$    &  0.07$^{+0.14}_{-0.14}$ \\ 
HD\,152989$^a$  & 2100$^{+400}_{-400}$    &  0.13$^{+0.06}_{-0.06}$ \\ 
HD\,152989$^b$  & 2000$^{+600}_{-600}$    &  0.10$^{+0.08}_{-0.08}$ \\ 
HD\,155853      & 3000$^{+600}_{-600}$    &  1.63$^{+1.07}_{-1.07}$ \\ 
HD\,159595      &  800$^{+300}_{-200}$    &  0.37$^{+0.34}_{-0.32}$ \\  
HD\,170116      & 11000$^{+4000}_{-4000}$ &  0.58$^{+0.49}_{-0.50}$ \\ 
HD\,176497      &  600$^{+400}_{-400}$    &  0.17$^{+0.30}_{-0.30}$ \\   
\hline
\end{tabular}
\tablefoot{Minimum disk mass ($M_\mathrm{disk,self-stir}$) and minimum planet mass ($M_\mathrm{pl,stir}$) 
required to excite the disk by self-stirring or secular stirring by an eccentric planet over the life of 
the system (for details see Sect.~\ref{sec:stirring}). In the case of HD\,152989, both available continuum 
models (a: Gaussian ring; b: double ring) were considered in the calculations.}
\end{center}
\end{table}


The resulting minimum disk masses, which range from 40 to 11\,000\,M$_\oplus$ are given in Table~\ref{tab:stirring}. 
A disk mass greater than 1000\,M$_\oplus$ is unlikely, as this would exceed the observed amount of solid material in protoplanetary 
disks \citep{krivov2021,pearce2022}. Self-stirring alone therefore  
is likely not a viable excitation mechanism for debris disks around HD\,152989, HD\,155853, and for HD\,170116.
It is worth considering, however, that there are still questions regarding the total mass of solids in protoplanetary 
disks \citep[e.g.,][]{liu2022,viscardi2025}, and the real uncertainties of the disks' dust mass estimates used as input data 
in the calculations can even significantly exceed the derived formal ones (Sect.~\ref{sec:dustmasses}),
meaning that this conclusion should be treated with caution.
A possible alternative is that the disk stirring is caused by the 
secular perturbation of an unseen, eccentric planet orbiting inside the disk \citep{mustill2009}. Using eq.~23 from 
\citet{pearce2022}, we can estimate the minimum mass of the planet ($M_\mathrm{pl,stir}$) that can excite the planetesimal 
population on a timescale shorter than the age of the system. This minimum value is calculated for all systems in 
Table~\ref{tab:stirring}, assuming, similarly to \citet{pearce2022}, that the eccentricity of the possible planet is 0.3. As 
the table shows, in most cases a relatively low mass planet is sufficient for the excitation, the only system that would require 
a planet with $>$1\,$M_\mathrm{Jup}$ is HD\,155853.  However, this calculation assumes a massless disk. If the self-gravity of 
the disk is taken into account, and if the mass of the disk exceeds that of the planet, the secular perturbation effect will be 
weaker, and a more massive planet than the ones inferred from the above approach is required for the sufficient excitation 
\citep{sefilian2024}. It is worth noting that several other avenues of planetary stirring have recently been proposed.
Resonance sweeping accompanied with the migration of a planet can excite the disk material igniting more violent, higher 
velocity collisions \citep{friebe2022}. \citet{costa2024} argue that massive projectiles scattered by a planet 
orbiting at the inner edge of the disk can also stir the orbit of other planetesimals. They also found that sufficiently massive 
planets ($\gtrsim$0.5\,$M_\mathrm{Jup}$) can excite a disk via broad mean motion resonances. Finally, the combination of 
self-stirring and the secular perturbation of a planet can be significantly more efficient than considering these mechanisms 
separately \citep{munoz-gutierrez2023}.

Thus, by investigating the possible dynamical excitation of our disks, we found that self-stirring is highly unlikely in at least 
three debris disks, HD\,152989, HD\,155853, and HD\,170116, and that for these systems, other mechanisms, most likely planetary 
stirring is required for the collision cascade to work. If debris material of the HD\,152989 system is indeed concentrated in 
two rings, this could be another strong indication of the presence of a planet in the gap between them, which would also explain 
how this disk is stirred. Such gaps, potentially carved by planets, have already been detected in several wide outer 
debris disks through mm continuum observations with high spatial resolution 
\citep{marino2018,marino2019,marino2020b,macgregor2019,nederlander2021}. We note that those gaps may also be present in the gas 
distribution and observable with ALMA for massive enough planets, as recently shown in \citet{bergez-casalou2024}.
It is worth mentioning that HD\,152989 has a wide, relatively massive companion (Appendix~\ref{sec:multiplicity}), which, if 
its orbit is sufficiently eccentric, could also contribute to the dynamical excitation of the planetesimals.
These findings make these three systems promising targets for future planet search programs.

\section{Summary and Conclusions}
Aiming to study the presence and evolution of gas in young dust-rich debris disks around intermediate-mass stars, we have carried 
out Band\,6 continuum and CO line observations with ALMA toward 12 systems that show strong excess emission based on their previous 
IR observations. The selected systems have received little or no attention so far, and our program has provided their first 
subarcsecond angular resolution measurements at millimeter wavelengths. We detected mm continuum emission from nine of our targets, 
and resolved the debris disk at least marginally in eight of them. In addition, we discovered CO gas in five disks. In two of them, 
HD\,152989 and HD\,170116, only emission from the most abundant main isotopolog $^{12}$CO was detected. In the other three, HD\,145101, 
HD\,9985, and HD\,155853, emission of $^{13}$CO gas was also detected, and in the latter two even the presence of the much rarer 
C$^{18}$O isotopolog was revealed. From the continuum and CO line measurements, we estimated the dust and gas masses of the disks 
and examined the radial distribution of dust and gas. In addition to these results, we also detected both continuum and CO line 
emission from the HD\,31305 system, but these detections are not associated with the targeted very young A-type star, but with a disk 
of its late-type companion, which is probably of protoplanetary origin.

While the CO content of the disks around HD\,152989 and HD\,170116 is $<$10$^{-4}$\,M$_\oplus$, the other three gas-bearing systems, 
HD\,9985, HD\,145101, and HD\,155853, have CO masses $>$10$^{-3}$\,M$_\oplus$, and are therefore classified as CO-rich debris disks. 
While the gas in the disks with low CO content is most likely of secondary origin, for the CO-rich objects a primordial origin may 
be a possible alternative.

The debris disk around HD\,152989, despite its outstandingly high fractional luminosity of $\sim$7$\times$10$^{-3}$, 
was unknown before our study. The position-velocity diagram of $^{12}$CO emission suggests that there is a gap in the 
radial distribution of the gas disk: the bulk of the CO is in two rings at radial distances of $\sim$75\,au and $\sim$130\,au. 
Based on the continuum data modeling, the distribution of large dust particles can also be described by a consistent 
two-ring model, but the solution is not unique. If this disk does indeed show such a gapped structure -- further higher spatial 
resolution, deeper ALMA measurements are needed to prove this -- then it is unique among known CO-bearing debris disks. 

The debris disk of HD\,170116 is one of the most distant that has ever been spatially resolved with ALMA. The disk has a 
peak radius of $\sim$200\,au, which makes it one of the largest known debris disks \citep[cf. the REASONS sample][]{matra2025}.
The spatial distribution of the relatively small amount of CO gas is highly asymmetric, with the detectable emission associated 
only with one side of the disk and showing a CO brightness peak located close to the inner edge of the dust disk. 
How this high degree of asymmetry has developed is not yet clear. However, the similarity of this feature to the disk around 
$\beta$\,Pic raises the possibility that a transient gas- and dust-producing event or a resonant planetary population may be 
behind the phenomenon \citep{matra2017b,cataldi2018}.

Of our sample, the debris disk around HD\,9985 is the most CO-rich with an estimated CO mass of $\sim$0.02\,M$_\oplus$. In many 
ways, this system resembles the previously discovered CO-rich debris system, HD\,121617 \citep{moor2017,cataldi2023}. The host 
stars have very similar luminosities, the gas and dust material is well co-located, the radial positions 
\citep[84$\pm$5\,au for HD\,9985 and 78$\pm$5\,au for HD\,121617, see Table~\ref{tab:contdiskprops} and][]{matra2025} as well 
as the fractional widths of the debris rings \citep[0.68$\pm$0.31 for HD\,9985 and 0.77$\pm$0.14 for HD\,121617, see Table~\ref{tab:contdiskprops} and][]{matra2025} are also quite similar to each other. Based on its estimated age of 44$^{+31}_{-16}$\,Myr, HD\,9985, along with 
49\,Cet and HD\,21997, is likely the oldest of the known CO-rich debris disks. 

The continuum observations of HD\,155853 reveal a broad dust belt extending from 60 to 175\,au. The disk is CO-rich, but unlike 
the case of HD\,9985, the spatial distribution of the gas component is different from that of the large dust grains, the former 
showing a more compact morphology. This feature makes this disk more akin to the CO-rich debris disks found around HD\,21997 and 
HD\,131488 \citep{kospal2013,pawellek2024}, whose gas and dust components are also not well co-located. Both the abovementioned 
HD\,152989 and HD\,155853 are the members of the UCL association, which has proven to be the richest reservoir of CO-bearing 
debris disks to date, with 7 of such objects identified so far in this group. 

HD\,145101 is a young ($\sim$7\,Myr), intermediate mass pre-main-sequence star that belongs to the US association. 
While in the vast majority of the young CO-bearing debris disks identified so far, including the 4 newly discovered ones above, 
the bulk of the gas and dust is located at radial distances larger than 40\,au, the disk around HD\,145101 is quite compact, with 
both components predominantly situated at $<$33\,au. Consistently, the analysis of the SED also indicates a rather high 
characteristic dust temperature of $\sim$160\,K. In this respect, this disk is more similar to the one around HD\,172555 
\citep{su2020,schneiderman2021}, which also contains warm dust only. However, the CO content of HD\,145101 is two orders of 
magnitude higher. If the gas is of secondary origin, it is possible that thanks to the higher temperature environment, thermal 
desorption of ices may play a larger role in the gas release than in the other known CO-bearing debris systems.

In a future paper, we plan for a more detailed analysis of the origin and evolution of gas in these five systems by taking into 
account the production and the subsequent photodissociation, shielding, and viscous evolution of the CO gas. 

By investigating the possible dynamical excitation of those debris disks from our sample that are spatially resolved in the 
continuum, we found that self-stirring of HD\,152989, HD\,155853, and HD\,170116 would require unreasonably large disk masses 
suggesting that their excitation is instead related to an unseen planet(s) or a companion. These systems could be promising 
targets for future planet search programs.

To study the general characteristics of young ($\leq$50\,Myr), dust-rich ($f_d>5\times10^{-4}$) CO-bearing debris disks and the 
environments in which they form, we combined our new results with data from the literature. In agreement with previous results 
\citep{lieman-sifry2016,moor2017}, we found that the occurrence rate of CO gas rises significantly for disks surrounding host 
stars with $L_*>6.5\,L_\odot$, corresponding to spectral types earlier than $\sim$A8. While in disks around stars with a luminosity 
lower than this value, the detection rate is 3/26 (12\%), in systems with higher luminosity hosts it is 18/26 (69\%). While none 
of the gas-bearing disks in the former subgroup are CO-rich, the latter contains 15 such objects. Interestingly, all the 9 gaseous 
disks around stars with luminosities from 13.2 to 21.9\,L$_\odot$ (A4--A0) are CO-rich. If we consider only the disks with 
$f_d > 10^{-3}$ in the higher stellar luminosity subsample, the CO detection rate is 15/18, i.e. these two easily measured 
parameters together provide a good indicator of the presence of CO gas in young debris systems. 

There is a general consensus that the gas in debris disks with CO masses of $<$10$^{-4}$\,M$_\oplus$ is derived from the erosion 
of planetesimals, similar to dust.  By comparing the estimated CO content and the mass loss rate of the solids in such disks -- 
the latter parameter is estimated by assuming an ideal collision cascade -- we found that the detected gas-bearing systems belong to 
the ones with the highest mass loss rates. This is an indication that collisions, in line with model predictions 
\citep{kral2017}, may play a role in gas production in debris disks with low CO content.
According to the recently developed shielded secondary gas disk model \citep{kral2019,marino2020}, 
even the large gas content of the CO-rich debris disks can be explained by second generation processes. Interestingly, when 
comparing low CO content and CO-rich disks, we do not see a significant 
difference in the estimated dust mass loss rates, and when looking only at the latter disks, we see no correlation between 
CO content and mass loss. This raises the question of what could be the difference that, at similar dust mass loss rates, lead to a 
CO-rich state in some of the debris disks but not in others. In addition to the possible explanation that the current dust mass 
loss rates do not reflect those characterising the initial stage of the CO-rich disks, due to the gas lifetime far exceeding the 
evolution of the collision cascade \citep{marino2020}, we also raise the possibility that in these particular systems the 
abundance of CO ices in the planetesimals may have been higher and/or other more efficient gas release mechanisms may have 
been active in addition to 
collisions. Alternatively, as \citet{nakatani2023} and \citet{ooyama2024} have shown, the CO-rich phenomenon can also be explained 
by that, under certain conditions, the gas material of massive protoplanetary disks depleted in small grains can persist for much 
longer than previously thought. By comparing the predictions of the latest model with the ages of known CO-rich disks, we found 
that for all of them it is conceivable that the gas is dominantly made by residual primordial material. 
Thus, the origin of the gas in CO-rich disks still remains an open question. 

\begin{acknowledgements}
We thank the anonymous referee for their useful suggestions. We are grateful to W. Ooyama for sending us 
their modeling data on the lifetimes of primordial gas disks as a function of stellar mass.
This paper makes use of the following ALMA data: ADS/JAO.ALMA\#2021.1.01487.S. ALMA is a partnership of ESO (representing its 
member states), NSF (USA) and NINS (Japan), together with NRC (Canada), NSTC and ASIAA (Taiwan), and KASI (Republic of Korea), 
in cooperation with the Republic of Chile. The Joint ALMA Observatory is operated by ESO, AUI/NRAO and NAOJ.  This work presents 
results from the European Space Agency (ESA) space mission Gaia. Gaia data are being processed by the Gaia Data Processing and 
Analysis Consortium (DPAC). Funding for the DPAC is provided by national institutions, in particular the institutions participating 
in the Gaia MultiLateral Agreement (MLA). The Gaia mission website is \url{https://www.cosmos.esa.int/gaia}. The Gaia archive website 
is \url{https://archives.esac.esa.int/gaia}. Funding for the DPAC has been provided by national institutions, in particular, the 
institutions participating in the Gaia Multilateral Agreement. This research has made use of the VizieR catalogue access tool, CDS, 
Strasbourg, France (DOI : 10.26093/cds/vizier). The original description of the VizieR service was published in A\&AS 143, 23. 
AMH gratefully acknowledges support from the National Science Foundation through Grant No. AST-2307920.
\end{acknowledgements}

%
%

\bibliographystyle{aa}

\begin{appendix} 
\section{Properties of the selected systems}  \label{sec:props}

\subsection{Fundamental stellar properties and photosphere models} \label{sec:stellarprops}
To estimate the basic parameters of the stars, such as their effective temperature ($T_\mathrm{eff}$) and 
luminosity ($L_*$), and to predict the photospheric contribution to the measured total flux at IR and mm wavelengths, 
we fitted the optical and near-IR broadband photometry available for the targets 
with ATLAS9 atmosphere models \citep{castelli2004}. The photometric measurements were taken 
from the TYCHO2 \citep{hog2000}, Hipparcos \citep{perryman1997}, Gaia~DR3 \citep{gaia2023}, APASS \citep{henden2016}, 
and 2MASS \citep{cutri2003} surveys. Where available, these data were complemented by $UBV$-band photometry from 
\citet{slawson1992}. 
According to our knowledge, HD~31305 is the only star in the sample having a close companion that could contaminate 
its photometry (see Appendix~\ref{sec:multiplicity}). In this case the pair of stars is not spatially resolved by 
Gaia either, but a high-resolution ground-based $K_\mathrm{p}$-band observation suggests that the brightness difference 
between the two stars is only 1\fm3, which, in addition to the stars' youth, is probably due to the fact that the 
late-type companion hosts a protoplanetary disk (see Appendix~\ref{sec:hd31305b}). Since the contamination can be significant at 
near-IR wavelengths, 2MASS measurements were ignored in the fitting of this target.

\begin{figure}[h!]
\centering
\includegraphics[bb=0 10 566 465,width=0.48\textwidth]{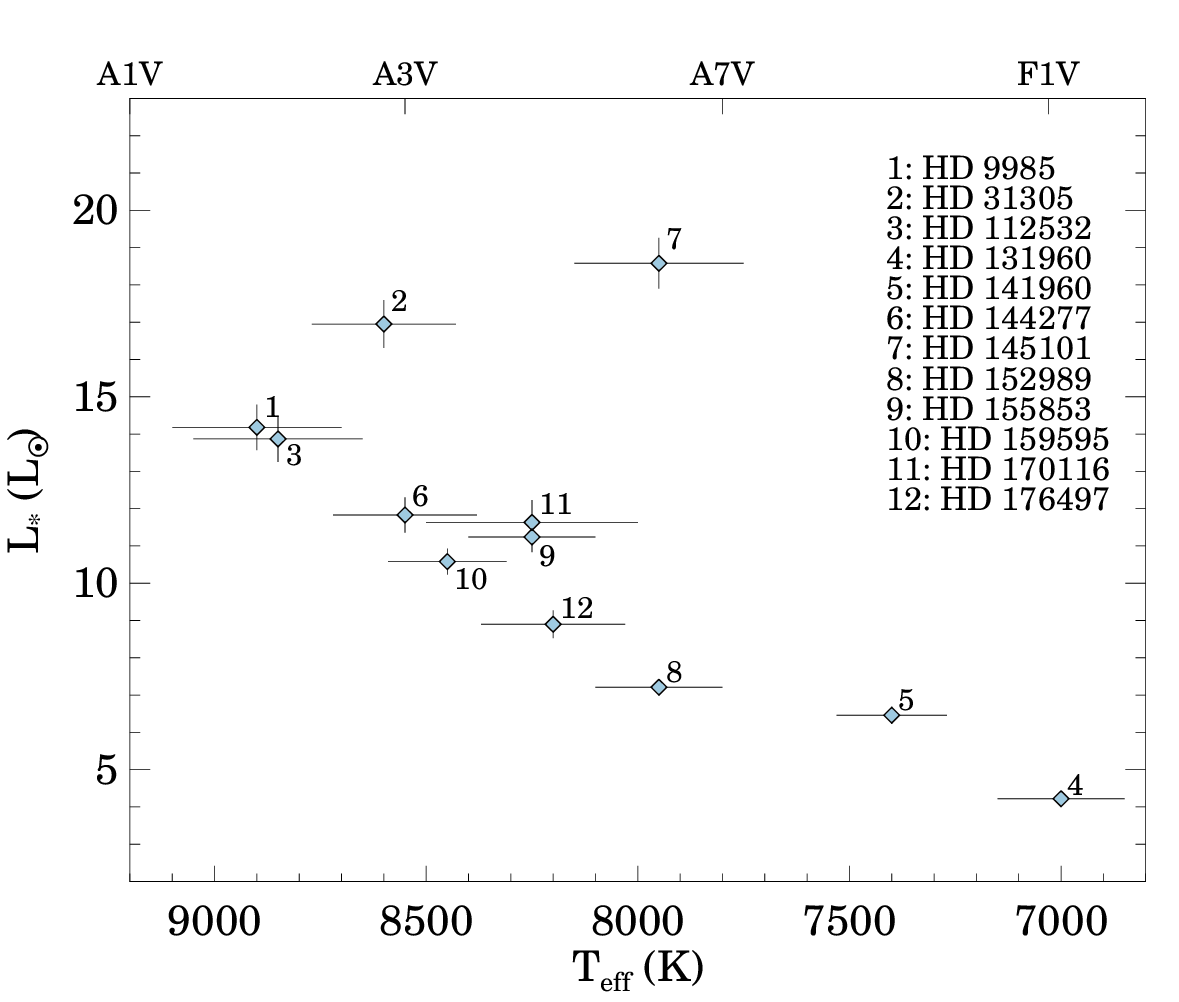}
\caption{Hertzsprung-Russel diagram of our targets. The effective temperature and luminosity values are taken from 
Appendix~\ref{sec:stellarprops}}
\label{fig:hrd}
\end{figure}

Synthetic photometry was derived from the model spectra for the given filter set and compared with 
the measured data in the fitting process. For the metallicity we adopted [Fe/H] = 0.0$\pm$0.1, while the estimates of 
the stars' interstellar reddenings ($E(B-V)$) were obtained using the \texttt{Stilism}~3D reddening map 
\citep{lallement2014,capitanio2017} and utilized as an a priori information. Distances were taken from \citet{bailerjones2021} 
adopting their geometric estimates. We used a Bayesian approach to estimate the posterior distributions of the model 
parameters, $T_\mathrm{eff}$ and the solid angle of the star. The derived $T_\mathrm{eff}$ and $L_*$ parameters with their 
uncertainties are summarized in Table~\ref{tab:targets}. The Hertzsprung--Russell diagram of the stars is shown in 
Figure~\ref{fig:hrd}.

\subsection{Multiplicity} \label{sec:multiplicity}
Using data from the Gaia~DR3 catalog, we searched for stellar companions of the target stars. By applying 
the method proposed by \citet{deacon2020} to Gaia objects in the vicinity of our targets, we identified two 
systems, HD~144277 and HD~152989, that have co-moving and co-distant pairs. 
The latter binary system was already known from the literature \citep{mason2001,dommanget2002,elbadry2018,tian2020}.
The projected separation is 12\farcs1 (1739\,au) for HD~144277\,B (Gaia~DR3 5997064901308024192) and 11\farcs2 (1288\,au) for 
HD~152989\,B (Gaia~DR3 6033260854136740992). 
By fitting the available Gaia~DR3 and 2MASS broadband photometric data of the revealed companions using stellar 
atmosphere models, ATLAS9 for HD~152989\,B and NEXTGEN \citep{hauschildt1999} for HD~144277\,B, we obtained effective 
temperature estimates of 5560$\pm$70\,K and 3320$\pm$60\,K, respectively. Based on the spectral type and $T_\mathrm{eff}$ 
sequences for young stars derived by \citet{pecaut2013} (see their table\,6), we found that these temperatures 
correspond to spectral types of G4 and M3.

In the Gaia~DR3 catalog, the renormalized unit weight error (\texttt{ruwe}) parameter shows how well the single 
star model fits to the astrometric observations. Therefore, the high value 
of \texttt{ruwe} \citep[$>$1.2, e.g.,][]{belokurov2020,bryson2021,wolniewicz2021} can be used as a robust 
indicator of the unresolved binarity for Gaia sources. Taking this into account, the \texttt{ruwe} value of 3.06 
obtained for HD~31305 strongly suggests the presence of a companion close to the star. This finding is further 
supported by the fact that the \texttt{ipd\_frac\_multi\_peak} parameter -- which shows the fraction of observing windows 
in which multiple peaks are detected and can thus also be used as a metric of binarity -- also has a high value of 36. 
In good agreement with this, as reported by \citet{cody2013}, a high-resolution $K_\mathrm{p}$-band image of HD~31305, 
obtained with the Keck/NIRC2 camera, reveals a nearby fainter star $\sim$0\farcs5 away, at a position 
angle of $-$128\fdg3, which is probably a late type companion of HD~31305. In the specific band, HD~31305 is only 1.3 
magnitudes brighter than its companion. The \texttt{ruwe} values of the other targets range between 
0.8 and 1.1, and their \texttt{ipd\_frac\_multi\_peak} is 0, i.e. there is no indication of close companions.
 
By comparing astrometric data from the Hipparcos and Gaia\,EDR3 catalogs, \citet{brandt2021} searched for signs of 
astrometric acceleration indicating the presence of companions. For HD~9985 and HD~170116, the two stars from our sample 
that are included in both catalogs, no significant acceleration was found. Consistently with this, the study of 
\citet{kervella2022} -- which is also based on these two catalogs -- did not reveal any significant proper motion anomaly 
for these two stars. Thus, these data do not indicate the presence of companion stars.
 
\subsection{Membership in young associations} \label{sec:membership}
We used the \texttt{BANYAN $\Sigma$} tool \citep{gagne2018}, a Bayesian classifier, to assess the membership 
probabilities of our target stars in 27 young associations situated within 150\,pc of the Sun. The input data for the 
calculations: stellar positions, proper motions, parallaxes and, where available, radial velocities (RVs), and the 
uncertainties of these parameters were taken from the Gaia~DR3 catalog. 
For HD~9985 there is no velocity data in the Gaia catalog, but \citet{gontcharov2006} lists an RV of 
$+$10.2$\pm$3.7\,km~s$^{-1}$. However, since the latter measurement has a rather large uncertainty we used its heliocentric 
systemic velocity ($6.3\pm0.3$\,km~s$^{-1}$) derived as the weighted center of the $^{12}$CO line instead 
(Sect.~\ref{sec:coimaging}). For two other targets, HD~31305 and HD~112532, no RV data are available, 
and therefore their derived membership probabilities are based only on the astrometric data. 
Table~\ref{tab:membership} shows the most likely young association, and the membership probability ($P_\mathrm{memb}$) 
for those 9 systems where the latter value is $>60$\%. For these objects, previous studies have already shown that 
they likely belong to the indicated young stellar groups (see Table~\ref{tab:membership}). 


\begin{table}
\begin{center}                                                                                                                                       
\caption{Membership of target stars in nearby young associations. 
 \label{tab:membership}}
\begin{tabular}{llrlc}                                                     
\hline\hline
Target name & Group & $P_\mathrm{memb}$ & Refs.     & SigMA cluster \\      
\hline
HD~31305    &  TAU  & 81.1              & 1,2,9,10  & \ldots	    \\
HD~112532   &  LCC  & 99.4              & 5         & $\sigma$ Cen  \\  
HD~131960   &  UCL  & 90.0              & 8         & Libra-South   \\  
HD~141960   &  US   & 63.1              & 3,6,7     & $\rho$ Sco    \\  
HD~144277   &  UCL  & 99.9              & 3,6       & $\eta$ Lup    \\  
HD~145101   &  US   & 99.9              & 6,7       & $\delta$ Sco  \\  
HD~152989   &  UCL  & 93.6              & 8         & $\eta$ Lup    \\  
HD~155853   &  UCL  & 71.5              & \ldots    & Scorpio-Body  \\  
HD~176497   &  UCRA & 100.0             & 3,4       & CrA-North     \\  
\hline
\end{tabular}
\tablefoot{Column 2 and 3 summarize the outcomes of
\texttt{BANYAN $\Sigma$}, showing the most probable young association to which the star may 
belong (Col.~2) and the obtained membership probability (Col.~3). Column 4 lists references for 
previous studies dealing with the star's membership in the respective group. Using a novel 
clustering method (\texttt{SigMA}) and harnessing the unique potential of Gaia~DR3, 
\citet{ratzenboeck2023a} proposed a new division of young stars in the Sco-Cen region. 
Column 5 shows the names of the subgroups to which the stars belong in this new classification. 
{References used in this table:} 
  (1): \citet{cody2013}; (2): \citet{gagne2018}; (3): \citet{gagne2018b}; (4): \citet{galli2020}; 
  (5): \citet{goldman2018}; (6): \citet{hoogerwerf2000}; (7): \citet{luhman2018}; (8): \citet{luhman2022}; 
  (9): \citet{luhman2023}; (10): \citet{mooley2013}. }
\end{center}
\end{table}


The \texttt{BANYAN $\Sigma$} algorithm, following the classical approach proposed by \citet{blaauw1964}, 
distinguishes three subgroups in the Scorpius-Centaurus (Sco-Cen) OB association: the Upper Scorpius (US), 
the Upper Centaurus Lupus (UCL) and the Lower Centaurus Crux (LCC). However, new studies, based on Gaia data, revealed 
a more complex morphology with more substructures in the Sco-Cen OB association \citep[e.g.][]{goldman2018}. 
Recently, by applying a new clustering algorithm, the significance mode analysis (\texttt{SigMA}), \citet{ratzenboeck2023a} 
identified 37 co-moving clusters in Sco-Cen. For the eight objects where this is relevant, the sub-clusters
corresponding to this new division have also been indicated in the table.
As for HD~31305, in their recent study of the Taurus star-forming region, \citet{luhman2023} proposed that it belongs 
to the L1517 subgroup.

\begin{figure}
\centering
\includegraphics[width=0.48\textwidth]{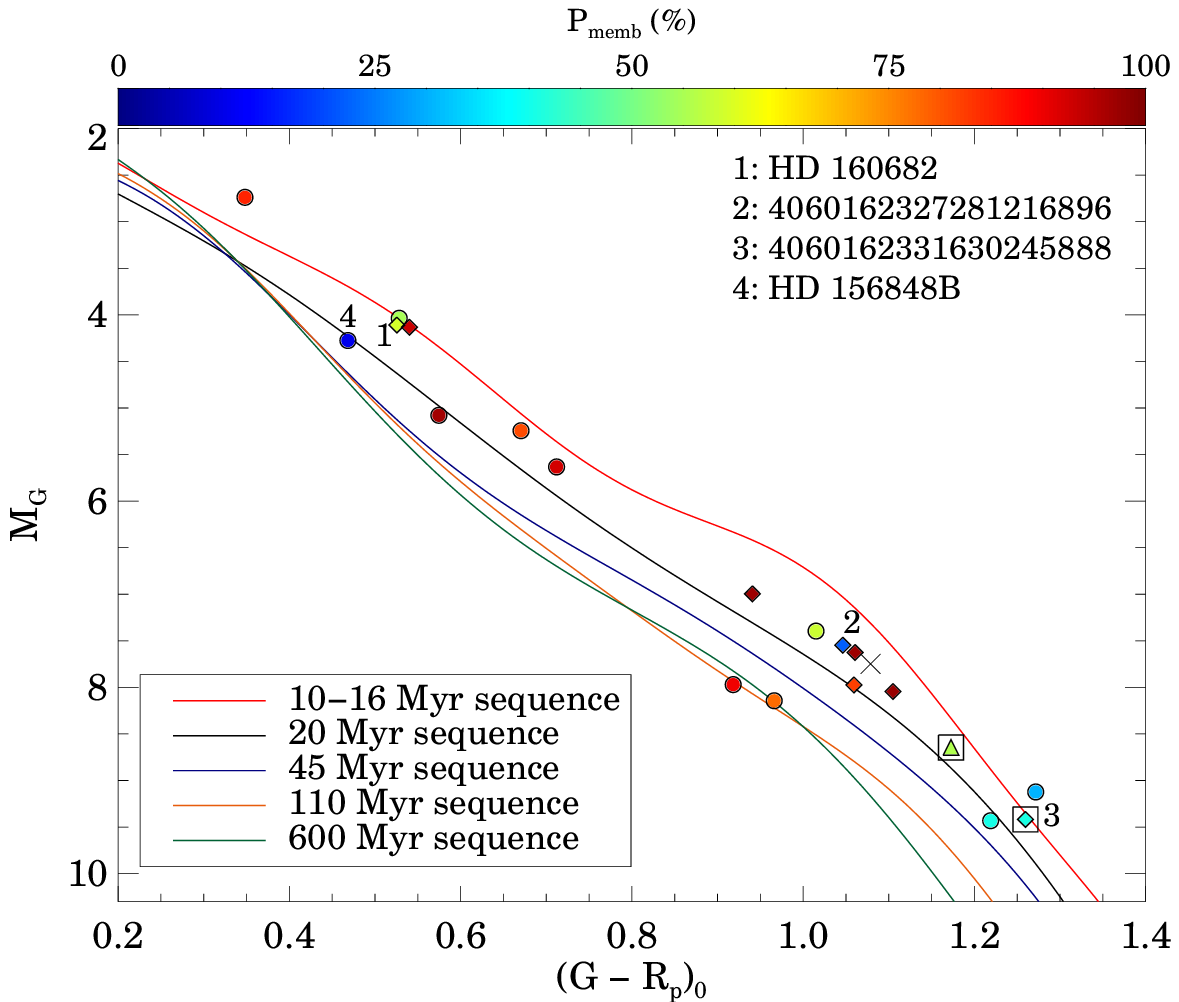}
\includegraphics[bb=0 30 566 465,width=0.48\textwidth]{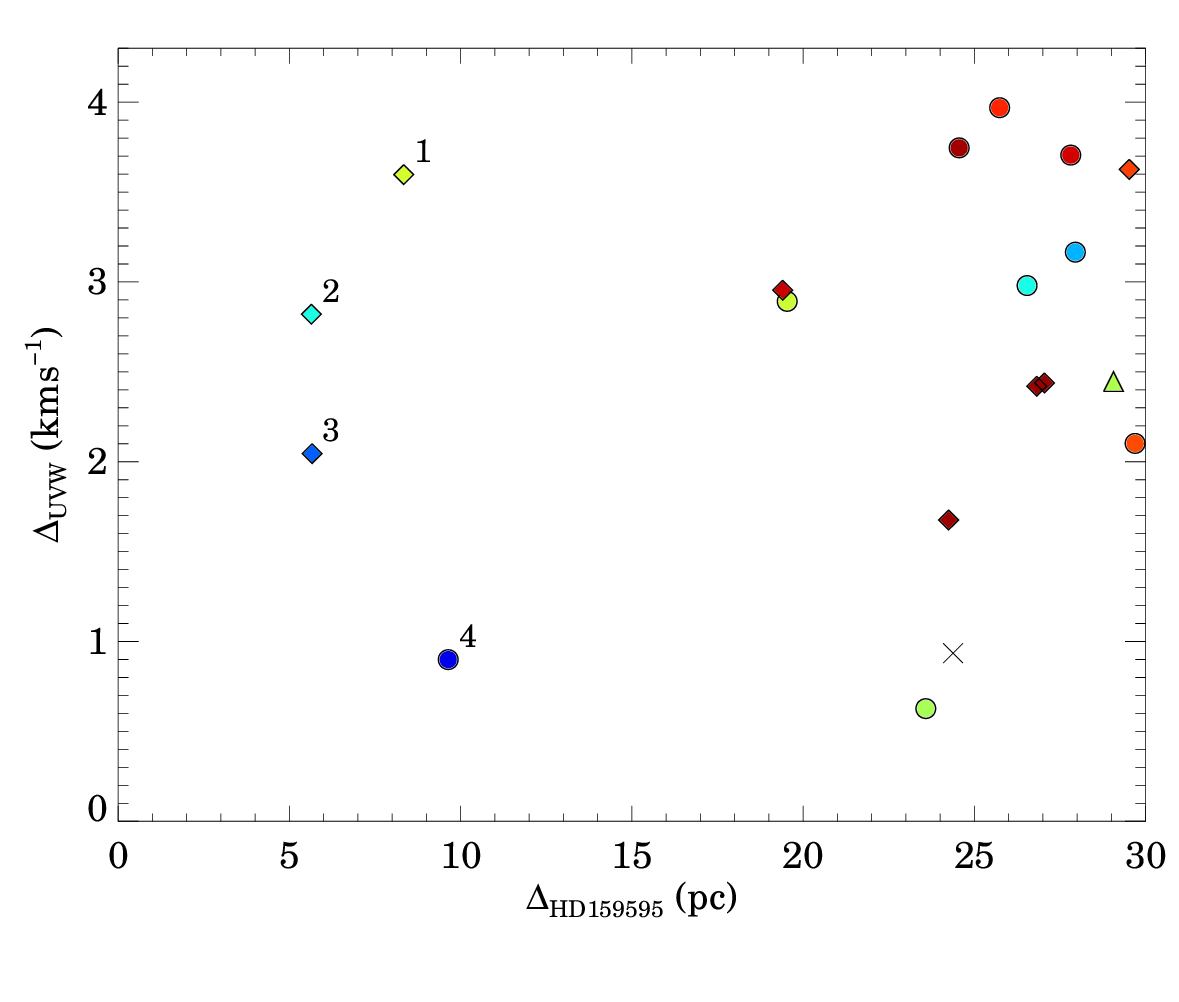}
\caption{Top: ($M_\mathrm{G}, (G-R_\mathrm{p})_0$) color-magnitude diagram for 21 Gaia~DR3 stars located 
within 30\,pc of HD\,159595 and having similar galactic space motion (Sect.~\ref{sec:membership}). 
Potential members of BPMG, UCL, and US are marked by filled diamonds, circles, and triangles, respectively. 
The color of the symbols indicates the membership probability. A star with a group membership probability of no more than 
10\% in any of these clusters is marked by a cross. Stars with Gaia \texttt{ruwe} $>$ 1.2
(possible binaries) are denoted with large squares. Empirical sequences of different ages are taken from \citet{gagne2021}. 
Bottom: Galactic space velocity differences ($\Delta_\mathrm{UVW}$) as a function of positional differences 
($\Delta_\mathrm{HD159595}$) with respect to HD\,159595 for 19 stars, found to be young 
from the above 21.}
\label{fig:hd159595}
\end{figure}

\paragraph{HD\,159595.} For HD~159595, \texttt{BANYAN $\Sigma$} returned membership probabilities of 24.7\% and 9.4\% for 
the $\beta$\,Pic moving group (BPMG) and UCL, respectively. The 
$U_0, V_0, W_0 = -10.11, -15.45, -8.43~(\pm0.65, \pm0.07, \pm0.03)$\,km\,s$^{-1}$ galactic space  
velocities of the star is in excellent agreement with the average space motion of the BPMG 
\citep[$\langle U \rangle, \langle V \rangle, \langle W \rangle = -10.9, -16.0, -9.10$\,km\,s$^{-1}$,][]{gagne2018}: the relatively low membership probability is rather due to the fact 
that the object is located outside the space region occupied by the BPMG as defined in the BANYAN model. However, some 
recent results suggest that this group may actually be more widespread than previously thought \citep[e.g.][]{hinkley2021}. 

Therefore, to explore this issue further, we have collected all stars located within 30\,pc of HD~159595 from the Gaia~DR3 catalog, 
that have 
1) \texttt{ruwe}$<$1.4, 2) \texttt{parallax\_over\_error}$>$10, and 3) radial velocity with an uncertainty $<$5\,km\,s$^{-1}$ and, 
to ensure the sufficiently good quality of the photometry, whose 4) corrected $BP$ and $RP$ flux excess $|C_*| < 5 \sigma_{C_*}$ 
\citep[see][for the definition of $C_*$ and $\sigma_{C_*}$]{riello2021}. After calculating their $U, V, W$ velocities we kept only 
those having velocities within 4\,km\,s$^{-1}$ of HD 159595 ($\Delta_\mathrm{UVW} = \sqrt{ (U-U_0)^2 + (V-V_0)^2 + (W-W_0)^2} < 4$~km~s$^{-1}$).
These selection criteria left us with 21 sources whose membership probabilities were also assessed using the \texttt{BANYAN $\Sigma$}. 
In Figure~\ref{fig:hd159595} (top) we show the $(G-R_\mathrm{P})_0$ color versus absolute $G$ magnitude diagram of 
the selected sources. To correct for interstellar extinction, the $E(B-V)$ reddenings of the objects were taken 
from \texttt{Stilism} 3D map \citep{lallement2014,capitanio2017}.  For those stars where \texttt{BANYAN $\Sigma$} yielded a 
probability of $>$10\% for a young association, we marked the most likely cluster and used colors to indicate the obtained 
probability. Empirical sequences of known coeval populations of stars that represent five different ages from \citet{gagne2021} 
were also added to the plot. The figure suggests that, with two exceptions, stars with similar galactic space motion and spatial 
position to those of HD~159595 are young ($\lesssim$50\,Myr). In some cases, other observations confirm this result. 
Seven out of the nineteen stars were included in the SACY survey \citep{torres2006} and their outstandingly high measured lithium 
equivalent line widths are clear indicators of their youth. Four other objects have counterparts in the ROSAT catalog 
\citep{boller2016} and their high X-ray luminosities suggest strong coronal activity that is most likely associated with young stars.
For the majority of these young stars, the most likely associated group is the BPMG (filled diamonds) or the UCL (filled circles).

Figure~\ref{fig:hd159595} (bottom) displays the $\Delta_\mathrm{UVW}$ velocity differences as a function of 
the distance from HD~159595 for the 19 young nearby co-moving stars. For most objects located more than 18\,pc away, 
\texttt{BANYAN $\Sigma$} returned membership probabilities $>$50\%. 
The five systems assigned to the BPMG,  all have $P_\mathrm{memb}>$82\%, including stars such as HD\,168210 and GSC~07396-00759, 
which have long been identified as secure members of this group \citep{torres2006}. Within 10\,pc, in the immediate 
vicinity of HD\,159595, there are four young stars that show similar galactic space motion. For HD~156848\,B 
(Gaia~DR3~4108078979665915648, indicated by 4 in Fig.~\ref{fig:hd159595}), \texttt{BANYAN $\Sigma$} yielded a probability of 
10.4\% to belong to the UCL. Apart from the fact that this indicates a rather low probability, it should be mentioned that the 
radial velocity of the star's companion, HD~156848\,A, is quite different from that of HD~156848\,B ($-4.56\pm$0.3\,km\,s$^{-1}$ 
compared to $-10.79\pm$1.5~km~s$^{-1}$) and thus for the former the $\Delta_\mathrm{UVW}$ is of 5.9\,km\,s$^{-1}$, which raises 
questions even how similar the space motion of this system to that of HD\,159595. The other three stars are candidate members of 
the BPMG. Gaia~DR3~4060162327281216896 (marked by number 2 in Fig.~\ref{fig:hd159595}, $P_\mathrm{memb,BPMG} = $22\%) 
and Gaia~DR3~4060162331630245888 (number 3, $P_\mathrm{memb,BPMG} = $41.1\%) are separated in the sky by $\sim$5\farcs1 
and, based on their very similar proper motions and parallaxes, they may form a wide binary system at a distance of ~5.7\,pc 
from HD\,159595. Finally, HD\,160682 (number 1, $P_\mathrm{memb,BPMG} = $58.8\%) is a well-known young star with a high lithium 
content \citep{torres2006,binks2020}. The proximity of the latter two systems (three stars) to HD~159595 suggests that they belong 
together to the BPMG, but are further away from its main region of the group. However, it cannot be excluded that, despite 
their very similar galactic space motion to the BPMG, these young stars form an independent small cluster locating not very 
far from this group.

\paragraph{HD~9985.} According to \texttt{BANYAN $\Sigma$}, HD~9985 does not belong to any of the 27 young clusters 
considered by the classification tool. However, \citet{oh2017} found that it has a co-moving pair, HQ~Psc (TYC~1204-171-1, 
Gaia~DR3~288826178310839936), a late-type star. In fact, based on Gaia~DR3, the two stars are 13.4\,pc apart, and their 
galactic space motion, $U, V, W = -10.1, -5.0, -5.1~(\pm0.2, \pm0.2, \pm0.2)$~km~s$^{-1}$ for HD~9985 
and $U, V, W = -11.0, -6.5, -4.3~(\pm0.8, \pm0.8, \pm1.0)$~km~s$^{-1}$ for HQ~Psc, are also very similar. 
\citet{binks2018} proposed that HQ~Psc, together with 13 other young late-type stars, form a spatially 
very extended kinematic association, the 30--50\,Myr old Pisces moving group. However, based on more recent and complete 
kinematic data, \citet{moranta2022} found that the galactic space velocities of the proposed members are quite different, 
making it unlikely that this is a single coeval group. Alternatively, they suggest that HQ~Psc may form a smaller group with 
three other previously proposed members of the putative Pisces MG, but the fact that HQ~Psc is 65\,pc away from even the 
nearest of these stars (and 123\,pc from the furthest one) makes this suggestion questionable as well, yet. 
So, although their similar motions and spatial positions suggest that HD\,9985 and HQ\,Psc belong to the same moving group 
further investigation is needed to see if there are other members in their vicinity.

\paragraph{HD~170116.} Due to its greater distance, the HD\,170116 is well outside the region considered by 
the \texttt{BANYAN $\Sigma$} model, so it is not surprising that the tool found no relationship between the 
listed young groups and the star. Other literature data do not suggest any group membership either. 
Using the same criteria as in the case of HD\,159595, our search for co-moving stars in the 30\,pc neighborhood of the target 
resulted in 9 objects. Of these, only one, Gaia~DR3~4270531719524839936, is likely a young star ($<$100\,Myr), based on its 
position on the color-magnitude diagram ($(G-R_\mathrm{P})_0$ vs. $M_\mathrm{G}$, CMD) and its strong X-ray emission 
($\log{L_\mathrm{X}/L_\mathrm{bol}} = -3.1\pm 0.2$). This star has a $UVW$ velocity of 
$-7.7, -22.5, -6.9~(\pm2.2, \pm1.2, \pm0.2)$~km~s$^{-1}$, which is almost identical with that of HD\,170116 
($U, V, W = -7.4, -22.2, -6.7~(\pm1.2, \pm0.7, \pm0.1)$~km~s$^{-1}$).

\subsection{Stellar ages and masses}
With the exception of HD~145101, stars that found to be members of young associations (Table~\ref{tab:targets}) were assumed to 
have the same age as that of the group. The ages for the US, UCL, and LCC subgroups of the Sco-Cen association, 
10$\pm$3, 15$\pm$3, and 16$\pm$2\,Myr, are taken from \citet{gagne2018}. For the L1517 subgroup of the Taurus, that includes HD~31305, 
\citet{luhman2023} obtained an age estimate of 2.5$_{-1.5}^{+3.8}$~Myr. To estimate the masses of these stars, their effective 
temperatures and luminosities obtained in Appendix~\ref{sec:stellarprops} were compared to the theoretical
predictions of the MESA Isochrones and Stellar Tracks \citep{choi2016,dotter2016}. In the isochrone fitting, we
followed the method proposed by \citet{pascucci2016}, by assuming a metallicity of [Fe/H] = 0.0$\pm$0.1 and taking 
into account the derived ages as an a priori information. Note that even if the new, finer SigMa-based subdivision of the 
Sco-Cen and the new subcluster ages \citep{ratzenboeck2023b} are used instead of the classical ones, no significant 
difference in the mass estimates is obtained, since the alternative age estimates are in good agreement with the 
used ones within the uncertainties. Based on its position on the HRD (Fig.~\ref{fig:hrd}), HD~145101 clearly appears to be 
a pre-main sequence star which is likely younger than the average of the US population, which group is found to exhibit 
quite wide age range between 3 and 19\,Myr by \citet{ratzenboeck2023b}. For this target, therefore we 
only assumed a priori that it is younger than 20\,Myr (as a member of the US group), and 
its final age estimate, 6.7$_{-0.7}^{+1.0}$\,Myr, was derived from the isochrone fitting. 
   
If HD~159595 belongs to the BPMG, then we can use the age of the group as a priori data in the isochrone fitting.
Using the average of the age estimates from the literature, \citet{gratton2024} derived an age of 21$\pm$4\,Myr for this group. 
However, it is possible that HD~159595, together with three other nearby stars (two of which, Gaia~DR3~4060162327281216896 and 
Gaia~DR3~4060162331630245888, constitute a wide binary), may form a small cluster separate from the BPMG 
(Appendix~\ref{sec:membership}). Taking into account the available age indicators, however, it can be concluded that this small 
group may not only have UVW velocity but also an age quite similar to that of the BPMG.
On the CMD shown in Figure~\ref{fig:hd159595} (top), HD~160682, Gaia~DR3~4060162327281216896, and Gaia~DR3~4060162331630245888, 
are all located between the 10--16 and 20\,Myr empirical sequences, although the latter star has a \texttt{ruwe} of 1.21, raising 
the possibility that it is not a single system. Comparing the lithium equivalent width measured in the SACY survey 
\citep{torres2006} for HD~160682 ($EW_\mathrm{Li}=260$\,m{\AA}, $(B-V)_\mathrm{0} = 0.070$) with the sequence of Li equivalent 
widths of BPMG stars \citep[][figure~5]{baffles} we find a good match. Using the \texttt{eagles} tool \citep{jeffries2023} to 
estimate the stellar age from the lithium observation we obtained an upper limit of 50\,Myr. Overall, this small group is also 
of similar age, or at most a few million years younger than the BPMG, so for the final age estimate of HD\,159595 we adopted the 
age of the BPMG by assigning a somewhat higher value to the lower age uncertainty (21$^{+4}_{-5}$\,Myr). We used this age as an 
a priori information in the mass estimate of the star.  

In age diagnostic of HD~9985, we assumed that the star is co-moving and coeval with HQ~Per. For the latter star, 
using the available broadband photometric data (from Gaia~DR3, TYCHO, APASS and 2MASS surveys) and the method described 
in Sect.~\ref{sec:stellarprops} we obtained $T_\mathrm{eff}$ of 5940$\pm$100\,K and $L_*$ of 1.27$\pm$0.06\,$L_\odot$. 
The posterior age probability distributions were then determined for both stars via isochrone fitting. 
Using the measured lithium equivalent width of HQ~Psc \citep[144$\pm$21\,m{\AA},][]{binks2015} and the derived $T_\mathrm{eff}$,
 we also obtained an age posterior probability distribution based on this indicator by applying the \texttt{eagles} package again. 
Finally, the fast rotation \citep[$P_\mathrm{rot}$=1.87\,d,][]{binks2015} and the strong X-ray emission 
\citep[$\log{L_\mathrm{X}/L_\mathrm{bol}}$ = -3.84,][]{binks2015} of this late-type star indicate an age of $<$100\,Myr. 
By combining all these age posterior distributions and determining the median and the 68\% confidence interval 
of the final distribution, we obtained an age estimate of 44$_{-16}^{+31}$\,Myr for HD\,9985 (and for HQ~Psc).   

HD\,170116 also has a potential young late-type co-moving pair, the Gaia~DR3~4270531719524839936. 
If we assume that the similarity in their kinematics is due to the fact that they were born in the same star formation region at 
roughly the same time, then their age analysis is worth combining. Based on optical and near-IR photometric data we 
derived $T_\mathrm{eff}$ of 4980$\pm$90\,K and $L_*$ of 0.44$\pm$0.04\,$L_\odot$ for the late-type star.
Its high X-ray luminosity (Sect.~\ref{sec:membership}) implies strong magnetic activity 
and fast spinning. To further study this, we downloaded its ASAS-SN $g$-band light curve using the 
\texttt{SkyPatrol} tool\footnote{https://asas-sn.osu.edu}. By applying the Generalized Lomb-Scargle algorithm 
\citep[GLS;][]{zechmeister2009} to the data we found that the star exhibits significant periodic variability with a period 
of $\sim$2.4\,$d$, likely due to rotational variations from surface spots. This rapid rotation and the strong X-ray emission 
both indicate that Gaia~DR3~4270531719524839936 is younger than 100\,Myr \citep{bouma2023,stassun2024}. By performing isochrone 
fitting for both stars -- assuming that neither has a close companion (in line with their Gaia \texttt{ruwe} of <1.2) - and 
combining the results, also taking into account the upper age limit derived above, we obtain an age estimate of 31$^{+9}_{-7}$\,Myr.  
We note that this result is mostly based on the isochrone fitting of Gaia~DR3~4270531719524839936. 


\begin{table}                                                                                                              
\begin{center}
\caption{Basic disk properties.  
 \label{tab:mbbprops} }
\begin{tabular}{lccc}                                                     
\hline\hline
Target name & $T_\mathrm{dust}$ (K) & $\beta$ &  $L_\mathrm{disk}/L_\mathrm{bol}$ \\
\hline
       HD\,9985 &        80$\pm$6 &   0.41$\pm$0.15 &  1.3e-03 \\
     HD\,112532 &      139$\pm$12 &            0.65 &  7.3e-04 \\
     HD\,131960 &      182$\pm$16 &            0.65 &  1.3e-03 \\
     HD\,141960 &       119$\pm$8 &            0.65 &  2.3e-03 \\
     HD\,144277 &      151$\pm$18 &            0.65 &  5.9e-04 \\
     HD\,145101 &      163$\pm$15 &            0.65 &  6.1e-04 \\
     HD\,152989 &        96$\pm$4 &   0.85$\pm$0.09 &  7.6e-03 \\
     HD\,155853 &      159$\pm$11 &            0.65 &  1.7e-03 \\
     HD\,159595 &        90$\pm$4 &            0.65 &  2.1e-03 \\
     HD\,170116 &        74$\pm$3 &   0.40$\pm$0.11 &  1.4e-03 \\
     HD\,176497 &       141$\pm$9 &   0.60$\pm$0.16 &  1.7e-03 \\
\hline
\end{tabular}
\end{center}
\end{table} 


\subsection{Fundamental disk properties} \label{sec:diskprops}
\begin{figure*}[h!]
\centering
\includegraphics[width=1.00\textwidth]{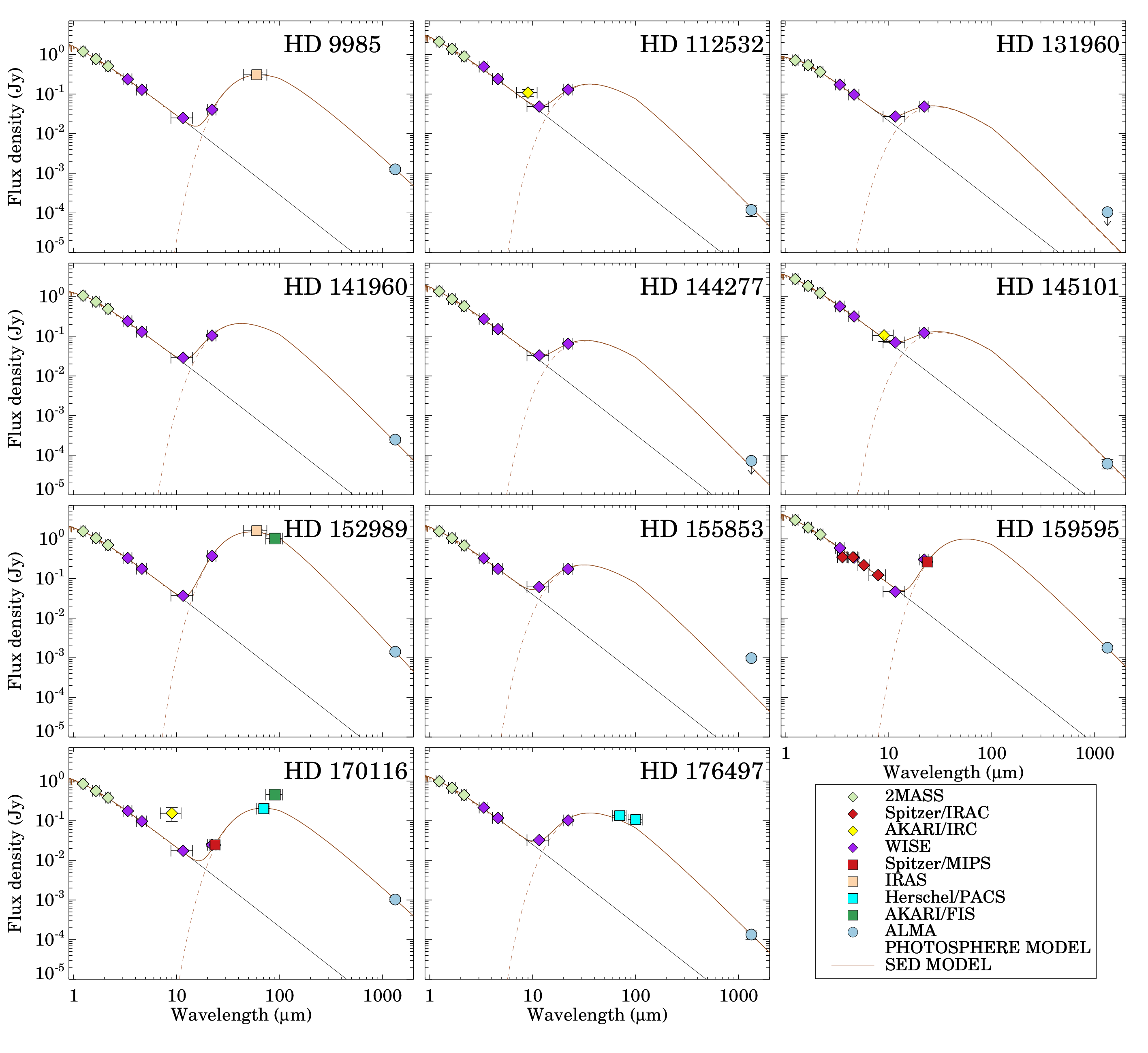}
\caption{Spectral energy distributions of the studied systems. The stellar photosphere and the best-fitting SED models 
(Sect.~\ref{sec:diskprops}) are also displayed. The horizontal bars show the width of the filters used to obtain the photometric 
data.}
\label{fig:sedplot}
\end{figure*}

Figure~\ref{fig:sedplot} shows the spectral energy distribution from 1\,$\mu$m to 2\,mm for all our targets 
except HD\,31305. The plotted IR photometry are taken from the 2MASS \citep{skrutskie2006}, AllWISE \citep{cutri2013}, 
{\sl AKARI} IRC and FIS \citep{ishihara2010,yamamura2010}, and {\sl IRAS} FSC \citep{moshir1990} all-sky survey catalogs.
In addition, three of our targets were serendipitously observed as part of larger surveys performed by the {\sl Spitzer} and/or 
the {\sl Herschel} space telescopes, allowing us to complement their SEDs by {\sl Spitzer} IRAC and MIPS photometry 
\citep{fazio2004,rieke2004} from the {\sl Spitzer} Enhanced Imaging Products \citep[SEIP;][DOI: 10.26131/IRSA3]{seip2021} and 
{\sl Herschel} PACS data \citep{poglitsch2010} from the Herschel/PACS Point Source catalogs \citep{herschel}.
To correct for potential overestimation of the true flux in the saturated WISE $W2$-band 
photometry at 4.6$\mu$m, we used the method proposed by \citet[][equation~5]{cotten2016}. 
The final photometric uncertainties were derived as quadratic sums of the measurement errors listed in the catalogs 
and the corresponding absolute calibrational uncertainties \citep[][The AllWISE Data 
Release\footnote{\url{https://wise2.ipac.caltech.edu/docs/release/allwise/}}, IRAC Instrument 
Handbook\footnote{\url{https://irsa.ipac.caltech.edu/data/SPITZER/docs/irac/iracinstrumenthandbook/}}, MIPS Instrument 
Handbook\footnote{\url{https://irsa.ipac.caltech.edu/data/SPITZER/docs/mips/mipsinstrumenthandbook/}}]{ishihara2010,yamamura2010,balog2014}. 
The ALMA photometric data points obtained in our project at 1.33\,mm 
(Table~\ref{tab:targets}) are also added to the plots.

To estimate the basic disk properties, the characteristic dust temperature ($T_\mathrm{dust}$) and the 
fractional luminosity ($L_\mathrm{disk} / L_\mathrm{bol}$), the excess SEDs -- derived by subtracting the 
photospheric models (Sect.~\ref{sec:stellarprops}) from the measured flux densities -- are fitted by
a single temperature modified blackbody (MBB) model whose emissivity is 1 at wavelengths shorter than $\lambda_0$ 
and varies as $(\lambda/\lambda_0)^{-\beta}$ beyond that.
The rationale for using this type of model is that observations show that at wavelengths well above the typical grain 
size, the spectrum of debris disks falls off significantly steeper than that of a blackbody. 
To find the best-fitting model, we used the \textsc{MPFIT} IDL routine \citep{markwardt2009}.
For HD\,9985, HD\,152989, HD\,170116, and HD\,176497, where far-IR data points in the wavelength range 
between 25 and 200$\mu$m are available, 
we fixed the $\lambda_0$ break wavelength to 100$\mu$m \citep{williams2006}, while the dust temperature, 
the solid angle of the emitting region and the $\beta$ were free parameters. For the other objects, where the measurements,
except for the ALMA observation, are limited to wavelengths shorter than 25$\mu$m, the $\beta$ parameter was also fixed at 
0.65. The obtained disk parameters are summarized in Table~\ref{tab:mbbprops}.

The lack of far-infrared data points and, in more general, the sparse sampling of the SEDs, makes it difficult to assess 
for some of the sources how correct the adopted single temperature component model is, and whether a colder dust component 
is not actually present, as observed in many debris disks \citep{kennedy2014,chen2014}. The fact that, with the fixed disk 
parameters, most of these SED models agree reasonably well with the measured ALMA flux density at 1.33\,mm suggests that, 
if there is a colder dust component at all, its contribution is not particularly pronounced. HD\,155853, however, is an 
exception: its MBB model significantly underestimates the millimeter brightness of the disk and even if a pure blackbody 
model were used, the measured flux density would not match. We also know from the ALMA observations (Sect.~\ref{sec:continuum}) 
that the star is surrounded by a spatially extended, presumably cold, outer disk. So for this object, with a proper sampling, 
we would probably see an excess SED that is better modeled by two dust components with different temperatures. Assuming that 
this cold component can also be described by a modified blackbody model with the same fixed $\lambda_0$ and $\beta$ parameters 
as we used above, the fractional luminosity can be estimated from the ALMA measurement. The temperature of the cold dust is 
not known, but it is reasonable to assume that it should be higher than 47\,K, which is the blackbody temperature corresponding 
to the measured radius of the disk (Table~\ref{tab:targets}). Taking these into account, we obtain a lower limit of 
3.3$\times$10$^{-4}$ for $f_\mathrm{d}$ of the cold component.

\setcounter{section}{1}
\section{Overview of the imaging parameters} \label{sec:impars}
Table~\ref{tab:imagingpars} present the relevant imaging parameters for our continuum and line observations. 

\begin{table*} 
\begin{center}                                                                                                                                       
\caption{Overview of the imaging parameters for continuum and CO line observations. \label{tab:imagingpars}}
\begin{tabular}{lcccc}                                                     
\hline\hline
    & Continuum   &   $^{12}$CO    &    $^{13}$CO    &    C$^{18}$O  \\
\hline
    &  \multicolumn{4}{c}{HD\,9985}        \\ 
\hline    
Beam size (arcsec$\times$arcsec) & 0.58$\times$0.45  &  0.58$\times$0.46   &  0.61$\times$0.47   &  0.61$\times$0.47   \\ 
Beam PA (\degr)                  &   36.6            &      38.1           &       38.1          &      38.4           \\
rms noise ($\mu$Jy~beam$^{-1}$ / mJy~beam$^{-1}$~ch$^{-1}$)  &    15.9     &   1.6               &   0.8    & 0.6          \\
\hline
    & \multicolumn{4}{c}{HD\,31305}        \\
\hline 
Beam size (arcsec$\times$arcsec) & 0.42$\times$0.25  &  0.42$\times$0.25   &  0.45$\times$0.27   &  0.47$\times$0.27   \\ 
Beam PA (\degr)                  &   -0.5            &      0.1            &       5.6           &      4.9           \\
rms noise ($\mu$Jy~beam$^{-1}$ / mJy~beam$^{-1}$~ch$^{-1}$)    &    21.3       &   2.7       &   1.4        & 1.2          \\
\hline
    & \multicolumn{4}{c}{HD\,112532}        \\ 
\hline
Beam size (arcsec$\times$arcsec) & 0.57$\times$0.41  &  0.58$\times$0.42   &  \ldots   &  \ldots   \\ 
Beam PA (\degr)                  &   -66.4           &      -66.7          &  \ldots   &  \ldots   \\
rms noise ($\mu$Jy~beam$^{-1}$ / mJy~beam$^{-1}$~ch$^{-1}$)    &    19.0       &   2.3   &   \ldots    &  \ldots  \\
\hline
    & \multicolumn{4}{c}{HD\,131960}        \\
\hline 
Beam size (arcsec$\times$arcsec) & 1.16$\times$1.05  &  1.25$\times$1.15   &  \ldots   &  \ldots   \\ 
Beam PA (\degr)                  &   81.8            &     76.8            &  \ldots   &  \ldots   \\
rms noise ($\mu$Jy~beam$^{-1}$ / mJy~beam$^{-1}$~ch$^{-1}$)    &    27.3       &   4.1   &   \ldots    &  \ldots  \\
\hline
    & \multicolumn{4}{c}{HD\,141960}        \\ 
\hline
Beam size (arcsec$\times$arcsec) & 0.51$\times$0.40  &  0.53$\times$0.40   &  \ldots   &  \ldots   \\ 
Beam PA (\degr)                  &   -77.6           &     -76.4           &  \ldots   &  \ldots   \\
rms noise ($\mu$Jy~beam$^{-1}$ / mJy~beam$^{-1}$~ch$^{-1}$)    &    14.9       &   1.8   &   \ldots    &  \ldots  \\
\hline
    & \multicolumn{4}{c}{HD\,144277}        \\ 
\hline
Beam size (arcsec$\times$arcsec) & 0.63$\times$0.54  &  0.65$\times$0.54   &  \ldots   &  \ldots   \\ 
Beam PA (\degr)                  &   -81.6           &     -82.4           &  \ldots   &  \ldots   \\
rms noise ($\mu$Jy~beam$^{-1}$ / mJy~beam$^{-1}$~ch$^{-1}$)    &    18.0       &   2.2   &   \ldots    &  \ldots  \\
\hline
    & \multicolumn{4}{c}{HD\,145101}        \\ 
\hline
Beam size (arcsec$\times$arcsec) & 0.51$\times$0.37  &  0.52$\times$0.37   &  0.54$\times$0.42   &  0.54$\times$0.42   \\ 
Beam PA (\degr)                  &   -74.2           &      -77.9          &       -71.2         &      -71.0          \\
rms noise ($\mu$Jy~beam$^{-1}$ / mJy~beam$^{-1}$~ch$^{-1}$)    &    15.1       &   1.9       &   1.0        & 0.8          \\
\hline
    & \multicolumn{4}{c}{HD\,152989}        \\ 
\hline
Beam size (arcsec$\times$arcsec) & 0.37$\times$0.28  &  0.38$\times$0.29   &  0.39$\times$0.32   &  0.40$\times$0.32   \\ 
Beam PA (\degr)                  &    88.5           &     84.6            &       83.0          &      84.9          \\
rms noise ($\mu$Jy~beam$^{-1}$ / mJy~beam$^{-1}$~ch$^{-1}$)    &    16.1       &  2.1       &   1.2        & 0.9          \\
\hline
    & \multicolumn{4}{c}{HD\,155853}        \\ 
\hline
Beam size (arcsec$\times$arcsec) & 0.57$\times$0.40  &  0.51$\times$0.38   &  0.57$\times$0.41   &  0.58$\times$0.41   \\ 
Beam PA (\degr)                  &    -71.7          &     -72.6           &      -73.7          &      -74.5          \\
rms noise ($\mu$Jy~beam$^{-1}$ / mJy~beam$^{-1}$~ch$^{-1}$)    &    15.3       &  2.4       &   1.0        & 0.9          \\
\hline
    & \multicolumn{4}{c}{HD\,159595}        \\
\hline
Beam size (arcsec$\times$arcsec) & 0.45$\times$0.39  &  0.42$\times$0.37   &  \ldots   &  \ldots   \\ 
Beam PA (\degr)                  &   -69.7           &     -73.9           &  \ldots   &  \ldots   \\
rms noise ($\mu$Jy~beam$^{-1}$ / mJy~beam$^{-1}$~ch$^{-1}$)    &    19.5       &   5.0   &   \ldots    &  \ldots  \\
\hline
    & \multicolumn{4}{c}{HD\,170116}        \\ 
\hline
Beam size (arcsec$\times$arcsec) & 0.49$\times$0.41  &  0.50$\times$0.41   &  0.52$\times$0.42   &  0.52$\times$0.42   \\ 
Beam PA (\degr)                  &    -70.3          &     -74.4           &      -74.0          &      -74.2          \\
rms noise ($\mu$Jy~beam$^{-1}$ / mJy~beam$^{-1}$~ch$^{-1}$)    &    13.0       &  1.7       &   0.9        & 0.7          \\
\hline
    & \multicolumn{4}{c}{HD\,176497}        \\ 
\hline
Beam size (arcsec$\times$arcsec) & 0.51$\times$0.32  &  0.49$\times$0.31    &  \ldots   &  \ldots   \\ 
Beam PA (\degr)                  &   81.7            &     82.5             &  \ldots   &  \ldots   \\
rms noise ($\mu$Jy~beam$^{-1}$ / mJy~beam$^{-1}$~ch$^{-1}$)    &    13.0       &   1.7   &   \ldots    &  \ldots  \\
\hline
\end{tabular}
\tablefoot{For systems where we detected circumstellar CO gas (even around the companion star, as in the case of HD\,31305), 
the imaging parameters (size and position angle of the synthesised beam, as well as 1 $\sigma$ rms noise levels) are provided 
for all three CO isotopologs in addition to the continuum, while for the other objects the parameters are listed only for the 
continuum and $^{12}$CO measurements. The quoted parameters refer to the naturally weighted data sets for HD\,144277 and HD\,176497, 
the tapered data for HD\,131960 (Gaussian taper with a size of 1\arcsec), and Briggs weighted data (robust=0.5) for the rest of 
the sample. For the $^{12}$CO measurements of HD\,155853 and HD\,159595, we consider the data cubes produced by neglecting the 
shortest baseline visibilities (see Sect.~\ref{sec:co}).
}
\end{center}
\end{table*} 

\FloatBarrier 


\setcounter{section}{2}
\section{The circumstellar disk of HD~31305~B} \label{sec:hd31305b} 
The ALMA continuum observation of HD\,31305 revealed a bright compact source (Fig.~\ref{fig:continuumplots}). To estimate 
its basic parameters we used \texttt{uvmultifit} as in Sect.~\ref{sec:uvmultifit}, adopting both a point source and an 
elliptical Gaussian model for the surface brightness distribution, the latter being proven more suitable based on the BIC 
test. The derived position differs by $\sim$580\,mas from that of the primary A-type component of this binary 
system (the intended target of the measurement). Based on the S/N and the beam size, the nominal expected absolute 
astrometric accuracy of the measurement is $\sim$25\,mas, i.e., even if the actual absolute positional accuracy may be 
twice this value or even worse\footnote{\url{https://help.almascience.org/kb/articles/what-is-the-absolute-astrometric-accuracy-of-alma}}, 
this positional difference is well determined. However, the detected source coincides well with the position of 
HD\,31305\,B. For the latter star, the separation of the millimeter source is 70\,mas or 40\,mas depending on whether the system's 
proper motion is taken from the Gaia\,DR3 or UCAC4 catalog in the calculations. Figure \ref{fig:continuumplots} and \ref{fig:co_31305b} 
display the UCAC-based position of the companion. The observed continuum emission thus probably comes from a disk around 
the companion. We obtained a flux density of 2.37$\pm$0.24\,mJy for the source, where the uncertainty is the quadratic sum 
of the measurement error (0.03\,mJy) and the calibration error (assumed to be 10\%). Assuming optically thin emission and that 
the temperature of the emitting grains is 20\,K, then Eq.~\ref{eq:mdust} yields a dust mass estimate of 1.42$\pm$0.15\,M$_\oplus$. 
Based on the FWHM of the major axis of the fitted Gaussian, we derived a characteristic disk size of 60$\pm$15\,mas, which 
corresponds to 8.6$\pm$2.2\,au at the distance of the system.

By examining the measured CO line data cubes, we found significant emission in the $v_\mathrm{LSR}$ interval from $-$10.4 
to $+$10.0\,km~s$^{-1}$ for all three isotopologs. We used this velocity range to construct moment 0 maps for all three 
lines. As demonstrated by the $^{12}$CO and $^{13}$CO moment 0 maps shown in Fig.~\ref{fig:co_31305b} (left and middle panels), 
the line emission originate from a compact source, which coincides with the position of the continuum source. To estimate the 
integrated line fluxes we used the same method as for HD\,145101 (Sect.~\ref{sec:coimaging}). By using elliptical apertures of 
different sizes, whose shape is identical to the actual beam, we determined the minimum aperture size that includes all the CO 
emission associated with the source and measured the line flux. The uncertainty of the derived flux was estimated by placing 
this aperture at random positions outside the source and calculating the standard deviation of the flux values measured in them. 
This method yielded line fluxes of $F_\mathrm{^{12}CO} = (30.6\pm3.3)\times 10^{-22}$\,W~m$^{-2}$, 
$F_\mathrm{^{13}CO} = (7.6\pm1.3)\times 10^{-22}$\,W~m$^{-2}$, and $F_\mathrm{C^{18}O} = (4.8\pm0.9)\times 10^{-22}$\,W~m$^{-2}$, 
where the quoted uncertainties are the quadratic sum of the measurement error and the calibration error. The spectra obtained 
using the given apertures for the three lines are shown in Fig.~\ref{fig:co_31305b} (right panel). Adopting abundance ratios 
typical of the local interstellar medium \citep[$^{12}$C/$^{13}$C = 77 and $^{16}$O/$^{18}$O = 560,][]{wilson1994}, the measured 
fluxes indicate that both the $^{12}$CO and the $^{13}$CO lines are optically thick. Based on the C$^{18}$O measurement, assuming 
that the line is optically thin and taking into account the above abundance ratio, we obtained a CO disk mass of 
2.05$\pm$0.42$\times$10$^{-2}$\,M$_\oplus$.

Although we know of some debris disks with millimeter-sized dust content only slightly lower than that of HD\,31305\,B, in 
those systems the dust is located at $>$50\,au from the star \citep[e.g.][]{matra2025}. Similar can be said for the gas component, 
as discussed in Sect.~\ref{sec:discussion}, in young debris disks with CO content similar to or even higher than that measured 
for HD\,31305\,B, the gas is located at radial distances of several tens of au. Furthermore, all CO-rich debris disks have been 
found around A-type stars. 

To further explore the nature of HD\,31305\,B, we examined the slope of its SED between 2 and 24\,$\mu$m. As a first step, we 
color corrected the photometric data gathered from the 2MASS, AllWISE, IRC, and SEIP catalogs. Since HD\,31305~A contributes to the 
measured emission in all photometric bands, we subtracted its stellar photosphere model from the measured flux densities. 
By fitting a line to the data points obtained this way, we then derived a mid-IR slope of $\alpha_\mathrm{IR} = -0.94\pm0.03$. This slope 
value, between $-$1.6 and $-$0.3, implies that HD\,31305\,B is probably a class\,II source \citep{lada1987,williams2011}, 
i.e. its disk is protoplanetary in nature. This classification is further supported by the fact that the system belongs to 
the 2.5\,Myr old L1517 subgroup of the Taurus star forming region, and that no such compact debris disk with such a 
high dust and gas content is known.

\setcounter{figure}{0}
\begin{figure*}[h!]
\centering
\includegraphics[width=0.33\textwidth]{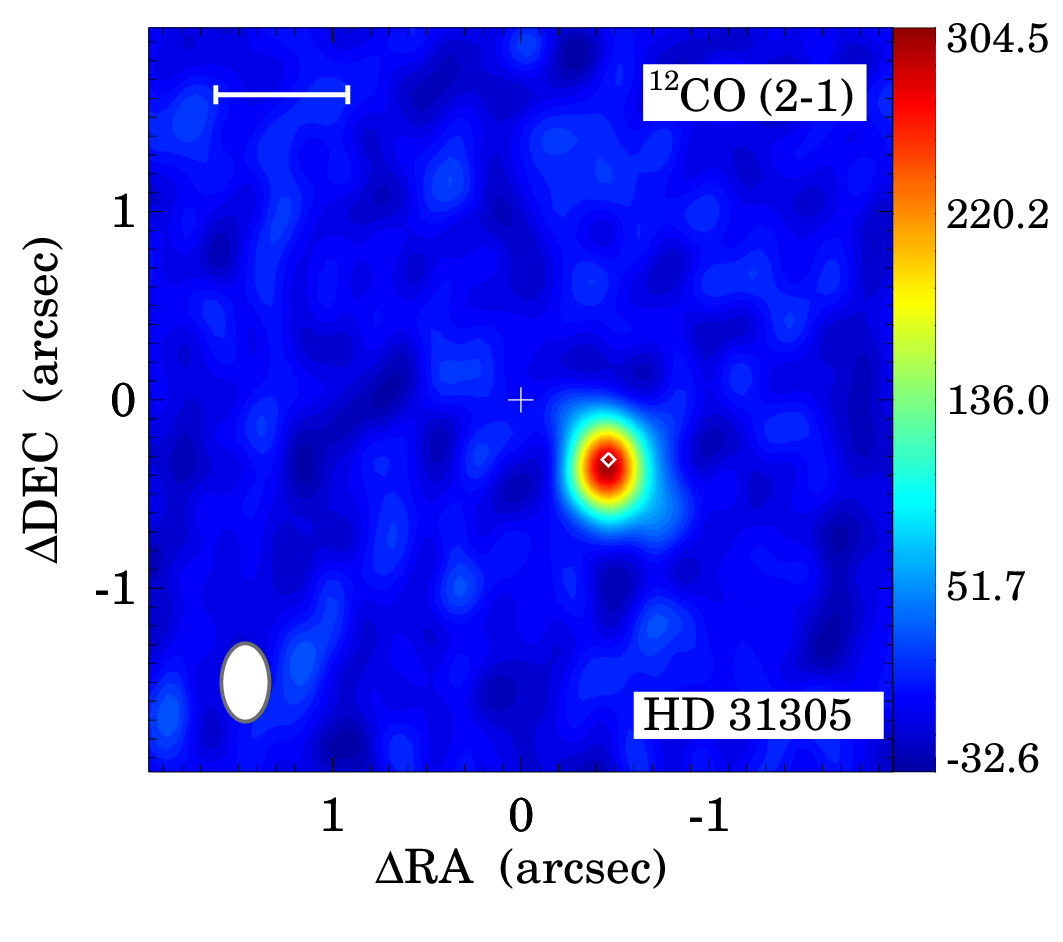}
\includegraphics[width=0.292967\textwidth]{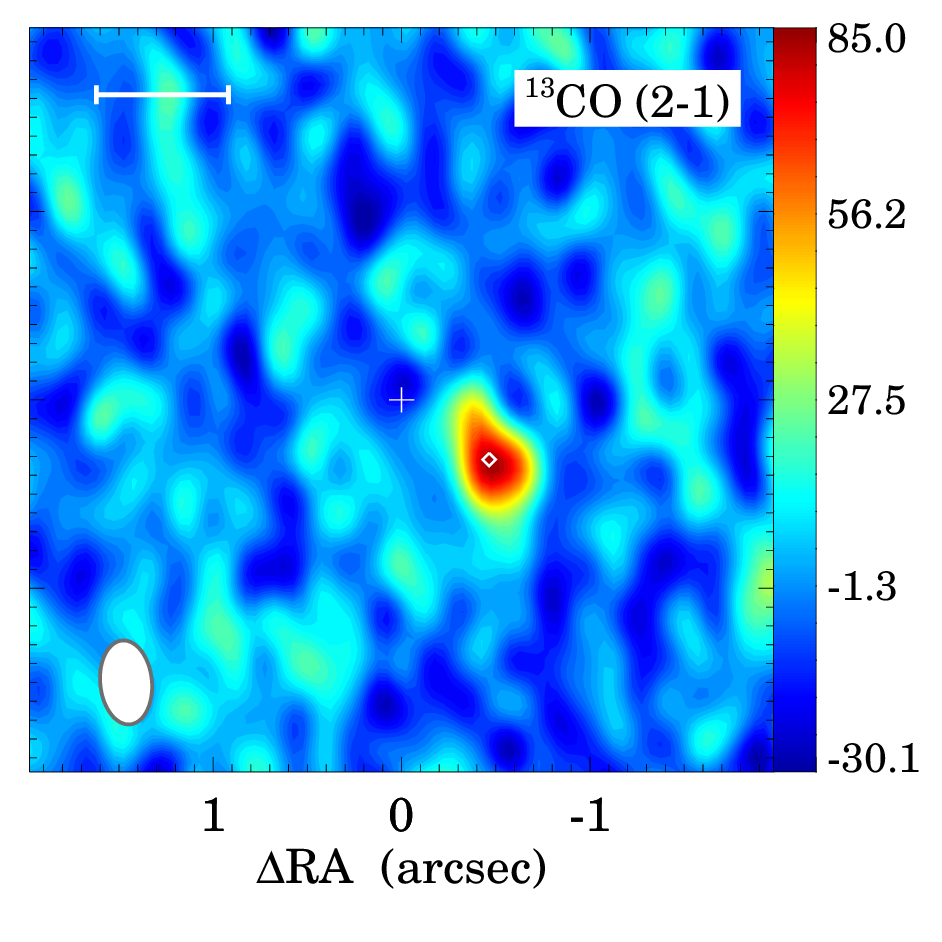}
\includegraphics[width=0.27\textwidth]{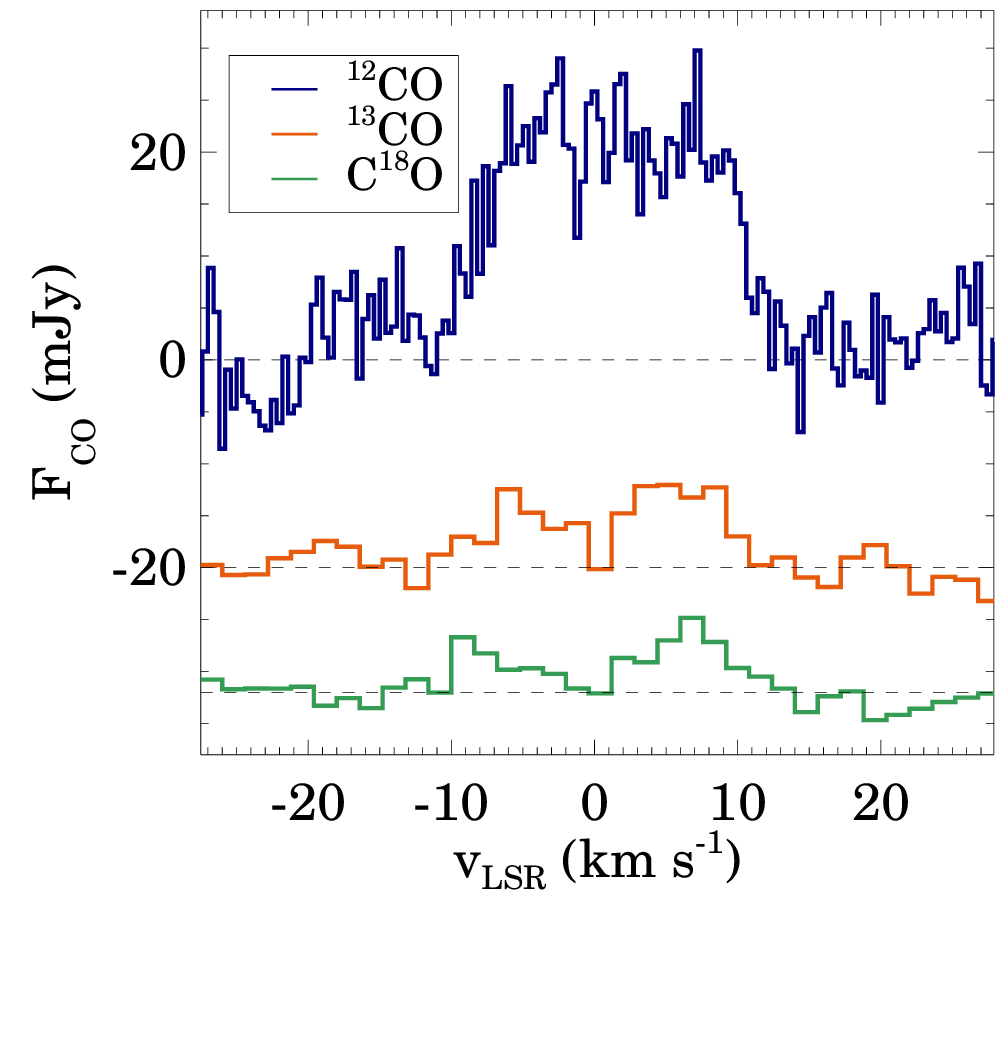}
\caption{Left and middle panels: $^{12}$CO and $^{13}$CO J=2--1 moment 0 maps for HD\,31305\,B, respectively. 
At the bottom left of each panel a filled white ellipse shows the beam size. The length of the horizontal white bars 
correspond to 100\,au. The white plus sign shows the position of HD\,31305\,A while the white diamond displays the position of 
HD\,31305\,B. The color bar units are mJy~beam$^{-1}$~km~s$^{-1}$. The right panel displays the obtained CO spectra in all 
three isotopologs. For clarity, the $^{13}$CO and C$^{18}$O spectra have been shifted downward. The horizontal dashed lines 
show the zero flux levels of the spectra.}
\label{fig:co_31305b}
\end{figure*}
\FloatBarrier

\section{Comparison sample} \label{sec:compsample}
Table~\ref{tab:codisksample} presents the main properties of those young ($<$50\,Myr) systems with dust-rich 
($f_\mathrm{d}>5\times10^{-4}$) debris disks for which CO line observations are available in the literature. 

\onecolumn
{\setlength{\tabcolsep}{1.5mm}
\begin{landscape}
\begin{longtable}{lcccccccccccccc} 
\caption{Comparison sample.}
\label{tab:codisksample} \\
\hline\hline
Identifier & D     &  Group &  Age &  SpT & $L_*$        & $M_*$       & $T_\mathrm{d}$ & $L_\mathrm{disk}$ / $L_*$  & $R_\mathrm{disk}$ & $W_\mathrm{disk}$ & $L_\mathrm{^{12}CO}$ & $M_\mathrm{CO}$ & Refs. & Label \\ 
           & (pc)  &   & (Myr) &   &  (L$_\odot$) & (M$_\odot$) &    (K)         &                            &    (au)           &    (au)           &   (W~m$^{-2}$) &  ($M_\oplus$) &  &   \\
\hline
\endfirsthead
\caption{continued}\\
\hline
Identifier & D     &  Group &  Age &  SpT & $L_*$        & $M_*$       & $T_\mathrm{d}$ & $L_\mathrm{disk}$ / $L_*$  & $R_\mathrm{disk}$ & $W_\mathrm{disk}$ & $L_\mathrm{^{12}CO}$ & $M_\mathrm{CO}$ & Refs. & Label \\ 
           & (pc)  &   & (Myr) &   &  (L$_\odot$) & (M$_\odot$) &    (K)         &                            &    (au)           &    (au)           &   (W~m$^{-2}$) &  ($M_\oplus$) &  &   \\
\hline	 
\endhead
\hline
\endfoot
\hline
\endlastfoot  
     RZ\,Psc & 184.7 &       \ldots &             30$\pm$10 &        G8V &     0.80 &     1.10 &                  500 &              7.0e-02 &     6.0 &       \ldots &             $<$3.23$\times10^{16}$ &              $<$1.0$\times10^{-5}$ &              43 &  \ldots   \\ 
     49\,Cet &  57.1 &     ARG &             48$\pm$10 &        A2V &    15.61 &     1.95 &               155/56 &      2.0e-04/7.0e-04 &   136.0 &   147.0 &       1.16$\pm$0.12$\times10^{18}$ &   7.1$^{+1.2}_{-1.1}\times10^{-3}$ &             2,22,26,30,48 &  a  \\ 
   HD\,15115 &  48.7 &     THA &             37$\pm$11 &        F4V &     3.60 &     1.40 &                   62 &              5.1e-04 &    93.0 &    21.0 &             $<$4.80$\times10^{15}$ &              $<$1.2$\times10^{-6}$ &            17,22,35,36,48 &  \ldots   \\ 
   HD\,15745 &  71.5 &    BPMG &              21$\pm$4 &        F0V &     4.20 &     1.40 &               147/81 &      3.5e-04/1.9e-03 &    65.0 &   <50.0 &             $<$4.71$\times10^{16}$ &              $<$1.2$\times10^{-5}$ &                  10,22,24 &  \ldots   \\ 
   HD\,21997 &  69.6 &     COL &              36$\pm$8 &        A4V &    10.49 &     1.76 &                   61 &              5.7e-04 &    94.0 &    52.0 &       9.15$\pm$0.93$\times10^{17}$ &   4.8$^{+0.8}_{-0.9}\times10^{-2}$ &                2,22,26,48 &  b  \\ 
   HD\,32297 & 129.3 &       \ldots &                 $<$30 &        A7V &     7.98 &     1.65 &               234/82 &      8.7e-04/6.2e-03 &   122.3 &    62.0 &       1.58$\pm$0.15$\times10^{18}$ &   5.0$^{+1.6}_{-1.5}\times10^{-2}$ &           1,2,16,22,30,48 &  c  \\ 
   HD\,36546 &  99.9 &  118TAU &              10$\pm$3 &        A2V &    13.86 &     1.86 &              570/135 &     \ldots/4.0e-03 &    70.0 &   150.0 &       2.45$\pm$0.25$\times10^{18}$ &   3.2$^{+1.2}_{-1.2}\times10^{-3}$ &               13,22,37,48 &  d  \\ 
$\beta$\,Pic &  19.4 &    BPMG &              21$\pm$4 &        A6V &     8.65 &     1.75 &                   85 &              2.6e-03 &   105.0 &    92.0 &       1.58$\pm$0.18$\times10^{17}$ &   4.8$^{+1.2}_{-1.4}\times10^{-5}$ &        2,4,11,21,22,38,47 &  e  \\ 
   HD\,48370 &  35.9 &     COL &              36$\pm$8 &        K0V &     0.77 &     0.96 &                   41 &              5.6e-04 &    88.3 &    37.1 &             $<$4.01$\times10^{16}$ &              $<$1.2$\times10^{-5}$ &                2,20,28,36 &  \ldots   \\ 
   HD\,61005 &  36.4 &     ARG &             48$\pm$10 &       G8Vk &     0.64 &     0.92 &               109/52 &      2.9e-04/3.0e-03 &    72.6 &    38.0 &             $<$1.05$\times10^{16}$ &              $<$4.0$\times10^{-6}$ &                 1,2,22,36 &  \ldots   \\ 
   HD\,95086 &  86.4 &     CAR &             28$\pm$11 &        A8V &     6.55 &     1.58 &               184/54 &      1.5e-04/1.3e-03 &   206.0 &   180.0 &             $<$1.43$\times10^{16}$ &              $<$4.0$\times10^{-6}$ &             2,19,22,26,48 &  f  \\ 
   HD\,98363 & 137.0 &     LCC &              15$\pm$3 &        A4V &    12.02 &     1.83 &              295/112 &      2.8e-04/6.4e-04 &       \ldots &       \ldots &             $<$6.22$\times10^{16}$ &              $<$9.5$\times10^{-6}$ &                   3,29,48 &   \ldots  \\ 
  HD\,106906 & 102.2 &     LCC &              15$\pm$3 &        F5V &     7.02 &     2.71 &               108/53 &      1.5e-03/3.5e-04 &   100.0 &    80.0 &             $<$3.84$\times10^{16}$ &              $<$8.3$\times10^{-6}$ &          1,10,22,34,40,48 &  \ldots   \\ 
    HR\,4796 &  70.7 &     TWA &              10$\pm$3 &        A0V &    21.42 &     2.10 &                  108 &              4.6e-03 &    77.8 &    14.8 &             $<$4.60$\times10^{16}$ &              $<$3.5$\times10^{-6}$ &             9,10,22,39,48 &  g  \\ 
  HD\,109832 & 107.1 &     LCC &              15$\pm$3 &        F1V &     5.37 &     1.46 &               186/92 &      2.5e-04/5.1e-04 &       \ldots &       \ldots &             $<$3.49$\times10^{16}$ &              $<$6.4$\times10^{-6}$ &                1,19,29,48 & \ldots    \\ 
  HD\,110058 & 129.7 &     LCC &              15$\pm$3 &        A5V &     8.71 &     1.67 &              499/112 &      2.4e-05/1.4e-03 &    30.9 &    48.9 &       1.41$\pm$0.18$\times10^{17}$ &   9.3$^{+3.5}_{-2.2}\times10^{-3}$ &                  2,3,6,48 &  h  \\ 
  HD\,111161 & 109.2 &     LCC &              15$\pm$3 &        A5V &     9.05 &     1.68 &                  142 &              7.2e-04 &    71.3 &       \ldots &             $<$4.61$\times10^{16}$ &              $<$1.2$\times10^{-5}$ &                   8,12,48 &  i  \\ 
  HD\,111520 & 107.8 &     LCC &              15$\pm$3 &      F5/6V &     2.66 &     1.32 &               107/45 &      1.4e-03/1.5e-03 &    76.0 &    50.0 &             $<$5.13$\times10^{16}$ &              $<$1.3$\times10^{-5}$ &             1,12,22,36,48 &  \ldots   \\ 
  HD\,112810 & 133.2 &     LCC &              15$\pm$3 &      F3/5V &     3.54 &     1.36 &               227/58 &      1.9e-04/1.1e-03 &    90.0 &  <200.0 &             $<$6.86$\times10^{16}$ &              $<$1.8$\times10^{-5}$ &             1,12,22,34,48 &  \ldots   \\ 
  HD\,113556 & 100.3 &     LCC &              15$\pm$3 &        F2V &     4.42 &     1.41 &               268/60 &      8.9e-05/4.4e-04 &   110.0 &  <400.0 &             $<$3.33$\times10^{16}$ &              $<$8.5$\times10^{-6}$ &             3,12,22,36,48 &  \ldots   \\ 
  HD\,113766 & 108.6 &     LCC &              15$\pm$3 &      F3/5V &     4.00 &     1.50 &               490/15 &      3.3e-02/7.0e-04 &    30.0 &   <60.0 &             $<$4.88$\times10^{16}$ &              $<$1.3$\times10^{-5}$ &         12,22,33,34,42,48 &  \ldots   \\ 
  HD\,114082 &  94.9 &     LCC &              15$\pm$3 &        F3V &     3.75 &     1.47 &                  110 &              4.3e-03 &    38.0 &   <40.0 &             $<$1.66$\times10^{16}$ &              $<$4.9$\times10^{-6}$ &                1,10,22,36 &  \ldots   \\ 
  HD\,115600 & 108.9 &     LCC &              15$\pm$3 &        F2V &     4.75 &     1.48 &                  109 &              2.5e-03 &    90.0 &  <200.0 &             $<$4.58$\times10^{16}$ &              $<$1.2$\times10^{-5}$ &                1,12,22,48 &  \ldots   \\ 
  HD\,117214 & 107.1 &     LCC &              15$\pm$3 &        F6V &     5.35 &     1.50 &                  111 &              3.1e-03 &    42.0 &   <50.0 &             $<$2.11$\times10^{16}$ &              $<$5.2$\times10^{-6}$ &                1,10,22,48 &  \ldots   \\ 
  HD\,121191 & 131.9 &     LCC &              15$\pm$3 &        A7V &     7.58 &     1.62 &              555/118 &      2.1e-03/2.6e-03 &    55.0 &    54.0 &       3.43$\pm$0.34$\times10^{17}$ &   5.6$^{+3.3}_{-2.8}\times10^{-3}$ &          2,10,22,29,44,48 &  j  \\ 
  HD\,121617 & 117.5 &     UCL &              16$\pm$2 &        A2V &    14.51 &     1.90 &                  105 &              4.8e-03 &    78.0 &    60.0 &       1.61$\pm$0.17$\times10^{18}$ &   2.6$^{+0.5}_{-0.4}\times10^{-2}$ &             2,22,25,29,48 &  k  \\ 
  HD\,129590 & 135.9 &     UCL &              16$\pm$2 &        G3V &     3.00 &     1.30 &                   89 &              7.7e-03 &    79.0 &    70.0 &       9.51$\pm$2.38$\times10^{16}$ &   9.5$^{+6.5}_{-6.5}\times10^{-5}$ &                   1,10,22 &  x  \\ 
  HD\,131488 & 151.4 &     UCL &              16$\pm$2 &        A2V &    13.64 &     1.87 &               570/94 &      2.8e-03/2.7e-03 &    92.0 &    46.0 &       1.64$\pm$0.17$\times10^{18}$ &   9.3$^{+1.2}_{-1.2}\times10^{-2}$ &               22,29,44,48 &  l  \\ 
  HD\,131835 & 129.1 &     UCL &              16$\pm$2 &        A4V &     9.81 &     1.72 &               176/71 &      8.2e-04/2.2e-03 &    83.7 &    87.0 &       1.22$\pm$0.13$\times10^{18}$ &   1.6$^{+0.2}_{-0.3}\times10^{-2}$ &             2,12,22,27,48 &  m  \\ 
  HD\,134888 & 111.8 &     UCL &              16$\pm$2 &      F3/5V &     3.27 &     1.36 &               166/67 &      3.0e-04/1.0e-03 &       \ldots &       \ldots &             $<$6.90$\times10^{16}$ &              $<$1.8$\times10^{-5}$ &                   1,12,48 & \ldots    \\ 
  HD\,135953 & 127.3 &     UCL &              16$\pm$2 &        F5V &     2.42 &     1.22 &                   61 &              9.0e-04 &       \ldots &       \ldots &             $<$8.05$\times10^{16}$ &              $<$2.1$\times10^{-5}$ &                   1,12,48 & \ldots    \\ 
  HD\,138813 & 135.9 &      US &              10$\pm$3 &        A1V &    21.74 &     2.11 &               153/75 &      7.4e-04/7.4e-04 &   120.0 &   130.0 &       2.39$\pm$0.27$\times10^{18}$ &   2.0$^{+0.4}_{-0.4}\times10^{-2}$ &               12,22,46,48 &  n  \\ 
  HD\,141569 & 111.3 &       \ldots &   5.9$^{+1.9}_{-0.6}$ &       A0Ve &    21.88 &     2.06 &          \ldots &      9.0e-03 &   130.0 &   240.0 &       1.14$\pm$0.11$\times10^{19}$ &   1.0$^{+0.5}_{-0.2}\times10^{-1}$ &                   5,23,45 &  o  \\ 
  HD\,142446 & 135.0 &     UCL &              16$\pm$2 &        F3V &     3.73 &     1.39 &               239/65 &      1.8e-04/7.7e-04 &   110.0 &    80.0 &             $<$9.05$\times10^{16}$ &              $<$2.3$\times10^{-5}$ &                1,12,22,48 &  \ldots   \\ 
     NO\,Lup & 132.8 &   Lupus &                  1--3 &         K7 &     0.29 &     0.70 &              117/$^{25-}_{120}$ &  3.4e-03/$^\mathrm{2.0e-07-}_\mathrm{4.0e-03}$ & $<$56.0 &       \ldots &       7.06$\pm$1.70$\times10^{17}$ &   4.9$^{+1.1}_{-1.1}\times10^{-5}$ &                   7,14,15 &  \ldots   \\ 
  HD\,143675 & 135.0 &     UCL &              16$\pm$2 &        A7V &     8.40 &     1.67 &              374/127 &      1.8e-04/4.1e-04 &    52.4 &       \ldots &             $<$1.68$\times10^{16}$ &              $<$5.0$\times10^{-6}$ &              3,8,10,29,48 &  p  \\ 
  HD\,145263 & 141.2 &      US &              10$\pm$3 &        F2V &     4.98 &     1.50 &                  265 &              1.5e-02 &       \ldots &       \ldots &             $<$1.05$\times10^{17}$ &              $<$2.7$\times10^{-5}$ &               12,32,34,48 &  \ldots   \\ 
  HD\,145560 & 120.8 &     UCL &              16$\pm$2 &        F5V &     3.27 &     1.35 &               103/53 &      1.8e-03/2.1e-03 &    76.0 &    50.0 &             $<$4.43$\times10^{16}$ &              $<$1.1$\times10^{-5}$ &                1,12,22,48 &  \ldots   \\ 
  HD\,145880 & 123.0 &     UCL &              16$\pm$2 &        A0V &    19.74 &     2.06 &               196/70 &      2.9e-04/8.3e-04 &       \ldots &       \ldots &             $<$9.88$\times10^{16}$ &              $<$2.0$\times10^{-5}$ &                1,19,29,48 & \ldots    \\ 
  HD\,146181 & 127.0 &     UCL &              16$\pm$2 &        F6V &     2.81 &     1.28 &               124/71 &      9.6e-04/2.0e-03 &    74.0 &   <90.0 &             $<$5.34$\times10^{16}$ &              $<$1.4$\times10^{-5}$ &             1,12,22,36,48 &  \ldots   \\ 
  HD\,146897 & 131.6 &      US &              10$\pm$3 &      F2/3V &     4.78 &     1.42 &               304/94 &      3.0e-04/5.3e-03 &    82.0 &   <90.0 &       9.57$\pm$2.39$\times10^{16}$ &  4.0$^{+59.1}_{-1.5}\times10^{-5}$ &              2,3,12,22,48 &  y  \\ 
  HD\,156623 & 108.0 &     UCL &              16$\pm$2 &        A2V &    13.28 &     1.86 &              605/123 &      1.6e-03/3.4e-03 &    77.8 &   119.0 &       1.27$\pm$0.13$\times10^{18}$ &   2.1$^{+0.4}_{-0.4}\times10^{-3}$ &               12,19,29,48 &  q  \\ 
  HD\,172555 &  28.8 &    BPMG &              21$\pm$4 &        A7V &     7.84 &     1.66 &                  250 &              7.5e-04 &     7.5 &       \ldots &       1.30$\pm$0.23$\times10^{16}$ &   8.3$^{+3.8}_{-3.8}\times10^{-6}$ &                  41,42,48 &  r  \\ 
  HD\,181327 &  47.7 &    BPMG &              21$\pm$4 &        F6V &     2.83 &     1.41 &                   78 &              2.6e-03 &    81.3 &    16.0 &       6.49$\pm$1.13$\times10^{15}$ &   1.9$^{+1.9}_{-0.6}\times10^{-6}$ &          2,18,22,35,36 &  z  \\ 
  HD\,191089 &  50.0 &    BPMG &              21$\pm$4 &        F5V &     2.73 &     1.20 &                   92 &              1.5e-03 &    44.8 &    16.0 &             $<$1.15$\times10^{16}$ &              $<$2.9$\times10^{-6}$ &               10,22,35,36 &  \ldots   \\ 
 CP-72\,2713 &  36.7 &    BPMG &              21$\pm$4 &       K7Ve &     0.19 &     0.80 &                   43 &              1.1e-03 &   130.0 &   100.0 &             $<$2.68$\times10^{16}$ &              $<$6.9$\times10^{-6}$ &               22,31,36,48 &  \ldots   \\ 
\hline
\end{longtable}
\tablefoot{With the exception of $\beta$\,Pic, whose distance data are from the Hipparcos catalog \citep{vanleeuwen2007}, the distance estimates are taken from \citet{bailerjones2021} using their geometric approach that is based only on the Gaia~EDR3 parallaxes. Group membership information is based on the BANYAN 
$\Sigma$ model \citep{gagne2018} in most cases, while for HD\,48370 and NO\,Lup are from \citet{torres2008} and \citet{lovell2021b}, respectively.
For stars known to be members of stellar associations, we have adopted the ages of the specific groups. The estimated ages of the TWA (TW\,Hya), 118TAU (118\,Tau), 
US (Upper Scorpius), UCL (Upper Centaurus Lupus), and LCC (Lower Centaurus Crux) groups are from \citet{gagne2018}. For the BPMG ($\beta$\,Pic moving group), 
THA (Tucana-Horologium association), CAR (Carina), and COL (Columba) moving groups, the ages are taken from the recent work of \citet{gratton2024}. 
For HD\,32297 and HD\,141569 we have used their individual age estimates from \citet{kalas2005} and \citet{wichittanakom2020}, respectively. 
In the case of RZ\,Psc, the quoted age combines the results of \citet{punzi2018} and \citet{potravnov2019}. The $^{12}$CO line luminosities 
($L_\mathrm{^{12}CO}$) given in the 12$^\mathrm{th}$ column refer to the J=2--1 rotational transition, except for NO\,Lup where the 
3--2 line luminosity is quoted. The last column shows which label we use to mark the given object in Figs.~\ref{fig:rdiskls}--\ref{fig:primordial}.
References to the stellar and disk parameters presented are summarized in the 14$^\mathrm{th}$ column of the table.
\tablebib{1. \citet{ballering2013} - 2. \citet{cataldi2023} - 3. \citet{chen2014} - 4. \citet{dent2014} - 5. \citet{difolco2020} - 6. \citet{hales2022} - 7. \citet{hardy2015} - 8. \citet{hom2020} - 9. \citet{kennedy2018} - 10. \citet{kral2020} - 11. \citet{lacour2021} - 12. \citet{lieman-sifry2016} - 13. \citet{lisse2017} - 14. \citet{lovell2021} - 15. \citet{lovell2021b} - 16. \citet{macgregor2018} - 17. \citet{macgregor2019} - 18. \citet{marino2016} - 19. \citet{marino2020} - 20. \citet{marshall2021} - 21. \citet{matra2017b} - 22. \citet{matra2025} - 23. \citet{meeus2012} - 24. \citet{moor2011a} - 25. \citet{moor2011b} - 26. \citet{moor2015a} - 27. \citet{moor2015b} - 28. \citet{moor2016} - 29. \citet{moor2017} - 30. \citet{moor2019} - 31. \citet{moor2020} - 32. \citet{moor2021} - 33. \citet{olofsson2013} - 34. \citet{paegert2021} - 35. \citet{pawellek2021} - 36. \citet{pearce2022} - 37. \citet{rebollido2022} - 38. \citet{rhee2007} - 39. \citet{riviere-marichalar2014} - 40. \citet{rodet2017} - 41. \citet{schneiderman2021} - 42. \citet{su2020} - 43. \citet{su2023} - 44. \citet{vican2016} - 45. \citet{wichittanakom2020} - 46. Xiaoming et al. (in prep.) - 47. \citet{zwintz2019} - 48. This work.}}
\end{landscape}
}


\end{appendix}
\end{document}